\documentclass{aa}

\usepackage{newtxtext,newtxmath}

\usepackage[T1]{fontenc}

\DeclareRobustCommand{\VAN}[3]{#2}
\let\VANthebibliography\thebibliography
\def\thebibliography{\DeclareRobustCommand{\VAN}[3]{##3}\VANthebibliography}

\usepackage{graphicx}	
\usepackage{amsmath}
\usepackage{caption}
\usepackage{subcaption}
\usepackage[pdfencoding=auto,psdextra]{hyperref}
\hypersetup{
    colorlinks=true,
    linkcolor=blue,
    filecolor=magenta,      
    urlcolor=blue,
    citecolor=blue
}
\usepackage{color}

\newcommand{\dif}{{\mathrm d}}
\newcommand{\sn}{{\epsilon}}

\newcommand{\Mpc}{\, h^{-1} \, {\rm Mpc}}

\begin{document}

   \title{Towards an optimal marked correlation function analysis for the detection of modified gravity}
   
    \titlerunning{Optimal marked correlation function analysis}

   \author{M. K\"archer\inst{1,2}\thanks{\email{martin.karcher@lam.fr}},
          J. Bel\inst{2},
          S. de la Torre\inst{1}
          }

   \institute{Aix Marseille Univ, CNRS, CNES, LAM, Marseille, France
         \and Aix Marseille Univ, Universit\'e de Toulon, CNRS, CPT, Marseille, France
        }

  \abstract{Modified gravity (MG) theories have emerged as a promising alternative to explain the late-time acceleration of the Universe. However, the detection of MG in observations of the large-scale structure remains challenging due to the screening mechanisms that obscure any deviations from General Relativity (GR) in high-density regions. The marked two-point correlation function, which is particularly sensitive to environment, offers a promising approach to enhance the discriminating power in clustering analysis and potentially detect MG signals. This work investigates novel marks based on large-scale environment estimates but also that exploit the anti-correlation between objects in low- and high-density regions.  This is the first time that the propagation of discreteness effects in marked correlation functions is investigated in depth, as in contrast to standard correlation functions, the density-dependent marked correlation function as estimated from catalogues is affected in a non-trivial way by shot noise. We assess the performance of various marks to distinguish GR from MG. This is achieved through the use of the ELEPHANT suite of simulations, which comprise five realisations of GR and two different MG theories: $f(R)$ and nDGP. In addition, discreteness effects are thoroughly studied using the high-density Covmos catalogues. We establish a robust method to correct for shot-noise effects that can be used in practical analyses. This methods allows the recovery of the true signal, with an accuracy below $5\%$ over the scales of $5\Mpc$ up to $150\Mpc$. We find that such correction is absolutely crucial to measure the amplitude of the marked correlation function in an unbiased manner. Furthermore, we demonstrate that marks that anti-correlate objects in low- and high-density regions are among the most effective in distinguishing between MG and GR, and uniquely provide visible deviations on large scales, up to about $80\Mpc$. We report differences in the marked correlation function between $f(R)$ with $|f_{R0}|=10^{-6}$ and GR simulations of the order of 3-5$\sigma$ in real space. The redshift-space monopole exhibits similar features and performances as the real-space marked correlation function. The combination of the proposed $\tanh$-mark with shot-noise correction paves the way towards an optimal approach for the detection of MG in current and future galaxy spectroscopic surveys.}

   \keywords{large-scale structure of Universe}

   \maketitle
\section{Introduction}

The seminal works of \citet{Riess1998AJ} and \citet{Perlmutter1999ApJ} revived the cosmological constant $\Lambda$ as a form of dark energy to explain the late-time accelerated expansion of the Universe. Together with cold dark matter (CDM), this settled the $\Lambda$CDM model as the current concordance model of cosmology. Upon closer examination, however, the $\Lambda$CDM model is found to exhibit certain inherent problems. On the theoretical side, the fine-tuning problem of $\Lambda$, as extensively studied in \citet[][]{Martin2012CRPhy}, represents a significant challenge. On the observational side, most recent cosmological results show a growing tension between early- and late-time measurements of the Hubble constant $H_0$, respectively extracted from the cosmic microwave background (CMB) anisotropies \citep{Planck2020A&A} and local distance ladders \citep[][]{Riess2022ApJ}. Another source of contention, although not as significant, comes from an apparent mismatch in the measured variance of matter fluctuations $\sigma_8$, between early- \citep{Planck2020A&A} and late-time large-scale structure measurements \citep{Troester2020A&A}.

In order to address the aforementioned issues, numerous attempts have been made at the theoretical level. Of particular interest to circumvent the introduction of a cosmological constant or dark energy component in the first place, are modified gravity (MG) theories \citep[see][]{Clifton2012PhR}. One of the most popular modifications is the theory of inflation by \citet{GuthPhysRevD} accommodated with scalar-tensor models to resolve the classical flatness and horizon problem in $\Lambda$CDM. MG theories are commonly defined and compared through their respective action. Possibly the most straightforward extension to the Einstein-Hilbert action in GR consists in replacing the Ricci scalar by a free function of it, dubbed $f(R)$ theories \citep[see][for a review]{DeFelice2010}. The latter have been subjected to comprehensive analysis, resulting in tight constraints on viable $f(R)$ functions to ensure the realisation of accelerated expansion without the necessity of a cosmological constant, while simultaneously satisfying solar system GR tests \citep{Cognola2008PhRvD}. Of particular importance is the $f(R)$ model proposed by \citet{Hu2007PhRvD}, which simultaneously realises accelerated expansion and evades solar system tests through the use of a so-called \textit{screening mechanism}

For MG theories to be a viable replacement or extension to GR they have to fulfil very stringent tests coming from solar system observations \citep[see][]{Bertotti2003Natur, Williams2004PhRvL, Williams2012CQGra_lunar_ranging}. On larger scales, the situation is more complex but there is an increasing effort to tighten constraints on MG theories by using CMB data \citep{Ade2016A&A_MG_Planck_constraints} in combination with large-scale structure and supernovae observables \citep{Lombriser2009PhRvD, Battye2018PhRvD_FofR_constraints}. 
In order for a modification to standard gravity to shroud its effects on small scales to recover GR, screening mechanisms are invoked. In general terms, the screening mechanism describes a suppression of any fifth force to a negligible level such that gravity follows GR in certain environments. Screening can happen in different ways such as the chameleon screening in scalar-tensor-theories of gravity \citep[see][]{Khoury2004PhRvD, Khoury2004PhRvL}. It also emerges in some $f(R)$ theories due to the equivalence between scalar-tensor and $f(R)$ theories \citep[][]{Satiriou2010}. The DGP gravity model, originally developed by \citet{Dvali2000PhLB}, exhibits the screening mechanism first introduced by \citet{Vainshtein1972PhLB}. A third popular screening mechanism by \citet{Damour1994NuPhB} is present in the symmetron model \citep{Hinterbichler2010PhRvL}. For a detailed description of the field of screening mechanisms we refer the reader to \citet{Brax2022PhRvD}. Intuitively, screening mechanisms in the cosmological context, particularly the chameleon one, can be understood as a density dependency, where modification to GR should appear only in low-density regions compared to the mean density of the Universe. In high-density regions, inside galaxies or stellar systems for instance, any modification should be negligible. This imprints a fundamental environmental dependency on the clustering of matter predicted in those theories.

Since modifications to GR are expected to be small, the observational detection of MG on cosmological scales poses a major challenge. \citet{Guzzo2008Nature} advocated the use of the growth rate of structure $f$ measured from redshift-space distortions (RSD) in the galaxy clustering pattern as an indicator of the validity of GR in the large-scale structure.
Since then, $f$ has become a quantity of major interest and has been measured in large galaxy redshift surveys \citep[e.g.][]{Blake2011MNRAS_WiggleZ_measure_f, Beutler2012MNRAS_6dF_measure_f, delaTorre2013A&A_VIPERS_measure_f, Bautista2021MNRAS_BOSS_measure_f}. It is now a standard probe that will be measured by ongoing surveys, in particular the dark energy spectroscopic instrument (DESI) \citep{DESI2016arXiv161100036D_initial_paper} and \emph{Euclid} mission \citep{2024arXiv_Mellier_Euclid_Overview} with an exquisite precision. It is worth mentioning the $E_g$ statistic developed by \citet{Zhang2007PhRvL}, a mixture of galaxy clustering and weak lensing measurements to probe the properties of the underlying gravity theory and that has been measured \citep[e.g.][]{Reyes2010Natur, delaTorre2017A&A_measure_EG, Jullo2019A&A_measure_EG, Blake2020A&A_measure_EG}. Other quantities that can in principle be measured from observations are the gravitational slip parameter $\eta$ and the growth index $\gamma$ \citep[see][for a review]{Ishak2019LRR}. At the present time, any of the aforementioned observables has enabled the detection of a deviation from the standard gravitational field.

In order to improve on existing approaches and to exploit the additional environmental dependency of MG in clustering analyses, \citet{White2016JCAP} proposed the marked correlation function as a tool to increase the difference in the clustering signal between MG and GR. In that case, the marked correlation function is a weighted correlation function normalised to the unweighted correlation function, and where object weights or marks, are a function of the local density. The latter is estimated from the density field inferred by dark matter or galaxies. With this methodology, \citet{Aguayo2018MNRAS}, \citet{Armijo2018MNRAS}, and \citet{Alam2021JCAP} investigated  marked correlation functions in N-body simulations of MG.  In addition to examining different mark functions based on density, they also considered marks based on the local gravitational potential or the host halo mass of the galaxy. They observed significant differences between MG and GR for marks based on density on small scales, below about $20~\Mpc$. \citet{White2009MNRAS} also showed the potential of marked correlation functions to break the degeneracy between of HOD and cosmological parameters.
Similar approaches using weighted statistics or transformation of the density field have further been proposed. \citet{Llinares2017MNRAS} used logarithmic transformations of the density field and computed power spectra of the transformed field in N-Body simulations to improve on the detection of MG. Boosting the constraining power in cosmological parameter inference using power spectra has been shown by using Fisher forecasts by \citet{Valogiannis2018PhRvD}, where they compare the Fisher boost using the field transformation of \citet{Llinares2017MNRAS}, the clipping strategy to mask out high-density regions \citep{Simpson2011PhRvL, Simpson2013PhRvD_clipping2} and the mark proposed in \citet{White2016JCAP}. Another application of clipping has been done by \citet{Lombriser2015PhRvL} to the power spectrum in order to better detect $f(R)$ theories with chameleon screening. Recently, the use of marked power spectra has been extended to constrain massive neutrinos \citep{Massara2021PhRvL} and tighten constraints on cosmological parameters \citep[][]{Yang2020ApJ, Xiao2022MNRAS_Cosmo_mCF}.

While a lot of effort on marked statistics for MG has been carried out on simulation, there have been several applications to observational data. \citet{Satpathy2019MNRAS}, for the first time, measured marked correlation functions from observations in the context of MG. They used the proposed original mark introduced by \citet{White2016JCAP} and investigated the monopole and quadrupole of the marked correlation function measured over the LOWZ sample of the Sloan Digital Sky Survey (SDSS) DR12 dataset \citep{Alam2015ApJS}. They could not detect significant differences between MG and GR and they attributed this to modelling uncertainties of the two-point correlation function (2PCF) on scales of $6\Mpc<s<69\Mpc$. \citet{Armijo2024MNRAS_HOD_mCF_paper2} applied the strategy introduced in \citet{Armijo2024MNRAS_HOD_mCF_paper1} to LOWZ and CMASS catalogues of SDSS thereby incorporating uncertainties of the HOD on the projected weighted clustering. They compare predictions from GR and $f(R)$ but find no significant differences, both fit the LOWZ data and are within the uncertainties for the investigated scales between $~0.5$$\Mpc$ and 40$\Mpc$. For the CMASS catalogue the predictions for both GR and $f(R)$ models fail to properly follow the data in the first place.

A number of the issues encountered in the literature regarding the use of marked correlation functions to distinguish MG from GR can be identified as arising from two main sources. The first is the choice of the mark function, which, in the majority of cases, results in significant differences on small scales only. On those scales, a thorough theoretical modelling is difficult as a proper inclusion of non-linear effects of redshift-space distortions is needed as well. The second issue is the propagation of discreteness effects in the mark estimation, i.e. computing the local density from a finite point set, into the measurement of the marked correlation function.  To the best of our knowledge, this has not been done so far and can lead to biased measurements if not accounted for. The present work therefore aims at identifying an optimal mark function that is able to significantly discriminate GR from MG on larger scales where theoretical modelling is more tractable. For this, we develop new ways to include environmental information into weighted statistics as well as investigating new algebraic functions of the density contrast to be used as a mark. Furthermore, we investigate the discreteness effects and devise a new methodology to correct marked correlation function measurements for the bias induced by estimating density-dependent marks on discrete point sets. We demonstrate that by applying this methodology we are able to robustly measure the amplitude of marked correlation function and mitigate possible artefacts in the subsequent analysis of MG signatures.

This article is structured as follows. Section~\ref{sec:modified_gravity} describes the $f(R)$ and nDGP gravity models that are later investigated and tested. Section~\ref{sec:basics} introduces the basics of weighted two-point statistics and marked correlation function. Section~\ref{sec:sim_baseline} presents the MG simulations used in this work and measurements of unweighted statistics, which serve as a reference for comparison with the marked correlation function. Section~\ref{sec:what_mark} presents new marks to be used in the analysis of MG. This is followed in Section~\ref{sec:imapct_sn} by the study of the effects of shot noise in weighted two-point statistics. Section~\ref{sec:Results} shows the main results of this article, which are obtained by applying the previously-defined methodology to MG simulations. Section~\ref{sec:discussion} comprises a discussion on the optimal methodology for marked correlation function and conclusions are provided in Section~\ref{sec:conclusion}.

\section{Modified Gravity}\label{sec:modified_gravity}

We provide in this section a brief review of the theory behind the two classes of MG models that are used later in this work. In particular, we report the respective actions alongside with the equation of motion for the additional scalar degree of freedom, which elucidates the different screening mechanisms incorporated in those gravity theories.

\subsection{f(R) Gravity}
A general extension to the Einstein-Hilbert action in GR is accomplished by adding a general function of the Ricci scalar $f(R)$, which then takes the form
\begin{equation}
    S = \int \text{d}^4x \sqrt{-g} \left\{\frac{R+f(R)}{16\pi G}+\mathcal{L}_m\right\},
\end{equation}
when including a matter Lagrangian $\mathcal{L}_m$. This leads to the field equations
\begin{equation}
    G_{\alpha\beta} + f_R R_{\alpha\beta} - \left(\frac{f}{2}-\Box f_R\right)g_{\alpha\beta} - \nabla_{\alpha}\nabla_{\beta} f_R = 8\pi G T_{\alpha\beta},
\end{equation}
where $\Box$ denotes the d'Alembertian operator and Greek indices are running from 1 to 4. From these field equations an equation of motion for the scalaron field $f_R = \partial f/\partial_R$ can be deduced by taking the trace. The Ricci scalar is given by
\begin{equation}
    R = 12 H^2 + 6HH',
\end{equation}
where a prime denotes a differentiation with respect to the natural logarithm of the scale factor. In a $\Lambda$CDM universe, today's Ricci scalar is 
\begin{equation}
    R_0 = 12H_0^2 - 9 H_0^2 \Omega_{\textrm{m}}^0.
\end{equation}

Although the $f(R)$ function being completely general, there are several constraints concerning its derivatives with respect to $R$ to obtain a theory that is free from ghost instabilities \citep[see][for a derivation of those stability conditions]{Tsujikawa2010LNP}. Furthermore, specific functions can be chosen depending on the context of the theory.
Here we focus on a cosmological model with a late-time accelerated expansion for which the Hu-Sawicki theory \citep{Hu2007PhRvD} is the most promising. 
The $f(R)$ function in this model takes the form
\begin{equation}
    f(R) = -m^2 \frac{c_1 (R/m^2)^n}{c_2(R/m^2)^n+1},
\end{equation}
with $m=8\pi G \rho_0/3$, and $c_1$, $c_2$, $n$ being constants. In the simulations presented in the next sections, a value of $n=1$ was used. To produce a background expansion as dictated by $\Lambda$CDM, the ratio between $c_1$ and $c_2$ has to be chosen such that
\begin{equation}
    \frac{c_1}{c_2} = 6 \frac{\Omega_{\Lambda}^0}{\Omega_{\textrm{m}}^0}.
\end{equation}
From this follows a Lagrangian of the form $\mathcal{L} = R/16\pi G - \Lambda$ for the gravitational sector in the $R>>m^2$ limit where $f(R)\approx -m^2 c_1/c_2$, which corresponds to the well-known Einstein-Hilbert action with cosmological constant. Furthermore, by expanding the $f(R)$ function in the aforementioned limit but keeping the next-to-leading order term we arrive at
\begin{equation}
    f(R) = -\frac{c_1}{c_2} m^2 \left(1-\frac{m^2}{Rc_2}\right) = -m^2 6\frac{\Omega_{\Lambda}}{\Omega_{\textrm{m}}} - f_{R0} \frac{R_0^2}{R},
\end{equation}
where in the second equality we used the expression of the scalaron field 
\begin{equation}
    f_R = -\frac{m^4}{R^2} \frac{c_1}{c_2^2}
\end{equation}
evaluated for the background Ricci scalar value today ($f_{R0}$). We replaced $c_1/c_2$ with the previous expression to obtain a $\Lambda$CDM background. In this approximation, and by fixing $n=1$, the $f(R)$ function depends solely on the cosmological parameters and $f_{R0}$, the latter encoding the strength of the modification to GR.

Having an $f(R)$ modification in the Lagrangian will introduce additional force terms into the Poisson equation in the quasi-static and weak-field limit, as can be derived from perturbed field equations \citep{Bose2015JCAP}
\begin{equation}
    \nabla^2\Phi = 4 a^2\pi G (\rho - \bar{\rho}) -\frac{1}{2}\nabla^2 f_R
\end{equation}
and
\begin{equation}
    \nabla^2f_R = -\frac{a^2}{3} (R - \bar{R}) - \frac{8\pi G}{3} a^2 (\rho-\bar{\rho}),
\end{equation}
where $\bar{\rho}$ and $\bar{R}$ are the matter density and Ricci scalar at the background level.
These additional terms should be suppressed in the vicinity of massive objects, otherwise solar system tests might have detected the fifth force.
When $f(R)$ gravity is rewritten as a scalar-tensor gravity, the potential of the scalar field receives a contribution from the matter density \citep{Khoury2004PhRvD} as
\begin{equation}
    V_{\textrm{eff}}(\varphi) \equiv V(\varphi) + \rho e^{\varphi \beta /M_{pl}}.
\end{equation}
This leads in turn to a modified equation of motion for the scalar field $\varphi$ that includes density-dependent potential.
In this context a thin-shell condition can be derived, stating that the difference between the scalar field far away from the source $\varphi_{\infty}$ and inside the object $\varphi_c$ should be small compared to the gravitational potential on the surface of the object \citep{Khoury2004PhRvD}. Exterior solutions for $\varphi$ around compact objects satisfying the thin-shell condition will reach the solution $\varphi_{\infty}$ at larger distances, thereby suppressing the effect of the scalar field close to the object.

\subsection{nDGP Gravity}
The modification to standard gravity devised by Dvali, Gabadadze and Porrati \citep{Dvali2000PhLB}, hereafter DGP gravity, is of a radically different kind compared to $f(R)$ gravity. 
The setup is a 4D brane embedded in a 5D bulk and the modification to gravity comes from the fith dimensional contribution.
The action is given by \citep{Clifton2012PhR}
\begin{equation}
\begin{split}
    S = &M_5^3 \int \text{d}^5 x \sqrt{-g_5}\, R_5 \\
    & + \int \text{d}^4 x \sqrt{-g_4} \left\{-2M_5^3 K + \frac{M_4^2}{2}R_4 - \sigma + \mathcal{L}_m\right\},
\end{split}
\end{equation}
where $g_5$ and $g_4$ are the 5D and 4D metric, respectively.
The matter Lagrangian $\mathcal{L}_m$ does live on the 4D brane as well as the brane tension $\sigma$, which can act as a cosmological constant. Furthermore, there is both a 5D Ricci scalar $R_5$ and its 4D counterpart $R_4$, and the brane has an extrinsic curvature term $K$. Generally, both the brane and bulk have their individual mass scales $M_4$ and $M_5$ and they give rise to a specific cross-over scale $r_c$ defined as
\begin{equation}
    r_c = \frac{M_4^2}{2M_5^3},
\end{equation}
which regulates the contribution of 4D with respect to 5D gravity.

The modified Poisson equation for the gravitational potential and the equation for the additional scalar degree of freedom $\varphi$ (also called brane-bending mode as it describes the displacement of the brane) lead to the fifth force. They are given in the quasi-static approximation by \citep[see][]{Koyama2007PhRvD} 
\begin{equation}
\begin{split}
    &\nabla^2 \Phi = 4\pi G a^2 (\rho-\bar{\rho}) + \frac{1}{2} \nabla^2 \varphi \\
    &\nabla^2 \varphi + \frac{r_c}{3\beta a^2 }\left( (\nabla^2\varphi)^2 - (\nabla_i\nabla_j\varphi)^2\right) = \frac{8\pi G a^2}{3\beta}(\rho - \bar{\rho}),
\end{split}
\end{equation}
where $\beta$ is 
\begin{equation}
    \beta(t) = 1 \pm 2Hr_c \left(1+\frac{\dot H}{3H^2}\right).
\end{equation}
The dot refers to a derivative with respect to metric time $t$. One important feature of the DGP model is the existence of a normal branch and of a self-accelerating branch, indicated respectively by the $+$ and $-$ signs in the equation for $\beta$.
While the self-accelerating branch appears appealing for cosmology at first sight, as it can generate accelerated expansion without cosmological constant (the limit of vanishing brane tension), it contains unphysical ghost instabilities \citep{Clifton2012PhR}. Hence the model used in the simulation analysed in this work implements the normal branch, which does need a non-vanishing brane tension to produce accelerated expansion. It is interesting to study normal branch DGP models as it exhibits the Vainshtein screening mechanism \citep[see][]{Schmidt2009PhRvD,Barreira2015JCAP}. To illustrate that mechanism, the equation for the scalar field has to be studied around a mass source. Far away from the source, only the linear term $\nabla^2\varphi$ will dominate and this will contribute substantially to the usual gravitational force as it will also scale $\propto 1/r$.
However, non-linear terms start to dominate once we are closer to the source than to the Vainshtein radius $r_V$, defined by
\begin{equation}
    r_V \approx (r_sr_c^2)^{1/3},
\end{equation}
with $r_s$ being the Schwarzschild radius of the source. At some point, non-linear terms will dominate and the resulting force will scale as $\sqrt{r}$ and hence will be suppressed with respect to the gravitational force. A derivation of the solution for $\varphi$ can be found in \citet{Koyama2007PhRvD} for the general case, which includes linear and non-linear terms, and where the same scaling are recovered in the respective regimes.

At fixed Schwarzschild radius, the cross-over scale determines the Vainshtein radius, so by running simulations with different $r_c$ one will obtain different strengths of the Vainshtein screening. Therefore, varying $r_c$ allows the tuning of the amount of deviation to GR that is required.

\section{Weighted statistics and estimators}\label{sec:basics}

In Tab. \ref{tab:notation} we summarise the notation used throughout this work to ease distinguishing between the different discrete and continuous quantities.
\begin{table}
    \centering
    \begin{tabular}{c|c}
    \hline
    \hline
        $\langle \rangle$ & ensemble average (moment) \\
        $\langle \rangle_c$ & ensemble cumulant \\
        $\delta_M(\mathbf{x})$ & continuous weighted density contrast \\
        $\delta(\mathbf{x})$ & continuous density contrast \\
        $\delta_{f}(\mathbf{x})$ & discrete density contrast \\
        $\delta_R(\mathbf{x})$ & continuous smoothed density contrast \\
        $\delta_{Rf}(\mathbf{x})$ & discrete smoothed density contrast \\
        $m(\mathbf{x})$ & mark field \\
        $W(\mathbf{r})$ & weighted correlation function \\
        $W_f(\mathbf{r})$ & estimated weighted correlation function \\
        $\mathcal{M}(\mathbf{r})$ & marked correlation function \\
        $\mathcal{M}_f(\mathbf{r})$ & estimated marked correlation function \\
        $F(\mathbf{r})$ & smoothing kernel \\
        $\bar{m}$ & mean mark \\
        $\bar{m}_f$ & mean mark taken over a point set\\
        $\bar{n}$ & mean number of points per volume\\
        $\bar{N}$ & mean number of points per grid cell\\
        $a$ & size of one grid cell\\
    \hline
    \hline
    \end{tabular}
    \caption{Summary of notation used in this article.
    }
    \label{tab:notation}
\end{table}

\subsection{Unweighted Statistics}
The density contrast $\delta(\mathbf{x})$, which encodes the relative change of the density field $\rho(\mathbf{x})$, is defined as 
\begin{equation}
    \delta(\mathbf{x}) = \frac{\rho(\mathbf{x}) - \bar\rho}{\bar\rho},
\end{equation}
where $\bar\rho$ is the mean density. In order to study the matter clustering in the cosmological context, one of the most common summary statistic to characterise the density field, is the two-point correlation function $\xi(\mathbf{x},\mathbf{y})$ or its Fourier counterpart the power spectrum. The 2PCF is the cumulant $\langle \delta(\mathbf{x})\delta(\mathbf{y}) \rangle_c$ of the density contrast at positions $\mathbf{x}$ and $\mathbf{y}$. For two-point correlations, the cumulant and standard ensemble average are the same quantity. They remain the same up to three-point correlations but start to differ from four-point correlations onwards.
Due to the assumed statistical invariance by translation, the correlation function does only depend on the separation vector $\mathbf{r}=\mathbf{x}-\mathbf{y}$.
By inserting the definition of the density contrast we have that
\begin{equation}
\xi(\mathbf{r}) = \frac{\langle \rho(\mathbf{x}+\mathbf{r})\rho(\mathbf{x})\rangle_c}{\bar \rho^2}.
\end{equation}
From the last equation one can see that the 2PCF is zero if the field is totally uncorrelated at two different positions.

In order to estimate the 2PCF, we can deploy the commonly-used Landy-Szalay pair-counting estimator proposed by \citet{Landy1993APJ} to minimise the variance, and which takes the form
\begin{equation}
    \xi(\mathbf{r}) = \frac{DD(\mathbf{r})-2DR(\mathbf{r})+RR(\mathbf{r})}{RR(\mathbf{r})}.
\end{equation}
The terms $DD(\bf{r})$ and $RR(\bf{r})$ are the normalised pair counts measured in the data sample and a random sample following the geometry of the data sample, respectively.
In addition, a cross term with pairs consisting of one point in the data sample and the other in the random sample is given by $DR(\bf{r})$. 
In this work, we only compute two-point correlation functions in periodic boxes without selection function.
In this case, the term $DR$ converges to the term $RR$ in the limit of many realisations of random catalogues, and we can use the natural estimator given by \citet{Peebles1974ApJS}
\begin{equation}\label{eq:est_per}
    \xi(r) = \frac{DD(r)-RR(r)}{RR(r)}.
\end{equation}
The distribution of pairs in real space is isotropic, and together with periodic boundary conditions, lead the correlation function to only depend on the modulus $r$ of the pair separation vector.

In redshift space, it is useful to compute the anisotropic correlation function $\xi(s, \mu)$, binned in the norm of the pair separation vector $\mathbf{s}$ and the cosine angle between the line of sight (LOS) and the pair separation vector $\mu$. 
The 2PCF estimator for a periodic box is hence
\begin{equation}
    \xi(s, \mu) = \frac{DD(s, \mu) - RR(s, \mu)}{RR(s, \mu)},
\end{equation}
and normalised $RR$ counts are given by
\begin{equation}\label{eq:analytic_RR_counts}
\begin{split}
    RR([[s, s+\Delta s], [\mu + \Delta\mu]]) & = \frac{1}{L^3}\frac{4}{3}\pi \Delta \mu \{(s + \Delta s)^3 - s^3 \},
\end{split}
\end{equation}
which can be derived by calculating the volume covered by the respective bins in $s$ and $\mu$ relative to the total volume of the bin. For real space measurements, the $RR$ counts can be evaluated analytically in a similar fashion as in Eq.~\eqref{eq:analytic_RR_counts}.

The 2PCF in redshift space,  $\xi(s,\mu)$, can be decomposed into multipole moments, which is a basis encoding the different angle dependencies of the full 2PCF. Usually the decomposition is done into the first three non-vanishing multipole moments, being the monopole, quadrupole and hexadecpole. In the following, we will focus on the first two since the hexadecapole can be quite noisy for small point sets.
The multipole moment correlation functions are obtained by decomposing the $\xi(s,\mu)$ in the basis of Legendre polynomials as
\begin{equation}
        \xi_{\ell}(s) = \frac{(2\ell+1)}{2} \int_{-1}^1 \text{d} \mu \, \xi(s, \mu) P_{\ell}(\mu),
\end{equation}
yielding for the monopole and quadrupole to
\begin{equation}
    \begin{split}
        \xi_0(s) &= \frac{1}{2}\int_{-1}^1 \text{d} \mu \, \xi(s, \mu)\\
        \xi_2(s) &= \frac{5}{2}\int_{-1}^1 \text{d} \mu \, \xi(s, \mu) \frac{1}{2}(3\mu^2-1).
    \end{split}
\end{equation}
In practice, these integrals are discretised and we measure $\xi(s, \mu)$ in $100$ bins from $\mu=0$ to $\mu=1$ using the symmetry under interchange of galaxies for a given pair, which is fulfilled in our periodic box simulations. The discretised correlation function is then integrated by approximating the integral as a Riemann sum.

\subsection{Weighted statistics}
Let us now define the weighted density contrast 
\begin{equation}
    \delta_M(\mathbf{x}) = \frac{\rho_M(\mathbf{x}) - \overline{\rho_M}}{\overline{\rho_M}},
\end{equation}
where the weighted density field is given by $\rho_M(\mathbf{x}) = m(\mathbf{x}) \rho(\mathbf{x})$, $\overline{\rho_M}=\langle \rho_M(\mathbf{x})\rangle$, and $m(\bf{x})$ is the mark field. The weighted correlation function is the ensemble average of the weighted density contrast correlation,
\begin{equation}
    W(\mathbf{r}) = \langle \delta_M(\mathbf{x}) \delta_M(\mathbf{x}+\mathbf{r})\rangle,
\end{equation}
which when substituted with the definition of the density contrast, takes the form
\begin{equation}\label{eq:continuous_wCF}
    \begin{split}
        1+W(\mathbf{r}) & = \frac{\bar{\rho}^2}{\overline{\rho_M}^2}\langle m(\mathbf{x}) (1+\delta(\mathbf{x})) m(\mathbf{x}+\mathbf{r}) (1+\delta(\mathbf{x}+\mathbf{r})) \rangle.
    \end{split}
\end{equation}

The mark field $m(\mathbf{x})$ can be continuous in space, or discrete and defined on the point set (galaxy or halo catalogue).
Each object in the catalogue can be assigned a mark from the mark field, e.g. the $i$-th object has a mark $m_i$. The normalised weighted pair counts are  obtained as
\begin{equation}\label{eq:weightedpaircounts}
    WW(r) = \frac{\sum_{i\neq j}m_i m_j}{(\sum m_i)^2 - \sum m_i^2},
\end{equation}
where the sum is computed over all pairs with a separation inside the bin centred on $r$. The marked correlation function is then defined as \citep{Beisbart2000ApJ, Sheth2005MNRAS}
\begin{equation}\label{eq:mark_corr}
    \mathcal{M}(r) \equiv \frac{1+W(r)}{1+\xi(r)}.
\end{equation}
It converges to $\mathcal{M}(r)=1$ on large scales as $W(r)$ and $\xi(r)$ approach zero. 

In order to estimate the weighted correlation function from a catalogue, the natural estimator can be generalised to include weighted $DD(r)$ so that we can simply replace $DD(r)$ with $WW(r)$ counts, arriving at
\begin{equation}
    W(r) = \frac{WW(r)-RR(r)}{RR(r)}.
\end{equation}
Inserting this into the definition of the marked correlation function $\mathcal{M}(r)$ we have
\begin{equation}
   \mathcal{M}(r) = \frac{1+\frac{WW(r)-RR(r)}{RR(r)}}{1+\frac{DD(r)-RR(r)}{RR(r)}} = \frac{WW(r)}{DD(r)}.
\end{equation}
If the LS estimator is employed instead, one has to compute $WR(r)$ and $DR(r)$ terms in addition.

Computing the multipoles of the weighted correlation function is analogous to the unweighted case. However, the multipoles of the marked correlation function can be defined in two ways. The most intuitive definition is obtained by decomposing the marked correlation function $\mathcal{M}(s,\mu)$ in the basis of Legendre polynomials, yielding
\begin{equation}\label{eq:M_proper}
    \mathcal{M}_{\ell}(s) = \frac{(2\ell+1)}{2} \int_{-1}^1 \text{d}\mu \, \mathcal{M}(s, \mu) P_{\ell}(\mu).
\end{equation}
The second approach uses the following definition
\begin{equation}\label{eq:M_ratio_mult}
    \mathcal{M}_{\ell}(s) = \frac{1+W_{\ell}(s)}{1+\xi_{\ell}(s)},
\end{equation}
which is motivated by the fact that the denominator is the actual multipole of the unweighted 2PCF. This is not the case in the first definition in Eq.~\eqref{eq:M_proper}. The second definition has been used for instance by \citet{White2016JCAP} and \citet{Satpathy2019MNRAS}. Throughout this work we will use the form of Eq.~\eqref{eq:M_proper}.

\section{Simulations}\label{sec:sim_baseline}

\subsection{Characteristics}
In order to investigate different marked correlation functions and assess their discriminating power regarding GR and MG, we use the Extended LEnsing PHysics using ANalytic ray Tracing (ELEPHANT) simulation suite, thoroughly discussed in Sec. II B. of \citet{Alam2021JCAP}. We only provide a brief description of it in the following. This simulation suite consists of 5 realisations of GR with $\Lambda$CDM cosmology, $f(R)$ gravity with three different values of $|f_{R0}|=[10^{-6}, 10^{-5}, 10^{-4}]$, and nDGP gravity with $H_0r_c=[5.0, 1.0]$. Henceforth, we will refer to the different simulations as GR, F6, F5, F4, N5, and N1, respectively. The background cosmology is summarised in Tab. \ref{tab:cosmo_simulations} and resembles the best-fitting cosmology obtained from 9-year WMAP CMB analysis presented in \citet{Hinshaw2013ApJS}.
\begin{table}
    \centering
    \begin{tabular}{c|c|c}
    & ELEPHANT & DEMNUni/Covmos \\
    \hline
    \hline
         $\Omega_{b}$ & 0.046 & 0.05\\
         $\Omega_{cdm}$ & 0.235 & 0.27 \\
         $\Omega_{m}$ & 0.281 &  0.32 \\
         $\Omega_{\Lambda}$ & 0.719 & 0.68 \\
         $h$ & 0.697  & 0.67\\
         $n_s$ & 0.971 & 0.96\\
         $\sigma_8$ & 0.82  & - \\
         $A_s$ & -  & $2.1265 \times 10^{-9}$ \\
    \hline
    \hline
    \end{tabular}
    \caption{Reference cosmology of the ELEPHANT (first column) and DEMNUni (second column) simulations. The PDF of the dark matter particles in the DEMNUni simulation have been used to define the target PDF used to produce Covmos realisations.}
    \label{tab:cosmo_simulations}
\end{table}

Key simulation parameters are summarised in Tab. \ref{tab:simu}. The dark matter halos have been identified with the \texttt{ROCKSTAR} algorithm \citep{Behroozi2013ApJ} and have been subsequently populated with galaxies using the 5-parameter halo occupation distribution (HOD) model of \citet{Zheng2007ApJ}. For each realisation, redshift-space coordinates have been calculated by fixing the LOS to one of the three simulation box axes, individually, and by 'observing' the box from a distance equal to 100 times the box side length.
\begin{table}
    \centering
    \begin{tabular}{c|c}
    \hline
    \hline
        N-body code & ECOSMOG \\
        Box side length $L$ & 1024$\Mpc$ \\
        Particle number $N_p$  & $1024^3$ \\
        Number of grid cells $N_{grid}$  & $1024^3$ \\
        Initial redshift $z_{ini}$  & 49.0 \\
        Initial conditions  & Zel'dovich approx. (MPGrafic)\\
        Mass particle $M_{p}$ & $7.79854\times 10^{10} \textrm{M}_{\odot}$ \\
        Final redshift z & 0.506 \\
    \hline
    \hline
    \end{tabular}
    \caption{Characteristics of the ELEPHANT simulation suite.}
    \label{tab:simu}
\end{table}
One crucial property of this suite of simulations, which makes it particularly suitable for our studies, is the matching of the projected 2PCF of galaxies $w_p(r_p)$ predicted by GR in the MG simulations. The latter was done by tuning the HOD parameters of the MG simulations. For the GR simulation, the best-fit HOD parameters were taken from \citet{Manera2013MNRAS}.

We use a second set of simulations to assess discreteness effects in the estimation of the mark and how they propagate into the marked correlation function. For this, we make use of the Covmos realisations from \citet{Baratta2023A&A}.
These are not full N-body simulations, rather they reproduce dark-matter particle one- and two-point statistics following the technique described in \citet{Baratta2020A&A}. This procedure consists of applying a local transformation to a Gaussian density field such that it follows a target probability distribution function (PDF) and power spectrum. The point set is then obtained by a local Poisson sampling on the linearly interpolated density values. For the set of Covmos realisations, the target PDF and power spectrum were set by the DEMNUni N-body simulation \citep{Castorina2015JCAP} statistics. The DEMNUni simulation assumes a $\Lambda$CDM cosmology with parameters presented in Tab. \ref{tab:cosmo_simulations}. The Covmos catalogues contain about $20\times 10^6$ points in a box of 1 $h^{-1}$Gpc of side, resulting in a number density of about $0.02\,h^{3}~\mathrm{Mpc}^{-3}$. Such an high density allows treating those catalogues as being almost free from shot noise.

\subsection{Two-point correlation function}

Although the galaxy projected correlation function are matched in the ELEPHANT suite, it is instructive to assess residual deviations in other statistics, particularly for the interpretation of differences arising in the analysis of marked correlation functions.

\begin{figure}
    \centering
    \includegraphics[width=\columnwidth]{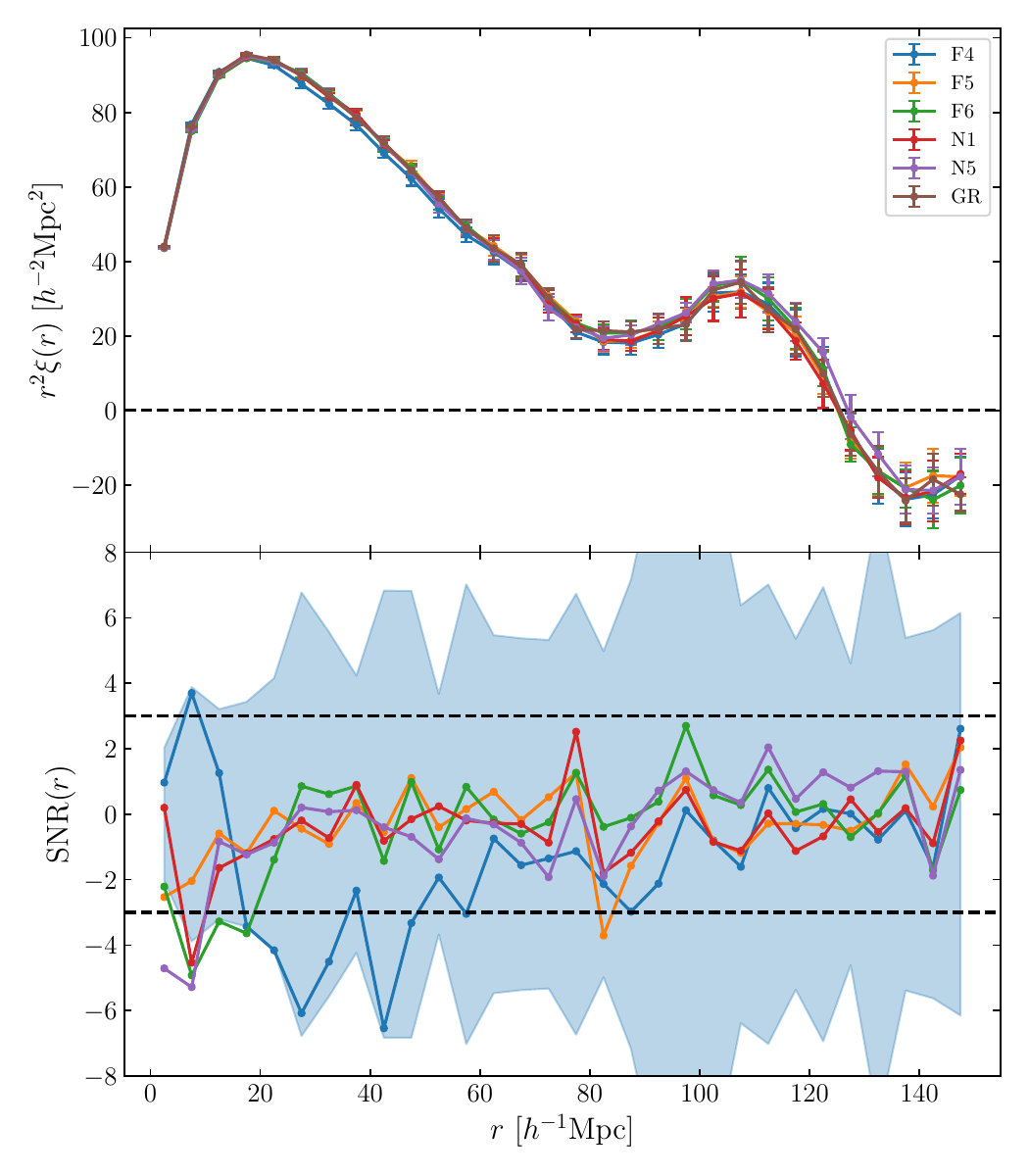}
    \caption{Difference in the measured standard correlation function $\xi(r)$ between GR and MG in real space.
    In the upper panel the correlation functions themselves are plotted where different colours indicate the underlying gravity theory.
    The curves show the average over 5 realisations and the errorbars correspond to the mean standard deviation over these realisations.
    The lower panel quantifies possible differences in terms of the SNR as introduced in Section~\ref{sec:sim_baseline}. Black dashed lines indicate a SNR of $\pm 3$. The shaded region refers to the error of a single measurement divided by the mean error of the difference as described in Eq.~\eqref{eq:alpha}.}
    \label{fig:baseline_diff}
\end{figure}

\begin{figure*}
    \centering
    \includegraphics[width=0.45\textwidth]{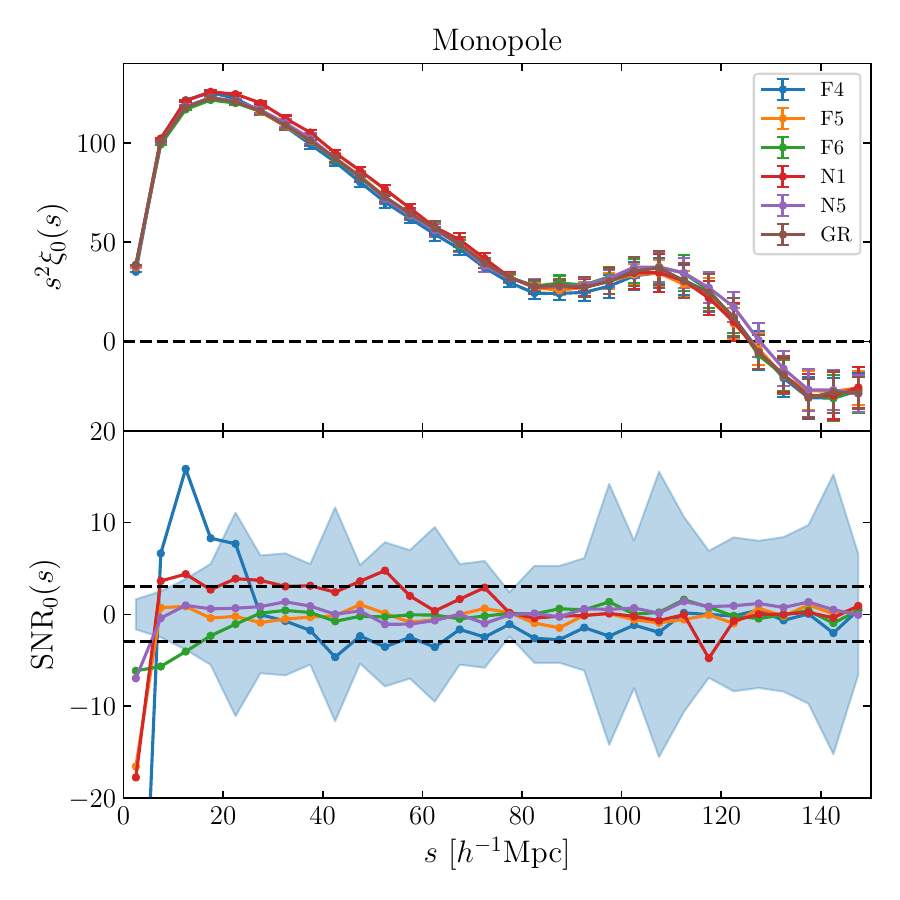}
    \includegraphics[width=0.45\textwidth]{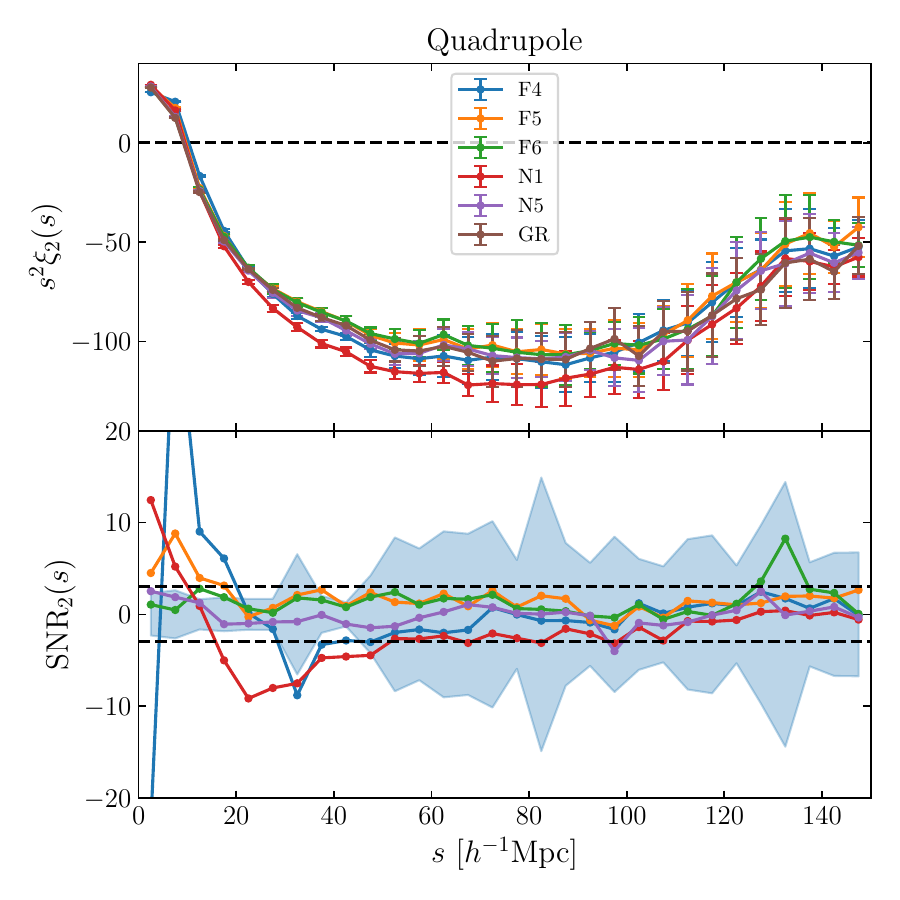}
    \caption{Differences in the standard measured correlation function multipoles in redshift space between GR and MG. The upper panels present the mean correlation function multipoles taken over 5 realisations with the monopole on the left side and the quadrupole on the right side. The errorbars correspond to the mean standard deviation over 5 realisations.
    The lower panels show the respective SNR with 3$\sigma$ indicated by the black dashed lines.
    The colour coding refers to different gravity simulations and the corresponding shaded regions refer to the error of a single measurement divided by the mean error of the difference (see Eq.~\eqref{eq:alpha}).
}
    \label{fig:baseline_diff_RSS}
\end{figure*}

We measured both the real- and redshift-space correlation functions in 30 linear bins in $r$ and $s$, respectively, ranging from $10^{-3}~\Mpc$ to $150~\Mpc$. For the redshift-space measurements, we used the ELEPHANT catalogues with the LOS fixed to the $x$-direction. All correlation function measurements in this work have been performed using the publicly available package \texttt{Corrfunc} \citep{Sinha10.1007_corrfunc, Sinha2020MNRAS_corrfunc}.
In the upper panel of Figure~\ref{fig:baseline_diff}, we show the standard correlation function in real space for the different gravity simulations. The measurements appear to be within the respective uncertainties over all scales, albeit on very small scales, a more careful assessment of possible deviations is advised as the error bars are very small.
In Figure~\ref{fig:baseline_diff_RSS}, the monopole (right) and quadrupole (left) of the anisotropic 2PCF in redshift space are presented in the upper panels. Similarly as for the real space correlation function, the multipoles are within the respective uncertainties on large scales, although the N1 measurement appears to deviate from the others in the quadrupole. On smaller scales, discrepancies seem to appear as uncertainties are getting very small and a visual inspection is not sufficient to quantify those differences.

In order to properly assess the difference between MG and GR in weighted or unweighted correlation functions, we define the difference between MG and GR as the mean
\begin{equation}
    \overline{\Delta \mathcal{M}}(r) = \frac{1}{5}\sum_{i=1}^{5}\{\mathcal{M}_{i,\text{MG}}(r)-\mathcal{M}_{i,\text{GR}}(r)\},
\end{equation}
where $i$ ranges over the number of realisations.
However, the mean differences alone does not tell about the significance as the data might fluctuate much more than differences. We therefore divide the mean difference by the standard deviation as
\begin{equation}
    \sigma_{\textrm{avg}}(r) = \sqrt{\frac{1}{N(N-1)}\sum_{i=1}^{5} \{\Delta_i \mathcal{M}(r) - \overline{\Delta\mathcal{M}}(r)\}^2}.
\end{equation}
The factor of $1/(N-1)$ is necessary in order to compute an unbiased standard deviation, since we only have 5 realisations at hand. Furthermore, the additional factor of $1/N$ comes from fact that we want the error on the mean and not of a single measurement. In a similar manner, we compute the standard deviation of a single marked correlation function as
\begin{equation}
    \sigma_{\textrm{s}}(r) = \sqrt{\frac{1}{N-1}\sum_{i=1}^{5} \{\mathcal{M}_{i}(r) - \overline{\mathcal{M}}(r)\}^2}.
\end{equation}
In the end, the ratio of interest is 
\begin{equation}\label{eq:signal_to_noise}
    \textrm{SNR}(r) = \frac{\overline{\Delta \mathcal{M}}(r)}{\sigma_{\textrm{avg}}(r)},
\end{equation}
giving directly the difference in terms of standard deviations. If the absolute value of this SNR is larger than 3 then we would advocate a significant deviation between MG and GR.

Another quantity of interest that we use throughout this work is the ratio between the error on a single measurement of the marked correlation function and the noise $\sigma_{\textrm{avg}}$, as used in the signal-to-noise ratio. We will refer to this ratio as
\begin{equation}\label{eq:alpha}
    \alpha(r) = \frac{\sigma_{\textrm{s}}(r)}{\sigma_{\textrm{avg}}(r)}
\end{equation}
and will include it the figures as shaded regions. The error of a single measurement $\sigma_s(r)$ is hereby taken for the GR case. This ratio gives an indication on the statistical significance of a difference, if we would have only one simulation/measurement at hand. To assess this, we have to compare $\textrm{SNR}(r)$ with $\alpha(r)$, and if $\textrm{SNR}(r)>3\alpha(r)$ then we can claim a $3\sigma$ difference to be detectable with a single measurement. Of course, care must be taken if the error of a single measurement is significantly different between GR and MG, since $\alpha(r)$ will differ depending on what simulations are used to estimate the error, therefore possibly affecting conclusions.

In the lower panel of Fig.~\ref{fig:baseline_diff}, we display the SNR$(r)$ as introduced above.
The differences rarely cross the limit of $3\sigma$ except for the very lowest scales, below 20$\Mpc$, or for F4 at intermediate scales where deviations can reach up to 6$\sigma$. However these large deviations happen only for single scales and there is no general trend. This suggests that the crossing of the 3$\sigma$ border might be caused by sample variance, as the statistical power is limited with only 5 realisations. In addition, when considering the error of a single realisation volume, as displayed by the blue shaded region, we can see that the deviations for F4 are within the uncertainty.

In the lower panels of Fig.~\ref{fig:baseline_diff_RSS} we show the SNR$(s)$ for the multipoles in redshift space. Except for the smallest scales, the simulations F5, F6 and N5 show generally no significant differences to GR. In the case of the monopoles, the SNR is close to 0 for almost all scales. For F4 and N1 significant differences are present although mainly on scales below $20~\Mpc$. On larger scales both simulations show SNR varying around 3$\sigma$. These differences are mostly within the uncertainty, if the error of a single volume is considered. This confirms that by considering the standard correlation function only, in real or redshift space, we cannot really distinguish between GR and those MG models.

\section{Marks for modified gravity}\label{sec:what_mark}

There is a vast space of possible marks that can be used and the specific choice strongly depends on the context in which the marked correlation function is studied. The most popular mark function $M[\rho(\bf{x})]$ in the literature within the context of detecting MG was introduced by \citet{White2016JCAP} and takes the form
\begin{equation}\label{eq:White_mark}
    m(\mathbf{x}) = M_W[\rho(\mathbf{x})] \equiv \left(\frac{1 + \rho_{*}}{\rho_{*}+\rho(\mathbf{x})}\right)^{p},
\end{equation}
where $\rho_{*}$ and $p$ are free parameters used to control the mark upweighting of low- versus high-density regions. We will refer to this mark in the following as the \textit{White mark} and be indicated via the $W$ in the subscript. The White mark can be seen as a \emph{local} transformation of the density field.
Choices for the free parameters range from $(\rho_{*}, p)=(10.0, 7.0)$ \citep{Aviles2020JCAP}, $(\rho_{*}, p)=(4.0, 10.0)$ \cite[][]{Alam2021JCAP, Valogiannis2018PhRvD}, to $(\rho_{*}, p)=(10^{-6}, 1.0)$ \citep{Aguayo2018MNRAS}. Upweighting galaxies in high-density regions using $(\rho_{*}, p)=(1.0, -1.0)$ has also been explored in \citet{Alam2021JCAP}. Other values have also been investigated by \citet{Satpathy2019MNRAS} and \citet{Massara2021PhRvL} for the marked power spectrum.
This underlines the wide range of possible mark functions and configurations to be used and the amount of freedom this can introduce in the analysis.

Marks based on the local density require an estimation of the latter from a finite point set in the first place and there exist several different approaches to do so. While we will use an estimation based on mass assignment schemes (MAS), adaptive approaches such as Delaunay \citep[][]{Schaap2000A&A} or Voronoi tessellations as used in void finders \citep[e.g.][]{Neyrinck2008MNRAS} could also be used. With a MAS applied to a discrete density field (subscript $f$ for \textit{finite}) 
\begin{equation}
    \rho_f(\mathbf{x}) = m \sum_{i=0}^{N-1} \delta_D(\mathbf{x}-\mathbf{x}_i),
\end{equation}
where the $i$-th point is located at position $\mathbf{x}_i$, the estimated density field on the grid takes the form \citep{Sefusatti2016MNRAS}
\begin{equation}
    \delta_{Rf}(\mathbf{x}) = \frac{1}{\bar N} \sum_{i=0}^{N-1} F\left(\frac{\mathbf{x}-\mathbf{x}_i}{a}\right) -1, 
\end{equation}
with $\bar N$ is the density of points per grid cell, $N$ is the number of points, $a$ is the size of one grid-cell, and $F(\mathbf{x})$ the MAS kernel. The coordinate $\mathbf{x}$ is only evaluated at grid points but in principle can be placed anywhere. For this derivation, we assumed all points to have the same mass $m$ and we made use of the simplified notation where $F(\mathbf{x}) \equiv F(x_1)F(x_2)F(x_3)$. The density field obtained in this way is related to the true density field by a convolution with the MAS kernel. In this work, we mainly use a piece-wise cubic spline (PCS) for the MAS but higher- and lower-order kernels are employed for specific tests. The explicit form of the used kernels up to septic order can be found in the appendix of \citet{Chaniotis2004JCoPh}\footnote{It has to be noted that there are two minor typos. Their octic spline is actually the septic spline and in its expression the term $s^7/20$ should be replaced by $s^7/720$.}.

\subsection{Beyond local density}
A way to include information beyond the local density field is by using the large-scale environment that can be divided into clusters, filaments, walls and voids. Generally, there are different ways to define these structures from a galaxy catalogue ranging from the sophisticated approach by \citet{Sousbie2011MNRAS} based on topological considerations to the work of  \citet{Falck2012ApJ} using phase-space information.
One of the most straightforward approaches utilises the T-web formalism \citep{ForeroRomero2009MNRAS} based on the Hessian of the gravitational potential. For a thorough comparison of the above mentioned cosmic web classifications and many more we refer the reader to \citet{Libeskind2018MNRAS}. In this analysis we will deploy the T-web classification that uses the relation of the eigenvalues $\lambda_1$, $\lambda_2$ and $\lambda_3$ of the tidal tensor to the density evolution as given in \citet{Cautun2014MNRAS}
\begin{equation}
    \rho(\mathbf{x}) = \frac{\bar \rho }{(1-D(t)\lambda_1)(1-D(t)\lambda_2)(1-D(t)\lambda_3)},
\end{equation}
where $D(t)$ is the growing solution to the growth factor. This expression can be derived from Lagrangian perturbation theory to linear order \citep{Zeldovich1970A&A}. The dimensionality of the structure then depends on the number of eigenvalues with positive sign. Three positive eigenvalues corresponds to a cluster as it encodes a collapse among all three spatial directions. Two or one positive eigenvalues result in a filament or wall, respectively. If all eigenvalue are negative then $\rho(\mathbf{x})$ will never diverge and we can interpret this as a void. A pitfall of this classification appears if some of the eigenvalues are very small but positive as the corresponding structure might not collapse in an Hubble time. To circumvent this issue while not having to rely on thresholds for the eigenvalues we use the scheme as proposed in \citet{Cautun2013MNRAS}. They give an environmental signature $\mathcal{S}$ for ordered eigenvalues $\lambda_1\leq \lambda_2 \leq \lambda_3$ in their Eq. 6 and 7 as signatures for clusters, filaments, walls and voids, respectively. We adapted this scheme to be used on the eigenvalues of the Tidal tensor instead of the Hessian of the density contrast as proposed in \citet{Cautun2013MNRAS}.

In order to obtain the tidal tensor in a simulation we follow the grid-based approach as used for the density field. With a density field on a grid at hand, the gravitational potential or tidal tensor can be straightforwardly deduced by a series of fast Fourier transforms (FT). For this we use the Poisson equation to relate the density field to the gravitational potential
\begin{equation}
    \nabla^2 \Phi(\mathbf{x}) = 4\pi G \delta(\mathbf{x}) \quad \overset{\textrm{FT}}{\Leftrightarrow}\quad  k^2 \Phi(\mathbf{k}) = -4\pi G \delta(\mathbf{k}).
\end{equation}
We absorb the constant $4\pi G$ into the definition of the gravitational potential. There exists a singularity when the wavevector is equal to zero, which we evade by simply setting the zeroth mode of $\Phi(\mathbf{k})$ to zero, as we expect the gravitational potential sourced by the density contrast to have a zero mean. The components of the tidal tensor $T_{ij}$ can then be derived by taking successive derivatives in the respective directions as
\begin{equation}\label{eq:tidal_tensor}
    T_{ij}(\mathbf{x}) = \partial_i\partial_j \Phi(\mathbf{x}) \quad \overset{\textrm{FT}}{\Leftrightarrow} \quad T_{ij}(\mathbf{k}) = -k_i k_j \Phi(\mathbf{k}).
\end{equation}
The off-diagonal terms suffer from a break in the Fourier symmetry pairs when evaluated on a finite grid.
This leads to a non-vanishing imaginary part once the tidal tensor in configuration space is obtained by inverse FT. We circumvent this issue by setting the imaginary part to zero via applying a filter that sets the symmetry breaking modes at the Nyquist frequency to zero. In our implementation we will evaluate the environmental signatures on the grid over which the eigenvalues of the tidal tensor have been computed.
For each grid cell the largest signatures defines the corresponding environment and if all signatures are zero then the environment is set to be a void.

Once each galaxy has been classified, this information can be used to enhance effects of MG in clustering measurements. The simplest use of the environmental classification is to divide the catalogue into sub-catalogues consisting of galaxies located in voids, walls, filaments and clusters, respectively and
compute auto-correlation functions. The difference in clustering amplitude in the different environments is expected to be stronger in MG, particularly in the correlation function of void galaxies.
In the work of \citet{Bonnaire2022A&A}, they computed the power spectra of density fields that have been obtained by splitting the original density field into respective contributions from galaxies in voids, walls, filaments and clusters. They applied this approach to the dark matter particles of the Quijote simulation \citep{Villaescusa-Navarro2020ApJS} thereby having a much larger set of points. 
A drawback in the analysis of our galaxy mock catalogues, but also when using limited survey samples, is the loss of information due to discarding many galaxies leading to an increase in uncertainty of the measurements.
Computing environmental correlation functions is the same as computing weighted correlation functions but with a mark set to one for all galaxies living in the respective environment and zero otherwise.
We refer the interested reader to Section~\ref{sec:app_cross} for some notes on the structure of weighted correlation functions for those kind of weights. In the following we will denote such marked correlation functions, consisting of the environmental weighted correlation function divided by the total unweighted correlation function as in Eq.~\eqref{eq:mark_corr}, via their respective environment, e.g. \emph{void marked correlation function}.

Conversely, we can use the full catalogue of objects and put larger weights to progressively more unscreened galaxies, as done by the following mark field
\begin{equation}\label{eq:void_peak}
       m(\mathbf{x})=\begin{cases} 4 \quad\text{if void} \\ 3 \quad\text{if wall} \\ 2 \quad\text{if filament} \\ 1 \quad\text{if cluster} \end{cases} \quad \text{Void$_{\rm LEM}$},
\end{equation}
where LEM stands for \emph{linear environment mark}. This approach is similar to the density split technique as used in \citet{Paillas2021MNRAS_density_split}, where cross-correlation functions of galaxies living in differently dense regions have been investigated. Our proposed mark can also be devided into a specific combination of auto and cross-correlations. This mark could in principle be extended to a $\textrm{Wall}_{\textrm{LEM}}$ mark, where wall galaxies get assigned a weight of 4 and voids galaxies a weight of 3, as well as similarly peaked functions for filaments and clusters.
However, as we expect MG to be the strongest in low-density regions we restrict ourselves to upweighing void or wall galaxies only.

Yet another idea of using the environmental classification of galaxies as a mark would be to further increase the anti-correlation present in low density regions. This can be accommodated with the following mark
\begin{equation}\label{eq:negative_void}
    m(\mathbf{x})=\begin{cases} -1 \quad\text{if void} \\ 1 \quad \text{else} \end{cases} \quad \text{Void$_{\rm AC}$},
\end{equation}
where we abbreviated \emph{anti-correlation} with AC. In principle, there is no difference if we switch signs of this mark because from Eq.~\eqref{eq:weightedpaircounts} it is clear that any overall factor of the marks would be cancelled by the division of the normalisation. This mark leaves galaxy pairs that are in voids unweighted as well as galaxy pairs not in voids. However, if one galaxy is in a void and the other is not, the weight will be -1 thereby creating an anti-correlation.

Marks based on the tidal tensor components may appear promising to go beyond the local density.
An interesting quantity first introduced by \citet{Heavens1988MNRAS} and then used by \citet{alam2019mnras}, is the tidal torque, defined as
\begin{equation}
    t(\mathbf{x}) = \frac{1}{2}\{(\lambda_3-\lambda_2)^2 + (\lambda_3-\lambda_1)^2 + (\lambda_2-\lambda_1)^2\},
\end{equation}
with $\lambda_1$, $\lambda_2$ and $\lambda_3$ are the eigenvalues of the tidal tensor.
The larger the difference between the eigenvalues the more anisotropic is the structure.
Hence we expect the tidal torque to be large for filaments and walls and small for clusters or voids.
Another field depending directly on tidal tensor components is the tidal field, also known as the second Galileon $\mathcal{G}_2$ \citep{Nicolis2009}, and which was used extensively for the emergence of non-local bias between galaxies and dark matter \citep{chan2012phrvd}.
The tidal field is defined as $\mathcal{G}_2 = (\partial_i\partial_j\Phi)^2-(\partial^{i}\partial_{i}\Phi)^2$, where we can identify the components of the tidal tensor as introduced in Eq.~\eqref{eq:tidal_tensor}.
In practise we investigate two separate marks consisting of the tidal field and tidal torque as they are, respectively. Using these fields in this way should give an insight on the suitability of the tidal field or tidal torque to disentangle MG from GR.

\begin{figure}
    \centering
    \includegraphics[width=\columnwidth]{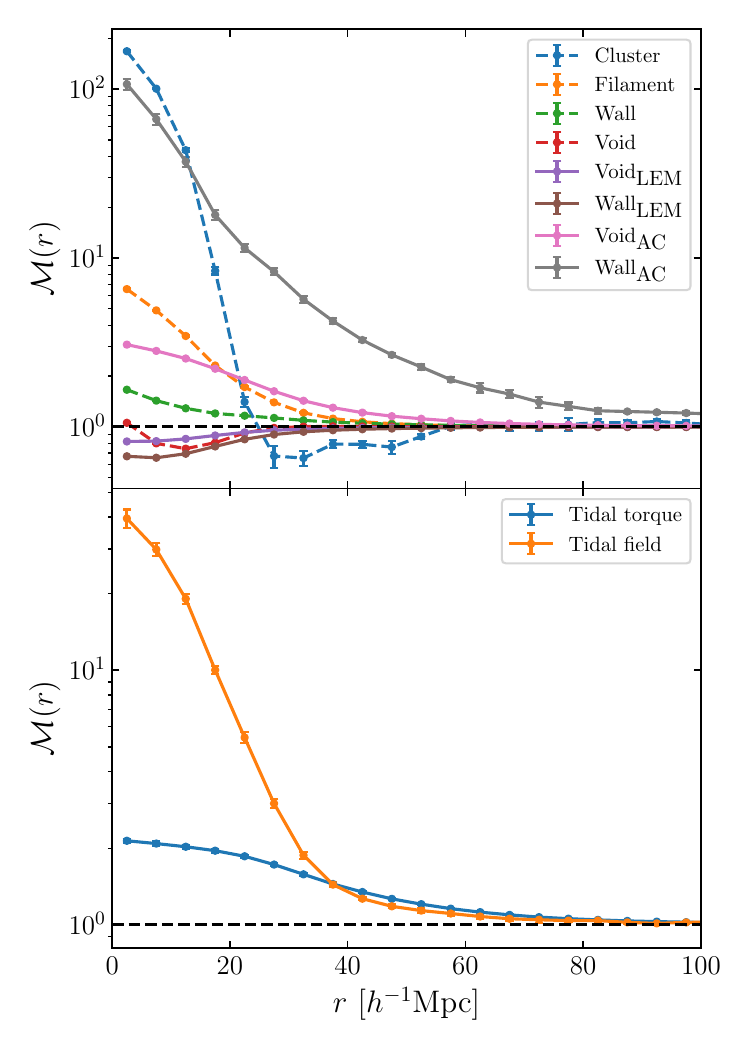}
    \caption{Summary of the different marked correlation functions using marks based on the tidal field, tidal torque (both in the lower panel) or large-scale environment (upper panel).
    The black dashed line indicates an amplitude of 1.
    Curves represent the mean taken over 5 realisations and the error corresponds to the mean standard deviation over 5 realisations.
    The measurements have not been corrected for any form of bias due to shot-noise effects.}
    \label{fig:env_mark_summary}
\end{figure}

In Figure~\ref{fig:env_mark_summary} we present the different marked correlation functions introduced in this section. For the cluster marked correlation function we see a strong signal on very small scales which relates to the correlation between galaxies insides clusters. The compensation feature on scales between 20$\Mpc$
and 60$\Mpc$ comes from less clustered regions around clusters and is similar, although reversed, to the compensation seen in the void-galaxy cross-correlation function \citep[e.g.][]{Aubert2022MNRAS_SDSS_VG, Hamaus2022A&A_Euclid_VG}.
The filament and wall marked correlation functions show progressively less signal as the clustering of galaxies inside walls and filaments are closer to the total clustering of all galaxies. Notably, if voids are considered, the observed signal below unity implies that void galaxies are less clustered compared to the total clustering. The large signal of the cluster marked correlation function comes at the cost of larger errors due to small amount of galaxies residing in clusters. In general we aim for mark correlation functions with a signal different from unity over a wide range of scales as this might lead to differences at those scales between MG and GR. On the other side, if the marked correlation function stays very close to unity on most scales, then any possible difference between MG and GR can only originate from the clustering itself.
The latter is matched between MG and GR in the ELEPHANT simulations to fit observations as described in Section~\ref{sec:sim_baseline}. For these reasons, the $\textrm{Void}_{\textrm{AC}}$ mark and particularly the $\textrm{Wall}_{\textrm{AC}}$ mark are of strong interest as they exhibit a signal up to large scales.
Looking at the lower panel of Fig.~\ref{fig:env_mark_summary}, we can see a strong signal when the tidal field is used, which extends to large scales. The same, although with considerably less amplitude on small scales, is found for the tidal torque. Hence, these marks are also interesting candidates to be investigated to discriminate between MG and GR. Although an impact of the mark is a necessary prerequisite, it is not sufficient to guarantee a disentanglement of GR from MG because the signal could be the same in MG and GR, nevertheless.

It has to be noted that the marked correlation functions shown in Figure~\ref{fig:env_mark_summary} are not corrected for a possible bias due to the estimation of the mark on a discrete catalogue, as we discuss in the next section. Hence, the exact amplitude of the measurements might be subject to changes if such a correction is applied. For the marks based on the environmental classification we do not expect this bias to be particularly strong because possible miss-classifications, originating from a biased estimate of the density field, should not affect every galaxy in a catalogue.

\subsection{Anti-correlating galaxies using local density}\label{sec:AC_delta}

Until now, we have seen that marks based on the local density are particularly simple and introducing an anti-correlation with the mark appears to be promising regarding the discrimination between GR and MG.
Therefore, we propose the following mark function based on the hyperbolic tangent, satisfying both aforementioned advantages, 
\begin{equation}\label{eq:tanh_mark}
    m(\mathbf{x}) = M_{\tanh}[\delta_R(\mathbf{x})] \equiv \tanh(a\,(\delta_R(\mathbf{x})+b)),
\end{equation}
where $a$ and $b$ are parameters controlling how steeply the transition from -1 to 1 takes place and where the transition happens, respectively. In general, we could use a third parameter $c$ as an overall factor in front of the hyperbolic tangent but constant factors can be pulled out of the mean and hence are cancelled by the normalisation in Eq.~\eqref{eq:weightedpaircounts}.
It is worth mentioning the fact that theoretical modelling of marks based on the environmental classification, such as the $\textrm{Void}_{\textrm{AC}}$ mark, might be particularly challenging as it is not straightforward to express the mark in terms of the density contrast. From a theoretical perspective, marked correlation functions with marks based on the density are more tractable. Furthermore, discreteness effects, arising in the density estimation itself can be more easily corrected for in the measurement of the marked correlation function as elucidated in the next section.

\section{Propagation of discreteness effects of the mark estimation into weighted correlation functions}\label{sec:imapct_sn}

When dealing with finite point sets we can assume the sampling process to be locally of Poisson nature \citep{Layzer1956AJ}. This means that the number of points found in some small-enough grid cells appears as being drawn from a Poisson distribution with some expectation value. However, the expectation value of the local Poisson process does have a PDF on its own. The PDF from which the expectation values is drawn is continuous and describes the density field globally. If this PDF is a Dirac delta function then the expectation value of the Poisson process is the same everywhere and the moments estimated from the sample points coincide with the moments from the continuous PDF. However, if the continuous PDF is not a Dirac delta function, as is the case for the cosmological density field, then the estimated moments contain a bias with respect to the true moments of the continuous PDF. This bias is usually called shot noise or Poisson noise in the literature. In the power spectrum estimation, the shot noise appears as an additive constant for all scales in $k$-space. In the 2PCF instead, the shot noise emerges only at zero lag, that is at a pair separation of zero. Hence, shot noise is inherently a problem of correlating a point with itself.
When we use the density field inside a mark function, we use a smoothed version of the true field. We spread points over a finite volume leading to self-correlations also at non-zero pair separation and in turn to shot-noise effects. Intuitively, this can be understood in the following manner: in the unsmoothed case all points are infinitely small dots, while in the smoothed case the points are represented by circles with a non-zero radius. Inside this radius one point can be correlated with itself.

In order to precisely understand how shot noise affects marked correlation functions we have to do a small detour and carefully distinguish between the statistical properties of the true density contrast $\delta(\mathbf{x})$, the smoothed true density contrast $\delta_R(\mathbf{x})$, and the respective quantities estimated from finite point sets, hereby denoted with an $f$ in the subscript $\delta_f(\mathbf{x})$ and $\delta_{Rf}(\mathbf{x})$. The weighted correlation function estimated from a finite point set can be written as
\begin{equation}
    1+W_f(\mathbf{r}) = \frac{w_f(\mathbf{r})}{\bar{m}_f^2},
\label{eq:weighted_corr}
\end{equation}
where we defined the quantity $w_f(\mathbf{r})$ as
\begin{equation}
w_f(\mathbf{r}) \equiv \langle M[\delta_{Rf}(\mathbf{x})] (1+\delta_f(\mathbf{x})) M[\delta_{Rf}(\mathbf{x}+\mathbf{r})] (1+\delta_f(\mathbf{x}+\mathbf{r}))\rangle,
\label{wf}
\end{equation}
and $\bar m_f = \frac{1}{V}\int_V m_f({\bf x})\rho_f({\bf x})/\bar{\rho}_f\dif^3 x$ is the mean mark taken over the points, i.e. weighted by the density. In Eq.~\eqref{eq:weighted_corr} both $w_f(\mathbf{r})$ and $\bar m_f$ are expected to be sensitive to the noise induced by the auto-correlation of objects with themselves, and we will denote the corresponding shot-noise free signals $w$ and $\bar m$. Indeed, Eq.~\eqref{wf} shows that there is a mark function $M$ of the smoothed density field that is multiplied by the density field itself. This constitutes the main source of shot noise that is expected to happen even at large separation $r$, where there is no overlap between the smoothing kernels. In this section, we first show the effect of shot noise on the marked correlation function for a specific mark and then devise a general method to correct for shot noise.

\subsection{A toy model}\label{sec:toy_model_SN}

In order to understand how the shot noise propagates into $w_f(\mathbf{r})$ and $\bar m_f$, we focus on a very simple mark function defined by $M[\delta_{Rf}({\bf x})] = \delta_{Rf}({\bf x})$ and $M[\delta_{Rf}({\bf x} + {\bf r})] = 1$. This corresponds to a marked correlation function in which only one point of the pair is weighted by the density contrast and the other point stays unweighted. In this instructive case we can split $w_f(\mathbf{r})$ into three terms
\begin{equation}
w_f(\mathbf{r}) = \sigma_{Rf}^2  + \xi_{Rf}({\bf r}) + \zeta_{Rf}({\bf r}),
\label{wf_split}
\end{equation}
where the individual contributions are given by
\begin{equation}\label{eq:toy_model_indiv_terms}
\begin{split}
    \sigma_{Rf}^2  & = \langle \delta_{Rf}(\mathbf{x}) \delta_f(\mathbf{x})\rangle \\
    \xi_{Rf}({\bf r}) &= \langle \delta_{Rf}(\mathbf{x}) \delta_f(\mathbf{x}+\mathbf{r})\rangle  \\
     \zeta_{Rf}({\bf r}) &= \langle \delta_{Rf}(\mathbf{x}) \delta_f(\mathbf{x}) \delta_f(\mathbf{x}+\mathbf{r})\rangle,
\end{split}
\end{equation}
and consist of correlators between the smoothed and unsmoothed density field estimated on a finite point set.
Given that the smoothed density field $\delta_{Rf}(\mathbf{x})$ is related to the density field $\delta_f(\mathbf{x})$ through the convolution 
\begin{equation}\label{eq:smoothed_deltaRf}
    \delta_{Rf}(\mathbf{x}) = \frac{1}{a^3}\int_{\mathbf{x}'} F\left(\frac{\mathbf{x}-\mathbf{x}'}{a}\right) \delta_f(\mathbf{x}') \text{d}^3x',
\end{equation}
one can immediately see that the three contributions in Eq.~\eqref{wf_split} are involving integrals over two- and three-point correlation functions of the density field. In general, $n$-point correlation functions are affected by shot noise as \citep[see][]{Chan2017PhRev}
\begin{equation}
\begin{split}\label{eq:general_SN}
    \Xi_f(\mathbf{r}_1,.., \mathbf{r}_{n-1}) &= \langle \delta_f(\mathbf{x}) \delta_f(\mathbf{x}+\mathbf{r}_1) ...\delta_f(\mathbf{x}+\mathbf{r}_{n-1}) \rangle_c \\
                                             &= \Xi(\mathbf{r}_1,...,\mathbf{r}_{n-1}) + \sum_{m=1}^{n-1}\frac{1}{\bar{n}^m} \mathcal{A}^{(n)}_{m}(\mathbf{r}_1,...,\mathbf{r}_{n-1})
\end{split}
\end{equation}
where the function $\mathcal{A}^{(n)}_m$ contains all the scale dependency of the shot-noise contribution to the $n$-point correlation function at the respective order in $\bar{n}$, the mean density of points in the volume $V$. Therefore, the shot noise takes the form of a power series in $1/\bar{n}$. 
In particular for the 2PCF we have 
\begin{equation}\label{eq:SN_2PCF}
    \xi_f(\mathbf{r}) = \xi(\mathbf{r}) + \frac{\delta_D(\mathbf{r})}{\bar{n}},
\end{equation}
and for the three-point correlation function
\begin{equation}\label{eq:SN_3PCF}
    \begin{split}
        \zeta_f(\mathbf{r}, \mathbf{s}) & = \langle \delta_f(\mathbf{x}) \delta_f(\mathbf{x}+\mathbf{r}) \delta_f(\mathbf{x}+\mathbf{s}) \rangle_c \\
                                        & = \zeta(\mathbf{r}, \mathbf{s}) + \frac{1}{\bar{n}}\left[\delta_D(\mathbf{r}) \xi(\mathbf{s}) \right.\\
                                        & \left. \quad + \delta_D(\mathbf{r} - \mathbf{s}) \xi(\mathbf{r}) + \delta_D(\mathbf{s}) \xi(\mathbf{r} - \mathbf{s})  \right] \\
                                        & \quad + \frac{1}{\bar{n}^2} \delta_D(\mathbf{r}) \delta_D(\mathbf{s}).
\end{split}
\end{equation}
As a result, one can express each individual term of Eq.~\eqref{wf_split} in terms of the true signal and a shot-noise contribution (depending on the number density of objects) as
\begin{equation}\label{eq:SN_individual_toy_model}
\begin{split}
\sigma_{Rf}^2 & = \sigma_{R}^2 + \frac{F({\bf 0})}{\bar N} \\
\xi_{Rf}({\bf r}) & = \xi_{R}({\bf r}) + \frac{F({\bf r}/a)}{\bar N} \\
\zeta_{Rf}({\bf r}) & = \zeta_{R}({\bf r} ) +  \frac{\xi({\bf r})}{\bar N} \left [F({\bf 0})+ F({\bf r}/a) \right ],   
\end{split}
\end{equation}
where we introduce $\bar N=a^3\bar n$ that corresponds to the mean number of objects per grid cell.
These noise contributions are obtained by using Eq.~\eqref{eq:smoothed_deltaRf} in Eq.~\eqref{eq:toy_model_indiv_terms}, inserting Eq.~\eqref{eq:SN_2PCF} and \eqref{eq:SN_3PCF}, and integrating out the Dirac delta functions where applicable. It has to be noted that we only report noise contributions in Eq.~\eqref{eq:SN_individual_toy_model} that are not proportional to Dirac delta functions, as these would appear at zero lag only and hence be irrelevant for our considerations.
By utilising the aforementioned splitting into signal and noise we can write  Eq.~\eqref{wf_split} as 
\begin{equation}
w_f(\mathbf{r}) = w(\mathbf{r}) + \sn_w(\mathbf{r}),
\label{wf_sn}
\end{equation}
where $w$ is the true signal and the shot-noise contribution is formally expressed as
\begin{equation}\label{eq:toy_model_SN_contr_w}
\sn_w(\mathbf{r})  =  \frac{1}{\bar N} \left ( 1+\xi({\bf r}) \right )\left [ F({\bf 0}) + F({\bf r}/a) \right ].
\end{equation}
Equation~\eqref{eq:toy_model_SN_contr_w} shows that even if on a scale $r$ larger than the smoothing scale (when $F({\bf r}/a)=0$) there is still a large-scale contribution to the shot noise due to $F({\bf 0})$.
In addition, the large-scale contribution is expected to decrease when increasing the order of the MAS ($F({\bf 0})$ is decreasing). That is the reason why, in general, increasing the order of the MAS is reducing the intrinsic shot-noise contribution to the signal.

Following the same reasoning, it is straightforward to show that with the toy model the shot-noise affected mean mark $\bar m_f$, estimated from a discrete set of objects, can be related to the true mean mark $\bar m$ via 
\begin{equation}
\bar m_f = \bar m + \sn_{\bar m},
\label{mf_sn}
\end{equation}
where 
\begin{equation}\label{eq:toy_model_SN_contr_mbar}
\sn_{\bar m} = \frac{F({\bf 0})}{\bar N}.
\end{equation}
Finally, by combining Eq.~\eqref{wf_sn} and \eqref{mf_sn} we can show that the shot-noise-corrected weighted correlation function $1+W(\mathbf{r})$ can be expressed as 
\begin{equation}\label{eq:toy_model_comb_SN}
1 + W({\bf r}) = \frac{w_f(\mathbf{r})-\sn_w(\mathbf{r})}{\bar{m}_f - \sn_{\bar m}}.
\end{equation}
This demonstrates that the shot-noise correction on the marked correlation function implies to correct both the numerator $w_f(\mathbf{r})$ and denominator $\bar m_f$ in order to properly extract the true signal.
This is at odd with usual shot-noise corrections on $n$-point correlation functions that are only additive. 

There is another subtlety due to the fact that we assign the mark field back on the galaxies to measure the weighted correlation function. This back-assignment is done with a specific scheme in the sense that we check in which grid cell a galaxy is located and assign the mark corresponding to that grid cell, thereby introducing another smoothing of the field with a NGP kernel.
In our computation this leads to an additional convolution for the field $\delta_{Rf}(\mathbf{x})$.
We show in Section~\ref{sec:app_SN} that this additional convolution is equivalent to a single one with a kernel that is a convolution of both a PCS and a NGP kernel, that is, a quartic kernel.
Hence, in the actual calculation of the analytic shot noise as in Eq.~\eqref{eq:toy_model_SN_contr_w} and \eqref{eq:toy_model_SN_contr_mbar}, a quartic kernel has to be used, which explicit expression can be found in the appendix of \citet{Chaniotis2004JCoPh}.

In order to validate the analytic prediction of the noise in $w_f(\mathbf{r})$ and $\bar{m}_f$ we use five realisations of Covmos as introduced in Section~\ref{sec:sim_baseline}. The goal is to have realisations at different number densities to assess the behaviour of shot noise as a function of $\bar{N}$.
Therefore, for each realisation we deplete the catalogue by randomly throwing away points down to the desired density. The exact densities are motivated by applying the shot-noise correction later to the ELEPHANT simulation suite, which has much lower point densities compared to Covmos. Hence by depleting the Covmos realisations down to $\{1.7\%, 1.53\%, 1.36\%, 1.19\%, 1.02\%, 0.85\%, 0.68\%, 0.51\% \}$ we generate catalogues with the same $\bar{N}$ as in the ELEPHANT suite if they were depleted down to $\{100\%,90\%,80\%,70\%,60\%,50\%,40\%,30\%\}$ with 64 grid cells per dimension.
The depletion is done to match the density of points in grid cells and not in the full volume because the shot-noise behaviour is a power series in $1/\bar{N}$.
The depletion is repeated 100 times followed by a mean to minimise the sample variance coming from the stochasticity of the random depletion process. We need to carefully distinguish the five independent realisations of Covmos from the depletion realisations used to get a converged result for a depleted catalogue, which has to be done for each of the five independent realisations.

\begin{figure}
    \centering
    \includegraphics[width=\columnwidth]{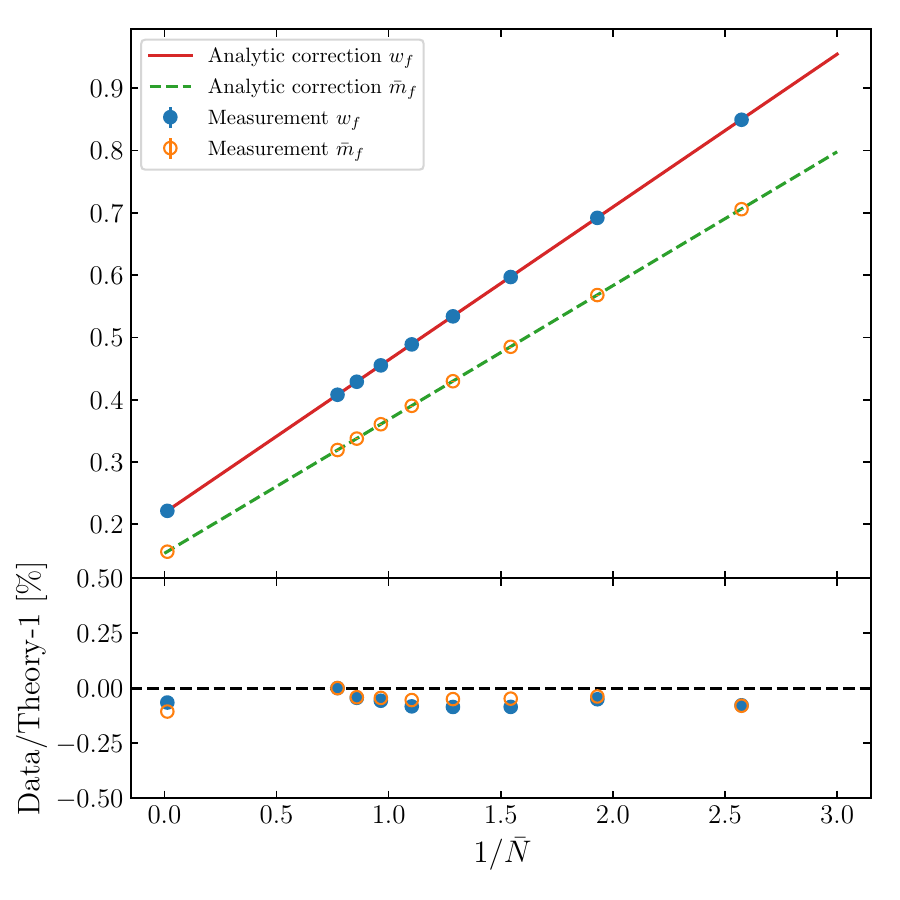}
    \caption{Analytical correction for $w_f(\mathbf{r})$ and $\bar{m}_f$ in case of the toy model where only one galaxy in the pair is weighted by the density contrast $\delta_{Rf}$.
    The points refer to the measured data from one of the Covmos realisations, in blue for $w_f$ and in orange for $\bar{m}_f$.
    The upper panel shows the measurements alongside the analytical prediction and the bottom panel presents the relative difference between the measurements and the theory.
    We show only one Covmos realisations for which we used 100 realisations of depletions to obtain the depleted catalogues and the scale bin in $r$ for $w_f$ is fixed to be close to 20$\Mpc$.
    For the analytic correction we use the depletion down to 1.7\% as an anchor and computed from there the expected signal using Eq.~\eqref{eq:toy_model_SN_contr_w} and \eqref{eq:toy_model_SN_contr_mbar}.}
    \label{fig:SN_toy_model_analytic_correction}
\end{figure}

As we have seen in Eq.~\eqref{eq:toy_model_comb_SN} we need to measure $w_f(\mathbf{r}) = (1+W_f(\mathbf{r}))\bar{m}_f$ as well as $\bar{m}_f$ and those can be straightforwardly computed from the weighted correlation function at each level of depletion.
In the upper panel of Figure~\ref{fig:SN_toy_model_analytic_correction} we present the measurements in one Covmos realisation of $w_f(\mathbf{r})$ (blue points) and $\bar{m}_f$ (orange points) as a function of $1/\bar{N}$. The scale at which we plot $w_f(\mathbf{r})$ is fixed to a bin of $20~\Mpc$.
We can already see that we there is a linear relation with $1/\bar{N}$ as predicted by the expression in Eq.~\eqref{eq:toy_model_SN_contr_w} and Eq.~\eqref{eq:toy_model_SN_contr_mbar}.
The solid and dashed curve refer to the analytical prediction using the depletion case to 1.7\% (the second data point from the left) as an anchor. That anchor is needed to obtain a noiseless signal by correcting for shot noise and then add to the true signal the noise contribution, as a function of $1/\bar{N}$, to obtain the curve. By doing so, the relative difference between the prediction and the measurement is exactly zero by construction for this depletion as can be seen in the lower panel of Figure~\ref{fig:SN_toy_model_analytic_correction}. Moreover, even for the other data points at different levels of depletion we can predict the expected signal with high accuracy. The relative difference is at the sub-percent level for both $w_f(\mathbf{r})$ and $\bar{m}_f$.

\begin{figure}
    \centering
    \includegraphics[width=\columnwidth]{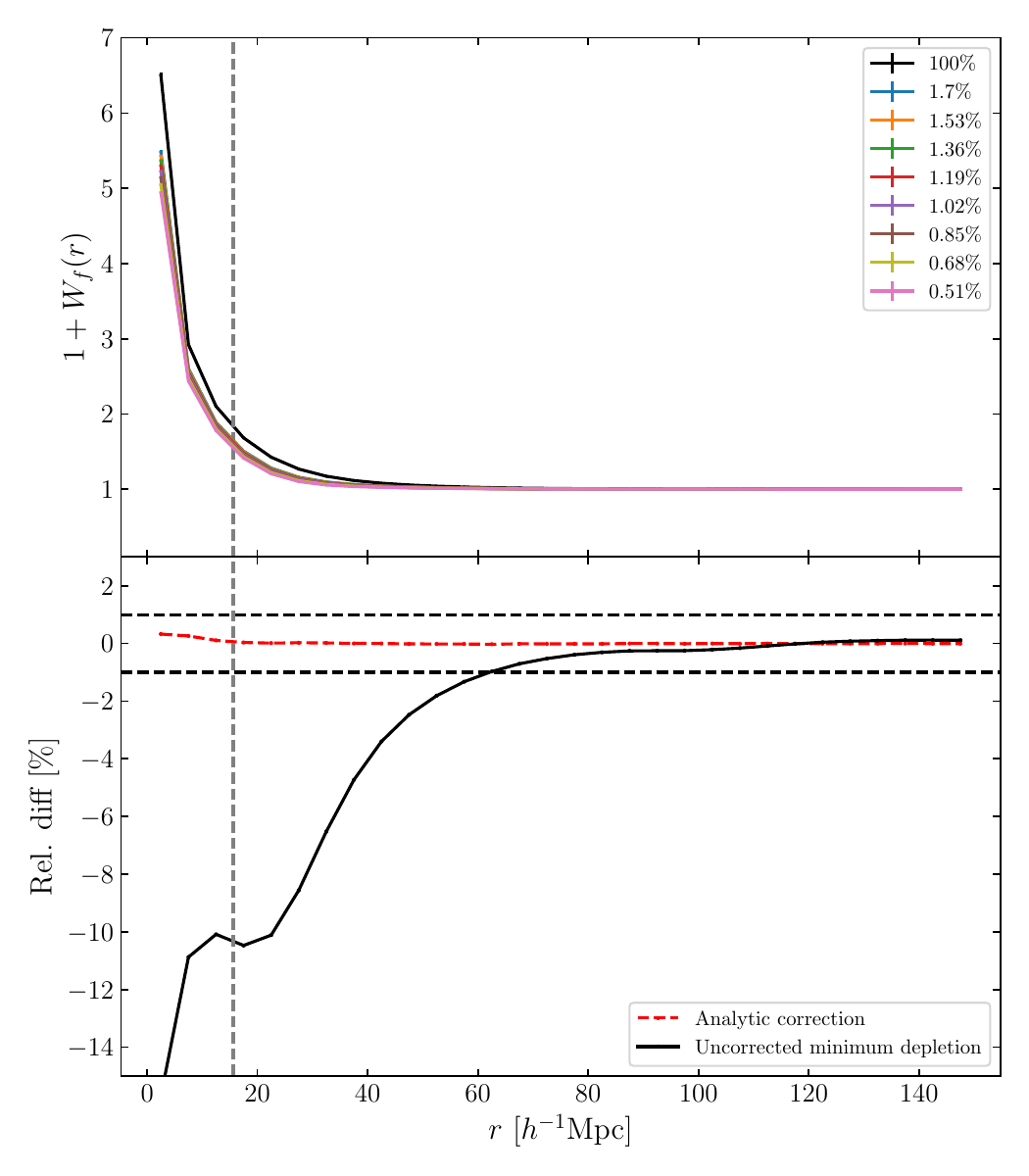}
    \caption{Results of the analytic shot-noise correction applied to the toy model for which only one point in a pair is weighted by the density contrast $\delta_{Rf}$.
    The upper panel shows the full weighted correlation function $1+W_f$ as a mean over the 5 Covmos realisations, where different colours denote different levels of depletion.
    Errorbars are computed by taking the mean standard deviation over 5 realisations.
    To obtain the depleted catalogues we took the mean over 100 depletions.
    The lower panel shows the relative difference of the analytically corrected result to the undepleted case as a red dashed line.
    The solid black line refers to the relative difference of the depletion level $1.7\%$ to the undepleted case, which illustrates the effect of no correction.
    Horizontal dashed lines in black indicate levels of relative differences of $\pm1\%$ and the vertical dashed line in grey refers to side length of one grid cell.
    We used 64 grid cells per dimension and a PCS MAS to obtain the density field on the grid.
    }
    \label{fig:SN_toy_model_results}
\end{figure}

Now that we have established the correctness of our analytical predictions for $w_f(\mathbf{r})$ and $\bar{m}_f$ individually, we can check how well they perform when combining them into $1+W_f(\mathbf{r})$, as shown in Figure~\ref{fig:SN_toy_model_results}. In the upper panel we present the mean over the five Covmos realisations at different levels of depletion as indicated with different colours in the legend.
As expected, since the kernel and 2PCF drop off at large separations, differences in the curves are only evident on smalle scales. This is further underlined by the lower panel where the relative difference between the depletion down to 1.7\% and the undepleted case is shown in black.
This curve refers to the difference between the two if we would not have applied any correction and only data with a depleted number density of 1.7\% would be available.
For $1+W_f(\mathbf{r})$ the relative difference can reach more than 10\% on small scales but at scales above around 60$\Mpc$ the depleted and undepleted case lay within 1\% and the effect of shot noise becomes negligible. This is somewhat expected due to the smoothing of the density field, as the quartic kernel decreases down to zero over the course of 2.5 grid cells, which in Covmos corresponds to $\approx$40$\Mpc$.
It is important to note here that, although we expect shot noise to be stronger when correlating within the volumes of the smoothing kernels, it is peculiar to the toy model that the shot-noise contribution does only contain linear factors of the kernel with and without the 2PCF.
It can be shown in the more general case that if one weights both galaxies in a pair by the associated density field, then the shot noise will contain contributions from a convolution of two quartic kernels resulting in a nonic kernel, which is much more extended in configuration space. In contrast to the black curve that has no correction, we show the relative difference of the analytical correction for $1+W_f(\mathbf{r})$ to the undepleted case in red. To have a fair comparison we corrected the 1.7\% case down to the density of the undepleted realisations, as these still have a finite, yet very high density. The analytical correction reproduces the undepleted measurements to within 1\% relative difference on all scales.
We conclude that for the toy model we are able to analytically predict the shot noise. Moreover, we show that even with this simple toy model the shot noise acquires a non-trivial scale dependency.
In the next section we extend this formalism to general weighted correlation functions and describe a procedure to estimate the signal without having to rely on an analytical model.

\subsection{A general model}

Building on top of the results obtained with the toy model, we can devise a general model that has a mark function expandable in powers of the density contrast as 
\begin{equation}\label{eq:mseries}
    M[\delta_{Rf}(\mathbf{x})]  = \sum_{i=0}^{\infty} \frac{c_i}{i!} \delta_{Rf}^i(\mathbf{x}),
\end{equation}
where $c_i$ are the coefficients of the Taylor series. Plugging the series expansion in Eq.~\eqref{eq:mseries} into Eq.~\eqref{eq:weighted_corr} we arrive at
\begin{equation}\label{eq:taylor_wf}
    w_f({\bf r}) =  \sum_{i,j} \frac{c_i c_j}{i!j!} \langle \delta_{Rf}^i(\mathbf{x}) (1+\delta_f(\mathbf{x})) \delta_{Rf}^j(\mathbf{x}+\mathbf{r})(1+\delta_f(\mathbf{x}+\mathbf{r}))\rangle
\end{equation}
and
\begin{equation}\label{eq:taylor_mbar}
    \bar{m}_f = \sum_i \frac{c_i}{i!}  \langle \delta_{Rf}^i(\mathbf{x}) \frac{\rho_f(\mathbf{x})}{\bar{\rho}_f}\rangle.
\end{equation}
It is evident from these expressions that by weighting both galaxies in a given pair, the resulting marked correlation function will contain auto-correlation contributions of the mark with itself.
If for example the weight is constructed by an external catalogue of voids then the weighted correlation function will consist of a smoothed version of the void auto-correlation and void-galaxy cross-correlation functions. In Section~\ref{sec:app_cross}, we give some further insights in how the weighted correlation function can be split up into two auto- and one cross-correlation function for certain weighting schemes.

In order to work out the shot-noise contribution to $w_f(\mathbf{r})$ and $\bar{m}_f$ we can use, analogously to Eq.~\eqref{eq:general_SN}, the relation
\begin{equation}\label{eq:general_SN_ij}
\begin{split}
    &\langle \delta_{Rf}^i(\mathbf{x})(1+\delta_f(\mathbf{x}))\delta_{Rf}^j(\mathbf{x}+\mathbf{r})(1+\delta_f(\mathbf{x}+\mathbf{r}))\rangle & \\
    &= \langle \delta_{R}^i(\mathbf{x})(1+\delta(\mathbf{x}))\delta_{R}^j(\mathbf{x}+\mathbf{r})(1+\delta(\mathbf{x}+\mathbf{r}))\rangle + \sum_{p=1}^{i+j+1} \frac{1}{\bar{N}^p}\mathcal{B}^{(i,j)}_p(\mathbf{r}),
\end{split}
\end{equation}
where $\mathcal{B}^{(i,j)}_p(\mathbf{r})$ contains the shot-noise contribution for a given $(i,j)$ proportional to the inverse of $\bar{N}$ to the power of $p$. Inserting this expression into Eq.~\eqref{eq:taylor_wf} we obtain
\begin{equation}
\begin{split}
    w_f(\mathbf{r}) = & \sum_{i,j} \frac{c_ic_j}{i!j!} \langle \delta_{R}^i(\mathbf{x})(1+\delta(\mathbf{x}))\delta_{R}^j(\mathbf{x}+\mathbf{r})(1+\delta(\mathbf{x}+\mathbf{r}))\rangle \\
                    & + \sum_{i,j} \frac{c_ic_j}{i!j!}\sum_{p=1}^{i+j+1} \frac{1}{\bar{N}^p}\mathcal{B}^{(i,j)}_p(\mathbf{r}) \\
                    = & w(\mathbf{r}) + \sn_{w}(\mathbf{r}),
\end{split}
\end{equation}
where we identified in the second equality the first sum to be the desired true signal $w(\mathbf{r})$ and the second sum to be the shot-noise contribution $\sn_w(\mathbf{r})$.
Similarly, for $\bar{m}_f$ we obtain
\begin{equation}
    \bar{m}_f = \bar{m} + \sum_i \frac{c_i}{i!} \sum_{p=1}^{i} \frac{1}{\bar{N}^p} \mathcal{B}^{(i)}_p = \bar{m} + \sn_{\bar{m}}.
\end{equation}
At this point it is clear that the double sum can be written as a power series of $1/\bar{N}$ such that 
\begin{equation}\label{eq:SN_poly}
    \sn_w(\mathbf{r}) = \sum_{p=1}^{\infty}\frac{1}{\bar N^p} \sn_{w,p}(\mathbf{r}) \qquad \sn_{\bar{m}} = \sum_{p=1}^{\infty}\frac{1}{\bar N^p} \sn_{\bar{m}, p},
\end{equation}
with correspondingly defined $\sn_{w,p}(\mathbf{r})$ and $\sn_{\bar{m},p}$.
The correction in the general case is therefore, analogously to Eq.~\eqref{eq:toy_model_comb_SN},
\begin{equation}\label{eq:main_SN_corr}
    1+W(\mathbf{r}) = \frac{ w_f(\mathbf{r}) - \sn_w(\mathbf{r})}{(\bar{m}_f - \sn_{\bar m})^2}.
\end{equation}
The power series in $1/\bar{N}$ (which in principle extends to infinite order) together with the fact that, following Eq.~\eqref{eq:SN_2PCF} and \eqref{eq:SN_3PCF}, shot noise of $n$-point correlation functions is scale-dependent and contains $(n-1)$-point correlation functions, makes an analytic correction for the general case untractable. It would require in particular to compute higher-order correlation functions, which are computationally expensive. One might think that a simple truncation of the Taylor expansion would solve the problem, but to avoid computing four-point correlators and above, the Taylor expansion would need to be cut already at linear order. Moreover, the conversion of moments into cumulants might lead to significant contributions from higher-order correlators at low order in $1/\bar{N}$. In the following we outline an approach to circumvent analytical computation and that uses the resummation of contributions into a power series in $1/\bar{N}$.

The quantities $w_f=(1+W_f(\mathbf{r}))\bar{m}_f^2$ as well as $\bar{m}_f$ are directly measurable from simulations for a given mark. We propose therefore an algorithm consisting of a polynomial fit through measurements of $w_f(\bar{N})$ and $\bar{m}_f(\bar{N})$ made at different levels of depletion, that is, at different values of $1/\bar{N}$. For $w_f(\mathbf{r})$, the fit is done with the same polynomial order for each bin in $r$ but with separate coefficients, which is necessary since the shot noise $\sn_w(\mathbf{r})$ is scale-dependent. With such a polynomial we can simply read off the noiseless signal from the $y$-axis intersection as this gives the extrapolation to $1/\bar{N}=0$, i.e. infinite densities.
It is important to note that truncating the fit at some polynomial order is not the same as truncating the Taylor expansion in $\delta_{Rf}$ as the linear coefficient in the power series contains the resummed contributions from all higher-order correlators as well. To test this approach and find the best order of polynomial to fit, we used the same depletion levels as described in the previous section.

\begin{figure*}
    \centering
    \includegraphics[width=0.9\textwidth]{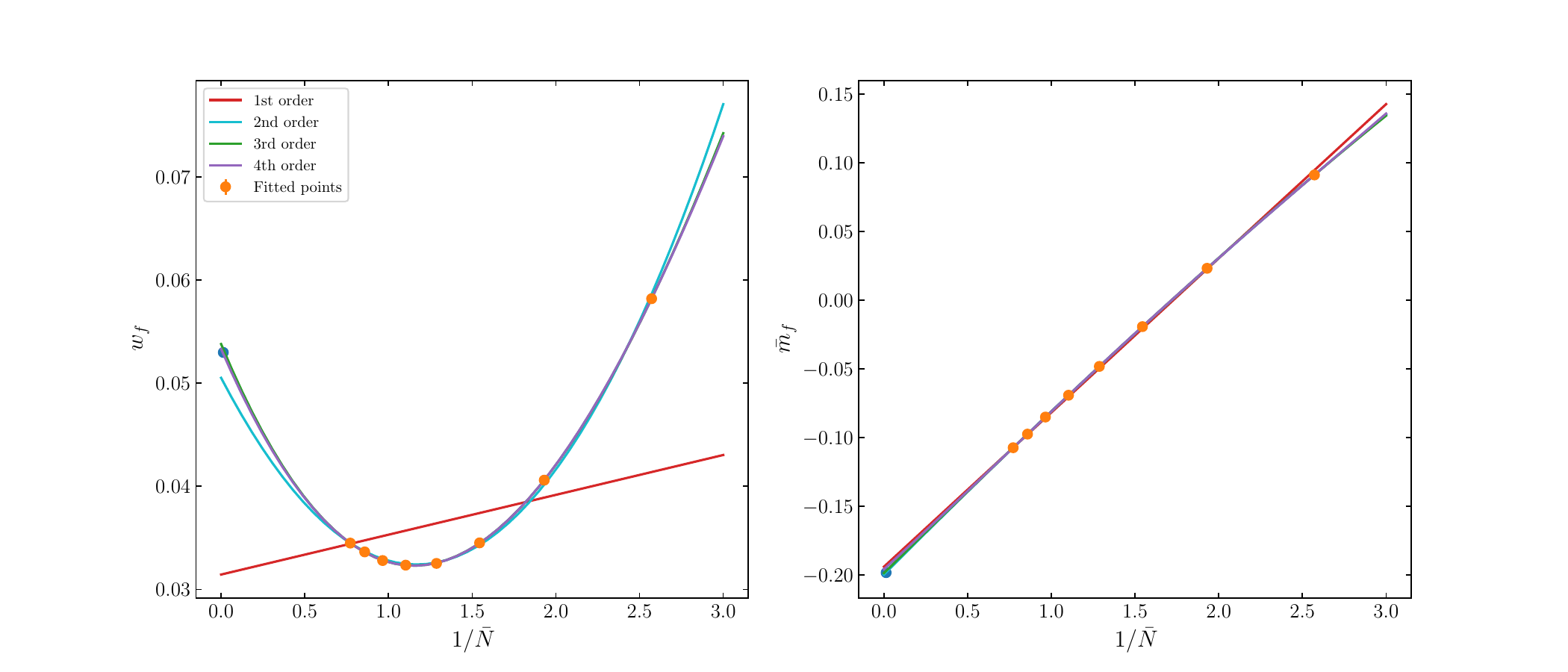}
    \caption{Fitting procedure to obtain the shot-noise-corrected signal in case of $w_f(\mathbf{r})$ (left panel) and $\bar{m}_f$ (right panel) for the $\tanh$-mark in the configuration $(a,b)=(0.6,-0.5)$.
    Following Eq.~\eqref{eq:SN_poly} we present the
    fits in dependency of $1/\bar{N}$, the reciprocal of the average number of points per grid cell.
    We show only one realisation of Covmos and the scale bin in $r$ for $w_f$ is fixed to be close to $20$$\Mpc$.
    The depleted measurements were obtained by taking the mean over 100 realisations of depletions.
    The orange points refer to the fitted measurements and the blue point is the undepleted reference (not included in the fit).
    Differently coloured lines refer to different orders in the polynomials we used to fit the data.
    Errorbars of the orange points are obtained by taking the mean standard deviation over the 100 realisations.
    Since the depletion down to 1.7\% (first orange point from the left) mimics the undepleted ELEPHANT density we use for this point an error that is 10\% of the minimum uncertainty over the remaining depletions.
    }
    \label{fig:SN_tanh_fit_results}
\end{figure*}

In Figure~\ref{fig:SN_tanh_fit_results} we present the results from polynomial fits to the quantities $w_f(\mathbf{r})$ and $\bar{m}_f$ as appearing in Eq.~\eqref{eq:main_SN_corr}. It is evident that for the general case, higher-order shot-noise contributions play an important role resulting in a more curved shape due to quadratic and cubic dependencies on $1/\bar{N}$. Therefore, a simple linear fit is not sufficient anymore and at least 2nd- or 3rd-order polynomials are to be used. Going to even higher orders, as we show with a 4th-order polynomial in purple, the behaviour outside of the fitted range becomes more unstable and can lead to severe over- or under-estimation of the true signal at the $y$-axis intersection.
Moreover, since we only employ 8 data points, it is crucial to keep the polynomial order as low as possible since otherwise an overfitting of the data might happen. In contrast to the behaviour for $w_f(\mathbf{r})$, the dependency of $\bar{m}_f$ on $1/\bar{N}$ appears to be much more linear and fits with 1st- or 2nd-order polynomials should be sufficient to recover the true signal accurately.

\begin{figure*}
    \centering
    \includegraphics[width=0.45\textwidth]{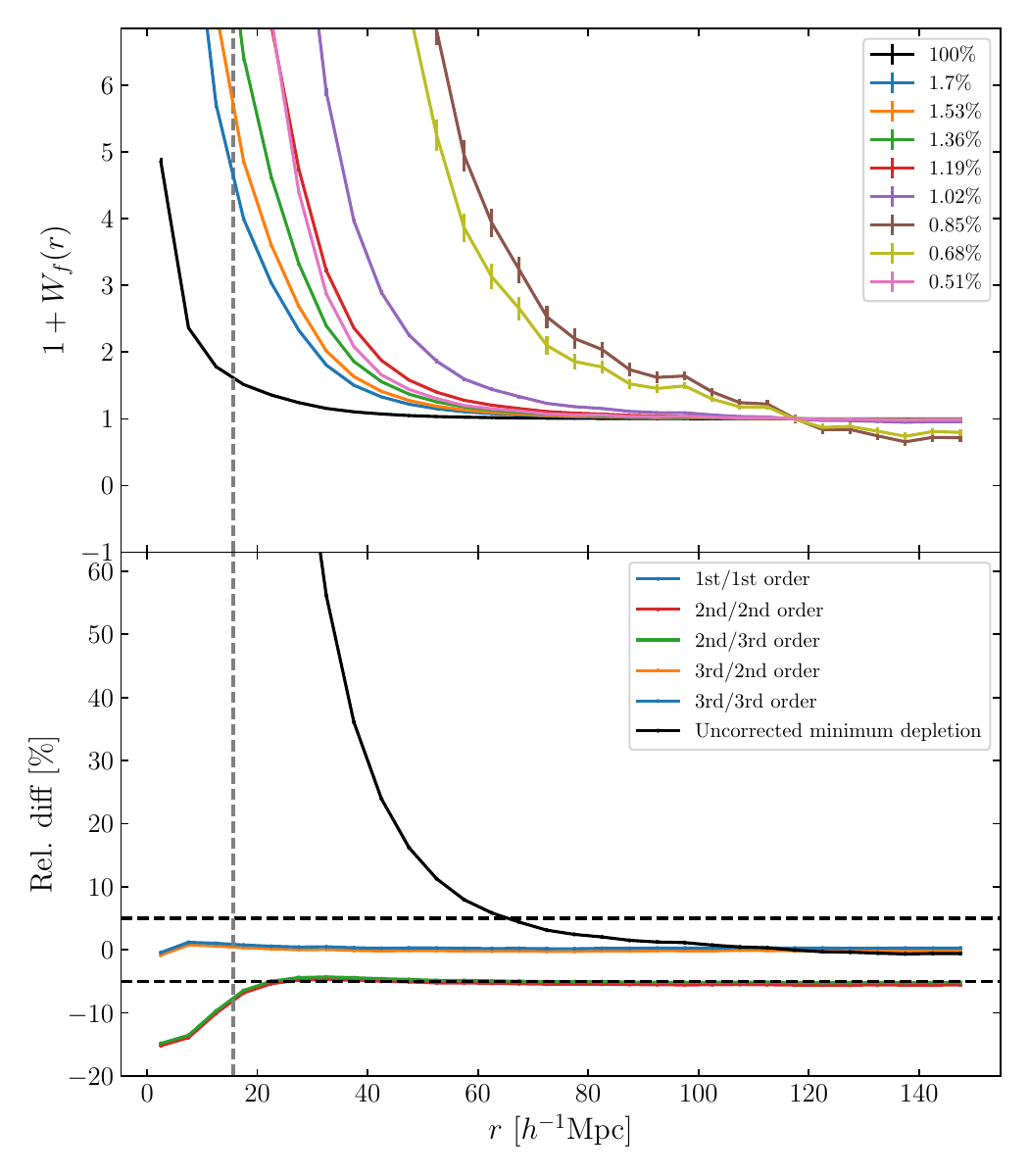}
    \includegraphics[width=0.45\textwidth]{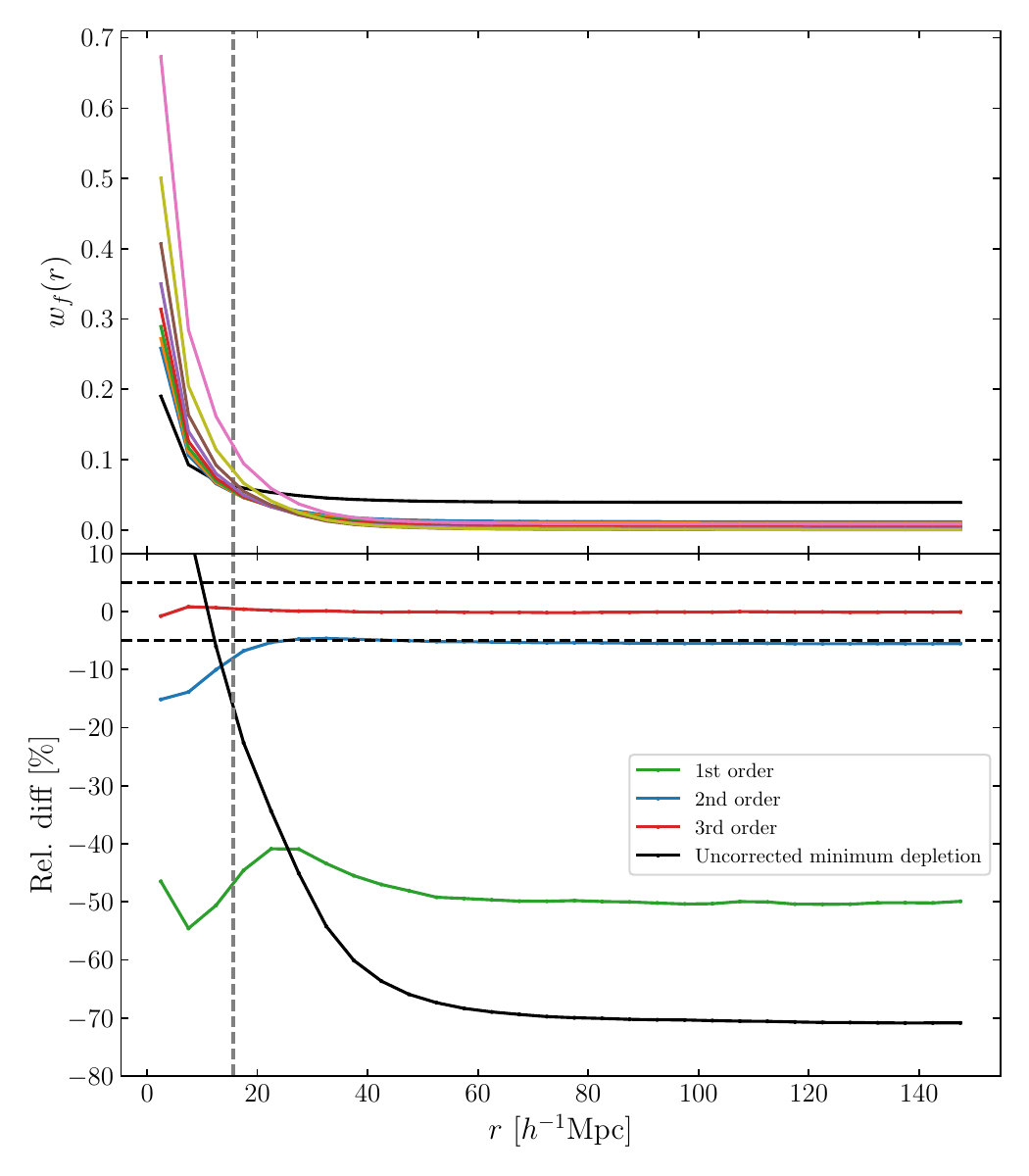}
    \caption{Results of the fitted shot-noise correction for the $\tanh$-mark with $(a,b)=(0.6, -0.5)$.
    The left panels present $1+W_f$ while the right panel shows $w_f = (1+W_f)\bar{m}_f^2$.
    Different colours in the upper panel refer to the mean over 5 Covmos realisations and errorbars correspond to the mean standard deviation over those 5 realisations.
    Depleted measurements were obtained by taking the mean over 100 realisations of depletions.
    The bottom panels show the relative differences of the corrected signal from the fit to the undepleted reference case.
    Different colours in the lower panels refer to different orders used in the polynomial fit and e.g. '3rd/2nd' indicates that a 3rd- and 2nd-order polynomial fit was used for $w_f$ and $\bar{m}_f$, respectively.
    The horizontal dashed lines in black corresponds to a relative difference of $\pm 5\%$ and the vertical dashed line in grey refers to the side length of one grid cell.
    }
    \label{fig:SN_tanh_results}
\end{figure*}

In Figure~\ref{fig:SN_tanh_results} the performance of the different correction orders as well as the weighted correlation function $1+W_f(\mathbf{r})$ and the quantity $w_f(\mathbf{r})$ are shown.
The diverging behaviour of $1+W_f(\mathbf{r})$ in the upper left panel at the depletions down to $0.68\%$ and $0.85\%$ (olive green  and brown lines, respectively) can be understood when looking at the corresponding points of the mean mark $\bar{m}_f$ in Figure~\ref{fig:SN_tanh_fit_results}, that is the second and third to last data point at $1/\bar{N}$ of around 2.0 and 1.6. The mean mark is very close to zero in that case and therefore $1+W_f(\mathbf{r})$ acquires a very large amplitude due to the division by $\bar{m}_f^2$. 
This behaviour is somewhat peculiar to marks that can switch signs, as the $\tanh$-mark, because in certain cases this can lead to a mean mark which is very close to zero.
Moreover, this can result in a turn-around in the dependency on $1/\bar{N}$ as seen for the last two depletions, $0.68\%$ and $0.51\%$, both in Figure~\ref{fig:SN_tanh_fit_results} and \ref{fig:SN_tanh_results}.
While a very small mean mark does not appear to be problematic for the fit, it can be an issue if the true mean mark is very close to zero.
Since we use only 8 data points in the fit, we have a limited accuracy on the recovery of $\bar{m}_f$.
This can become problematic as soon as the amplitude of the recovered mean mark approaches the accuracy of the fit, leading to very large relative uncertainties on $\bar{m}_f$ and on $1+W_f(\mathbf{r})$.
In this case, the accuracy of the polynomial fit is not enough to properly recover small mean marks and the results should not be trusted. Even though one could try to mitigate this issue and improve the accuracy of the fit with better estimates of the data points from larger sets of depleted catalogues, in general, we advise against using marks with a recovered mean mark being very close to zero.

In the upper panel to the right of Figure~\ref{fig:SN_tanh_results}, we show the measurements of $w_f(\mathbf{r})$. While the noise behaviour on small scales appears to be less severe compared to $1+W_f(\mathbf{r})$, on large scales we can observe a constant offset. Focusing on the relative differences in the lower left panel of Figure~\ref{fig:SN_tanh_results}, it is evident that for this particular mark the effect of shot noise diminishes at scales of around 60$\Mpc$ as the relative difference between the undepleted case and uncorrected measurement at 1.7\% depletion shrinks to below 5\%.
This is in contrast to the toy model in Figure~\ref{fig:SN_toy_model_results} as here the 5\% border is crossed already at scales of around 40$\Mpc$. This illustrates the fact that the shot-noise behaviour can have different amplitudes and scale-dependency that is subject to the chosen mark.
The coloured lines in the lower panels refer to different orders in the polynomial fit used for $w_f(\mathbf{r})$ and $\bar{m}_f$. From this we can conclude that fitting the behaviour of $w_f(\mathbf{r})$ with a 3rd-order and $\bar{m}_f$ with a 2nd-order polynomial results in a satisfactory performance and should  be used hereafter as the adequate shot-noise correction. With this choice, the relative difference in $1+W_f(\mathbf{r})$, depicted by the orange line in Fig.~\ref{fig:SN_tanh_results}, is within 5\% across all scales all the way up to 150$\Mpc$.

\begin{figure}
    \centering
    \includegraphics[width=\columnwidth]{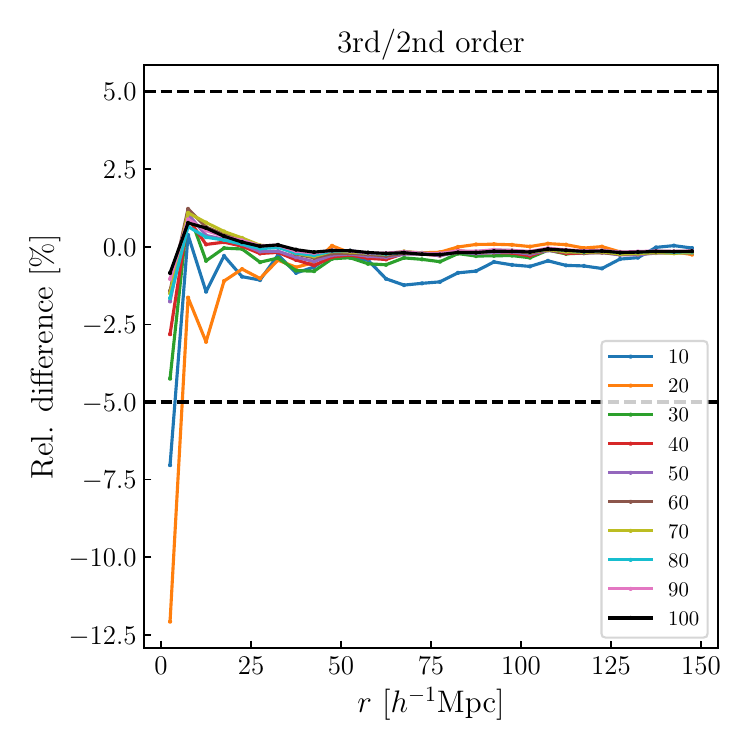}
    \caption{Test of convergence of the shot-noise correction for different amounts of depletion realisations. Black dashed lines refer to relative differences of $\pm5\%$ and colours denote the number of realisations for the depletions. Relative differences are shown for the total $1+W_f(r)$ where both $w_f(r)$ and $\bar{m}_f$ are corrected for the $\tanh$-mark with $(a,b)=(0.6, -0.5)$.}
    \label{fig:SN_tanh_convergence_results}
\end{figure}

Now that we have an optimal choice for the polynomial orders to describe the shot noise as a function of $1/\bar{N}$, we need to assess how many realisations of depletions yield converged results for the polynomial fits. Since the process of depletion consists in randomly throwing away a number of points such that we end up with some desired percentage of the original points, it is inherently noisy and should be repeated several times. The aim is to obtain a representation of the original catalogue with a lower density that looks like as if the simulation has been run with lesser points in the first place.
In Figure~\ref{fig:SN_tanh_convergence_results} we show the best correction as obtained from Figure~\ref{fig:SN_tanh_results}, being the 3rd/2rd-order polynomial, and compute relative differences for different amounts of realisations of depletion.
As we can see, using 30 realisations or more, the curves do not differ substantially and results can be considered converged. Even with only 10 or 20 realisations at hand the performance is nevertheless acceptable and well within 5\% except for the lowest bin in $r$. As a conservative choice, we will use 30 realisations in the following. This should allow mitigating sample variance and not affecting resulting corrections.

\subsection{Shot noise in redshift space}

In real observations, the quantity of interest in clustering analyses are usually the multipoles of the anisotropic 2PCF. In the following we show that the correction to the marked correlation function in redshift space is similar to that in real space. We indicate redshift-space quantities via their explicit dependency on the separation $s$ and angle $\mu$, but also the mean mark has to be understood as measured in redshift space.
For the full anisotropic marked correlation function $\mathcal{M}_f(s, \mu)$, the shot noise enters in the following way
\begin{equation}
\begin{split}
    \mathcal{M}_f(s, \mu) & = \frac{1+W_f(s, \mu)}{1+\xi(s, \mu)} = \frac{(1+W(s, \mu)) \bar{m}^2 + \sn_w(s, \mu)}{(1+\xi(s, \mu))(\bar{m}+\sn_{\bar{m}})^2} \\
    & = \mathcal{M}(s, \mu) \frac{\bar{m}^2}{(\bar{m}+\sn_{\bar{m}})^2} + \frac{\sn_w(s, \mu)}{(1+\xi(s, \mu))(\bar{m}+\sn_{\bar{m}})^2}, 
\end{split}
\end{equation}
where we used the fact that the unweighted 2PCF is not affected by shot noise and hence does not have the subscript $f$. Since the shot noise does contain $n$-point correlators, it will acquire an angle dependency as well. After decomposing the anisotropic marked correlation function into multipoles $\mathcal{M}_{f,\ell}(s)$ we obtain
\begin{equation}
    \begin{split}
        \mathcal{M}_{f,\ell}(s) = \mathcal{M}_{\ell}(s) \frac{\bar{m}^2}{\bar{m}_f^2} + \frac{2\ell+1}{2\bar{m}_f^2} \int_{-1}^1 P_{\ell}(\mu) \widetilde{\sn}_w(s, \mu) \text{d} \mu,
    \end{split}
\end{equation}
with 
\begin{equation}
\begin{split}
    \widetilde{\sn}_w(s, \mu) = \frac{\sn_w(s, \mu)}{(1+\xi(s, \mu))}.
\end{split}
\end{equation}
Solving the former expression for the true signal $\mathcal{M}_{\ell}(s)$, the shot-noise correction takes the same form analogous to Eq.~\eqref{eq:main_SN_corr}
\begin{equation}
   \mathcal{M}_{\ell}(s) = \frac{\mathcal{M}_{f,\ell}(s) \bar{m}_f^2 - \widetilde{\sn}_{w,\ell}(s)}{(\bar{m}_f - \sn_{\bar{m}})^2},
\end{equation}
with
\begin{equation}
    \widetilde{\sn}_{w,\ell}(s) = \frac{(2\ell+1)}{2} \int_{-1}^1 P_{\ell}(\mu) \widetilde{\sn}_w(s, \mu) \text{d} \mu,
\end{equation}
being the redefined shot-noise contribution to $w_f$ in redshift space. The shot noise will still be, under the assumption of a mark that can be Taylor expanded in powers of $\delta_{Rf}(\mathbf{x})$, a power series in $1/\bar{N}$ because the integral is additive. This means that the previously introduced methodology of fitting polynomials to $w_f(\mathbf{r})$ and $\bar{m}_f$ is applicable to redshift-space multipoles of the marked correlation function as well. In the case of redshift-space multipoles of the weighted correlation function the formulas are the same except the division by $1+\xi(s, \mu)$.
It is important to note here that, since the shot noise acquires a non-trivial angle dependency and hence has to be corrected for in each multipole, also in the marked power spectrum the shot noise will appear in higher multipoles. This is in contrast to the ordinary power spectrum with constant shot noise that will only affect the monopole.

\begin{figure}
    \centering
    \includegraphics[width=\columnwidth]{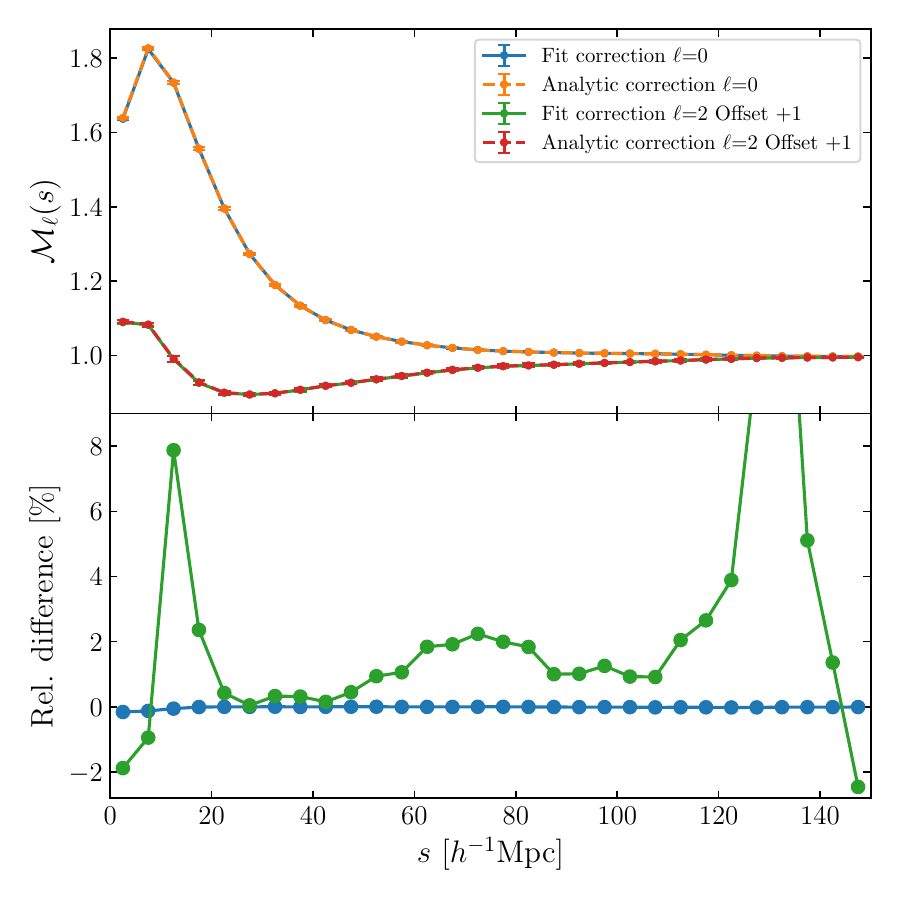}
    \caption{Redshift-space multipoles of the toy model, measured in the GR simulations of ELEPHANT, corrected for shot noise. The upper panel shows both the monopole and quadrupole in different colours corrected via the polynomial fit as solid lines and analytically corrected in dashed lines. In order for better visualisation we offset the quadrupole by +1. Errorbars refer to the mean standard deviation over 5 realisations. The lower panel shows the relative difference between the analytic and polynomial correction in percent.}
    \label{fig:RSS_toy_model}
\end{figure}

In order to test our derived shot-noise correction in redshift space we compute the analytical correction to the toy model in redshift space and compare with our fitting methodology.
As we have seen in Eq.~\eqref{eq:toy_model_SN_contr_w} and Eq.~\eqref{eq:toy_model_SN_contr_mbar} the shot noise in the toy model is described by a linear polynomial in $1/\bar{N}$.
It has to be noted that in this case, analogous to the real-space weighted correlation function, we correct $w_f$ with a linear factor of the mean mark due to only one of the galaxies in each pair being actually weighted. The results can be seen Figure~\ref{fig:RSS_toy_model}, where we plot both the result from the polynomial fit as well as the analytical correction for the monopole and quadrupole.
We find very good agreement between the two methods and the relative difference in the monopole is below 1\% over all scales up to 150$\Mpc$. For the quadrupole, the agreement is worse but still within around 2\% for most of the scales. There are specific spikes in the relative difference caused by the quadrupole crossing zero at about 15$\Mpc$ and approaching zero on large scales.

\subsection{Limits of the shot-noise correction}\label{sec:limits_SN}

It is important to assess where the shot-noise correction breaks down due to the assumptions not being valid anymore. While an exhaustive investigation is beyond the scope of this work, we discuss here several points in order to give conservative limits on when it is safe to apply the proposed method.
The most crucial assumption comes from approximating the mark functional as a Taylor expansion of the density contrast, as well as the derived power series in $1/\bar{N}$ to be approximated by a low-order polynomial.
A Taylor series of a function $f(x)$ has a convergence radius $B$ for which the series converges to the true function if it is evaluated inside the radius, that is, for $|x|\leq B$.
There is a straightforward way to compute the convergence radius of the Taylor series of the $\tanh(x)$, for which techniques such as the ratio test might not be applicable in certain cases.
We can simply compute the closest singularity to the point that we expand around ($x=0$), which gives us the convergence radius. While the $\tanh(x)$ is non-singular on the axis of real numbers, it has singularities in the complex plane. Since $\tanh(x)=\sinh(x)/\cosh(x)$ singularities appear when $\cosh(x)=0$, which is the case at $x=-b-1/a(1/2 i \pi+2 i \pi  c)$ with $c\in \mathbb{Z}$, if we have both a shift $b$ and factor $a$ as in the mark of Eq.~\eqref{eq:tanh_mark}. The closest singularity to $x=0$ is therefore obtained by setting $c=0$, leading to a convergence radius of $B = \sqrt{b^2 + (\pi/(2a))^2}$.
For the White mark the convergence radius is simply given by $B=1+\rho_{*}$ in the case of a positive exponent $p$ in Eq.~\eqref{eq:White_mark}.

In order to assess the effective validity of our correction based on the convergence radius, we can reformulate it into a criterion involving a measurable statistical quantity from the catalogues.
For this, we follow a similar approach as in \citet{Philcox2020PhRvD}.
Starting from the mathematical convergence criterion for the Taylor expansion $|\delta_{Rf}| \leq B$, we take the density-weighted average\footnote{That is $\langle g(\mathbf{x})\rangle_{\rho} = \frac{1}{V}\int g(\mathbf{x}) \frac{\rho_f(\mathbf{x})}{\bar{\rho}_f} \textrm{d}^3x$} of the square on both sides.
After taking the square-root we obtain
\begin{equation}
   \sqrt{\langle \delta_{Rf}^2\rangle_{\rho}} \leq B.
\end{equation}
The left-hand side quantity can be straightforwardly estimated by taking the arithmetic mean of the square of $\delta_{Rf}$ at galaxy positions.
For the Covmos realisations using 64 grid cells per dimension $\sqrt{\langle \delta_{Rf}^2\rangle_{\rho}}$ ranges from 0.61 to 1.01 over the different levels of depletion (1.7\% to 0.048\%).
In contrast, for the ELEPHANT simulations of GR, $\sqrt{\langle \delta_{Rf}^2 \rangle_{\rho}}$ takes values from 1.32 to 1.72 for no depletion down to 30\%, respectively. By choosing $(a,b)=(0.6, -0.5)$ for the $tanh$-mark we have a convergence radius of $B \approx 2.67$, which satisfies the convergence criterion and thus we can trust the Taylor expansion. Furthermore, the White mark with $(\rho_{*},p)=(4.0, 10.0)$ and $(\rho_{*},p)=(1.0,-1.0)$ have convergence radii of $B=5$ and $B=\infty$ therefore they do also fulfil our convergence criterion.

\begin{figure}
    \centering
    \includegraphics[width=\columnwidth]{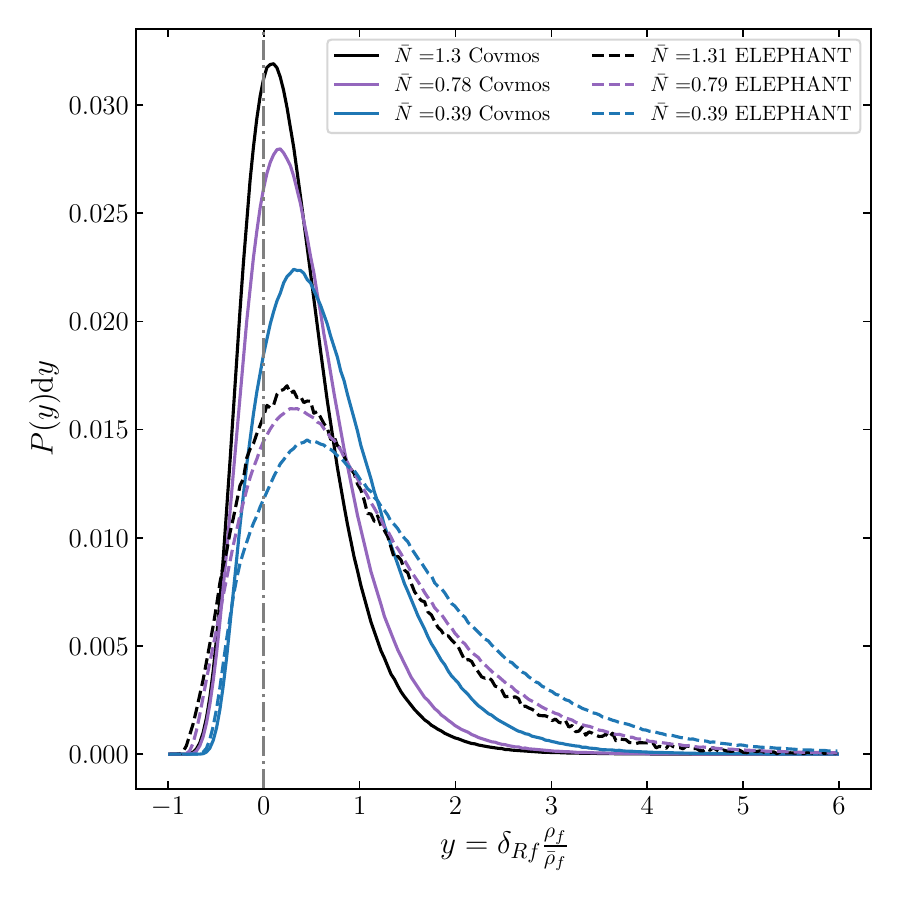}
    \caption{Comparison of the PDF for the density contrast weighted by the galaxie density as measured in Covmos and ELEPHANT (GR). Colours encode the amount of depletion resulting in number densities per grid cell as indicated in the legend. Dashed lines refer to the ELEPHANT simulation while solid lines refer to the Covmos catalogues. We show the mean over 5 realisations and depleted measurement were obtained by taking the mean over 30 realisations of depletions.}
    \label{fig:SN_PDF_comparison}
\end{figure}

The differences between the catalogues used in this analysis affecting the convergence criterion is illustrated in Figure~\ref{fig:SN_PDF_comparison} showing the density weighted PDF of $\delta_{Rf}$.
First thing that has to be noted is the fact that for the black dashed curve, depicting the undepleted ELEPHANT PDF, there is not depletion involved leading to more noise compared to the solid black curve for Covmos. For the latter, we took the mean over 30 realisations of depletion down to $1.7\%$ of the points of the full catalogue to match the undepleted $\bar{N}$ of ELEPHANT.
It is evident from the figure that the PDF for Covmos is more peaked around zero with a less pronounced tail to high densities as exhibited by the ELEPHANT PDF. That difference can be explained by the fact that the Covmos realisations are meant to reproduce the distribution of dark matter particles while the ELEPHANT catalogues are galaxies and hence contain bias. Due to the higher skewness for the PDF in the ELEPHANT simulation, care has to be taken to select appropriate marks not violating the convergence criterion of the Taylor expansion.

The careful reader might have noticed that for the White mark with $(\rho_{*}, p) = (10^{-6}, 1.0)$ the radius of convergence is only around unity and the above criterion would not be valid for the ELEPHANT simulation and only partially valid for the Covmos catalogues.
However, we do find good recovery of the undepleted signal via applying the shot-noise correction to this configuration in the Covmos catalogues.
Moreover, we find good recovery for the $\tanh$-mark with parameters $(a,b)=(10.6, -0.5)$ with a convergence radius of only $B\approx 0.52$ therefore breaking the criterion completely. 
This shows that even if the convergence criterion is not completely fulfilled, it does not directly imply a failure of the shot-noise correction.
Due to the strong skewness of the distributions as shown in Figure~\ref{fig:SN_PDF_comparison}, it might make sense to look directly at the percentage of points with assigned densities located inside the convergence radius.
While this percentage ranges from 70\% down to 46\% in the Covmos realisations for the $\tanh$-mark with $(a,b) = (10.6, -0.5)$, for the White mark with $(\rho_{*},p)=(10^{-6},1.0)$ in ELEPHANT it is still ranging from around 64\% down to 51\%, including at least half the amount of points.
Therefore, we would still trust the results of the applied shot-noise correction in this configuration as we analyse them in the next section.
After all, even without a Taylor expansion, the shot-noise behaviour might be well described with a polynomial since almost any function can be locally fitted with a polynomial.

As mentioned earlier, next to the Taylor expansion, another limiting factor are possible contributions of higher powers of $1/\bar{N}$, which are not captured by e.g. 3rd-order polynomial fits.
This is connected to the amplitude of Taylor coefficients, which do only grow if the convergence radius is smaller than 1.
In general, the coefficients of the polynomial are expected to decrease at some power as otherwise the shot-noise would look very noisy as a function of $1/\bar{N}$.
However, the coefficients might not be decreasing fast enough such that a low-order polynomial is sufficient.
This situation is expected to worsen when the Taylor coefficients are growing.
Non-accounted for higher-order polynomials in the fit should manifest as a bias in the recovered signal.
A way to circumvent this issue is by extending the polynomials to higher order, which, however, is not recommended in the case of only 8 data points due to overfitting.
Although we do find good performance of the shot-noise correction for some marks with growing coefficients we also find worse performance if e.g. the $b$ parameter is set to zero in the $\tanh$-mark.
In the latter case the Taylor coefficients are only non-zero for odd powers of $\delta_{Rf}$ hence we suspect that higher-order correlators are more important leading to higher-order polynomial contributions.
While a thorough assessment of the impact of Taylor expansion coefficients on the polynomial fit is not done in this work we conservatively advise to use only marks with a convergence radius larger than one to have decreasing Taylor coefficients.
Furthermore, the shift parameter $b$ in the $\tanh$-mark should be non-zero.
In Section~\ref{sec:discussion} we elaborate further on the connection between polynomial amplitudes and the smoothing induced by the MAS.

\subsection{Shot noise in the White mark}\label{sec:SN_white}

With the previously developed methodology for correcting for shot-noise effects in the weighted correlation function, we can evaluate the shot-noise contributions to marks used in the literature so far. One particular widely-used mark is the White mark given in Eq.~\eqref{eq:White_mark}, which has been used with the parameter combinations $(\rho_{*}, p)=(4.0, 10.0)$ and $(\rho_{*}, p)=(1.0, -1.0)$ in the work of \citet{Alam2021JCAP} and in the combination $(\rho_{*}, p)=(10^{-6}, 1.0)$ in the work of \citet{Aguayo2018MNRAS}.
Both studies used a count-in-cells scheme to compute a density field on the grid, which is the same as using a NGP MAS. In contrast, we used for most of our analysis a PCS MAS, which is much wider in configuration space and produces a smooth continuous and differentiable field.

\begin{figure}
    \centering
    \includegraphics[width=\columnwidth]{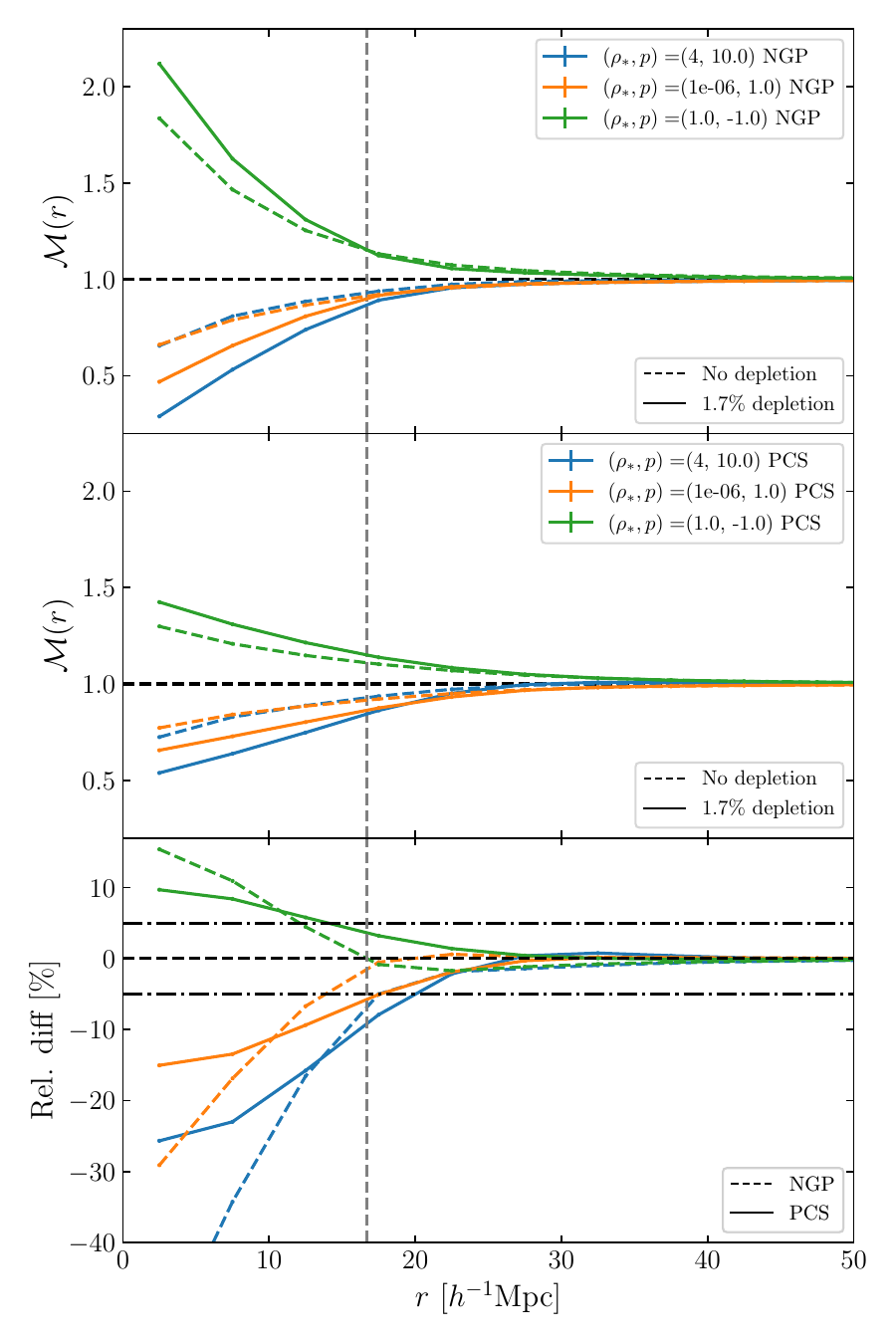}
    \caption{Marked correlation functions for the White mark measured in the Covmos catalogues.
    The upper and mid panel show the marked correlation functions for different configurations of the parameters $(\rho_*,p)$ using NGP and PCS MAS, respectively.
    The dashed line refers to the undepleted case whereas the solid line shows the measurement for a depleted catalogue down to 1.7\%, which corresponds to the same mean density of points per grid cell as in the undepleted ELEPHANT simulations.
    The lowest panel displays the relative difference between the marked correlation function as measured in the full data set and the depleted one.
    We used 60 grid cells per side length for the density field and the vertical dashed line in grey corresponds to the side length of a grid cell.}
    \label{fig:White_Mark_rel_difference_Covmos}
\end{figure}

Before we come to shot-noise-corrected differences between GR and MG in the White mark, it is instructive to check how the shot-noise behaviour changes when a different MAS is used. This can be investigated even without a shot-noise correction by studying marked correlation functions for the undepleted point set and the depleted point set down to 1.7\% in the Covmos catalogues as depicted in Figure~\ref{fig:White_Mark_rel_difference_Covmos}.
For this particular case, we use 60 grid cells per dimension, instead of 64, to mimic closer the procedure in \citet{Alam2021JCAP} and \citet{Aguayo2018MNRAS}.
By comparing the upper and middle panel in Figure~\ref{fig:White_Mark_rel_difference_Covmos} it can be immediately seen that a more extended MAS, meaning a more smooth density field, decreases the amplitude of the marked correlation function. Intuitively this makes sense as we can expect a possibly stronger small-scale correlation of the mark if the density field, used for the mark, is less smooth.
However, this comes at the price of stronger shot-noise effects on small separations as can be seen in the lowest panel.
Below scales of around 10$\Mpc$, the relative difference between the depleted and undepleted catalogue is larger if a NGP MAS is used as compared to a PCS scheme.
However, while the PCS scheme certainly reduces shot-noise effects on the smallest scales it does show extended effects up to larger scales, which can be seen as the solid lines (PCS) crossing the 5\% limit at larger scales compared to the dashed lines (NGP).
We have shown such a feature already in the toy model in Eq.~\eqref{eq:toy_model_SN_contr_w} and \eqref{eq:toy_model_SN_contr_mbar}, where the shot noise is regulated to some extend by the MAS kernels.
In our test on the Covmos realisations, we find the shot-noise correction to give biased results when a NGP MAS is used due to a very low smoothing size and shape of the NGP MAS.
In Section~\ref{sec:discussion} we give a more extended discussion on this aspect and in the following we refrain ourselves from using the NGP scheme in the White mark and use the PCS MAS throughout, unless otherwise indicated.

\begin{figure}
    \centering
    \includegraphics[width=\columnwidth]{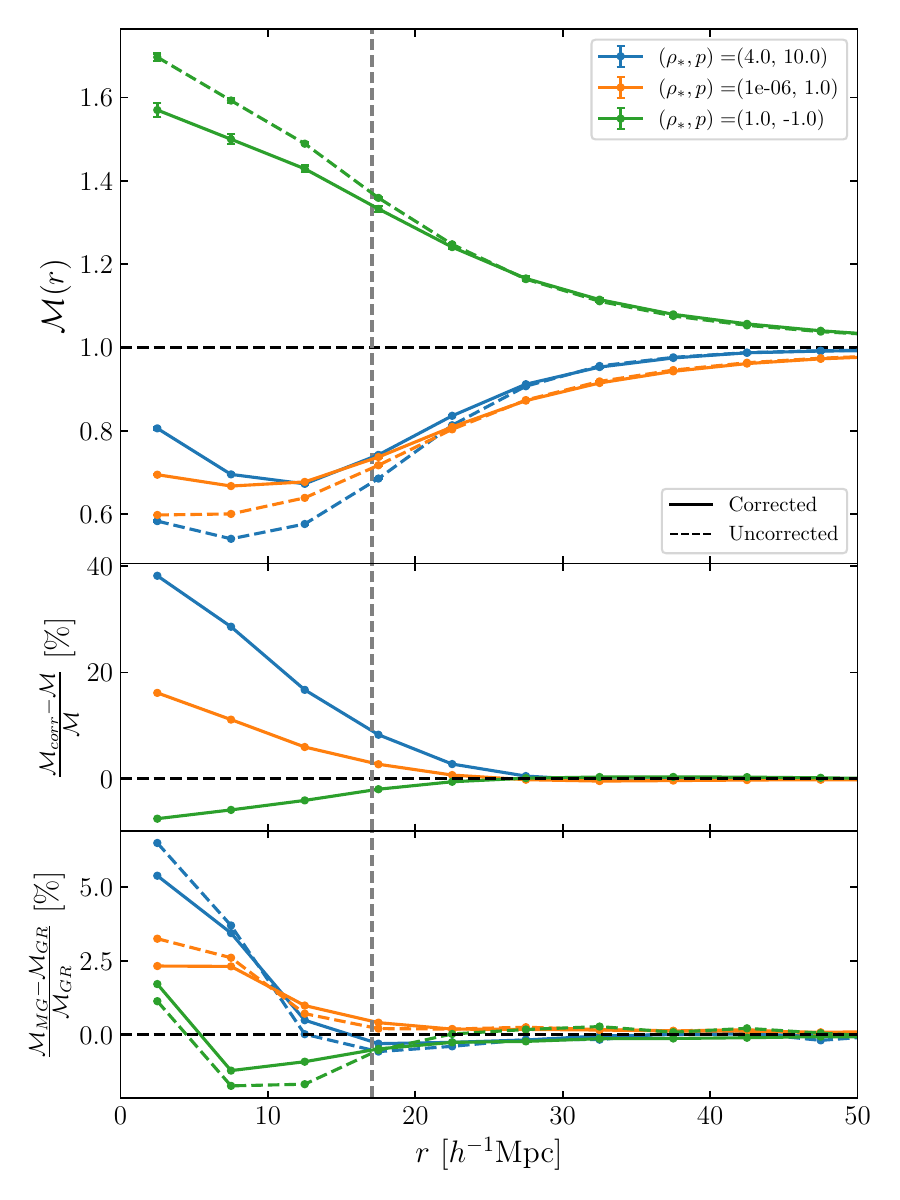}
    \caption{Shot noise in the White mark for the ELEPHANT suite. The different configurations of parameters are colour coded and the solid line refers to the corrected case while the dashed line refers to the uncorrected case. The upper panel displays the marked correlation function both corrected and uncorrected. The middle panel shows the relative difference between the corrected and the uncorrected marked correlation function in percent. The MG model is fixed to F4 in the upper and the middle panel. In the lower panel we show the relative differences for the different configurations between the F4 model and GR in percent.
    We used 60 grid cells per dimension and the vertical dashed line in grey refers to the side length of one grid cell.
    }
    \label{fig:White_shot_noise}
\end{figure}

Let us now focus on the impact of shot noise on differences between MG and GR in the ELEPHANT simulations.
We present in Figure~\ref{fig:White_shot_noise} the uncorrected marked correlation function as measured in ELEPHANT for different configurations of the White mark alongside with the shot-noise-corrected version.
The MAS is fixed to a PCS and we use again 60 grid cells per dimension.
The shot noise appears to only significantly affect the measurement below around 20 $h^-1$Mpc, but relative differences to the case with no correction can reach up to 40\% at these scales (middle panel).
This is similar to what we reported for the effect of shot noise in the Covmos simulations presented in Figure~\ref{fig:White_Mark_rel_difference_Covmos}.
The shot noise has the smallest contributions for the configuration $(\rho_{*}, p) = (1.0, -1.0)$ where galaxies in high density regions get upweighted as compared to the configurations with positive $p$ for which galaxies in low-density regions are getting upweighted.
In general, the effect of shot noise as shown in Figure~\ref{fig:White_shot_noise} is expected to be even stronger on the smallest scales when a NGP MAS is used.
In the studies of \citet{Alam2021JCAP} and \citet{Aguayo2018MNRAS} relative differences between MG and GR are particularly pronounced for scales below 20$\Mpc$ where we see that shot noise has a significant effect.
However, our findings do not nullify those claimed differences as they might still be present after correcting both GR and MG for shot noise.
Rather the amplitude of the marked correlation function itself will be different when correcting for shot noise, which is particularly important for the modelling of marked statistics as e.g. done in \citet{Aviles2020JCAP, Philcox2020PhRvD, Philcox2021JCAP}.
Indeed, as can be seen in the lowest panel of Figure~\ref{fig:White_shot_noise} the relative differences are largely unaffected regarding a correction for shot noise.
Although caution is advised as this does not have to be universally valid for every mark.
Furthermore, as we have shown in the last paragraph about shot-noise effects in the Covmos catalogues (see Figure~\ref{fig:White_Mark_rel_difference_Covmos}), if a NGP MAS is used, the larger contributions at small scales might impact relative differences stronger on those scales.

\section{Results} \label{sec:Results}

\subsection{Performance of marks based on large-scale environment}

It has been already argued in Section~\ref{sec:what_mark} that, even though we do not have a method for correcting the bias introduced by shot noise in this case, it is instructive to assess the performance of marks based on the environmental classification and tidal field/torque regarding distinguishing GR from MG.

\begin{figure}
    \centering    
     \includegraphics[width=0.9\columnwidth]{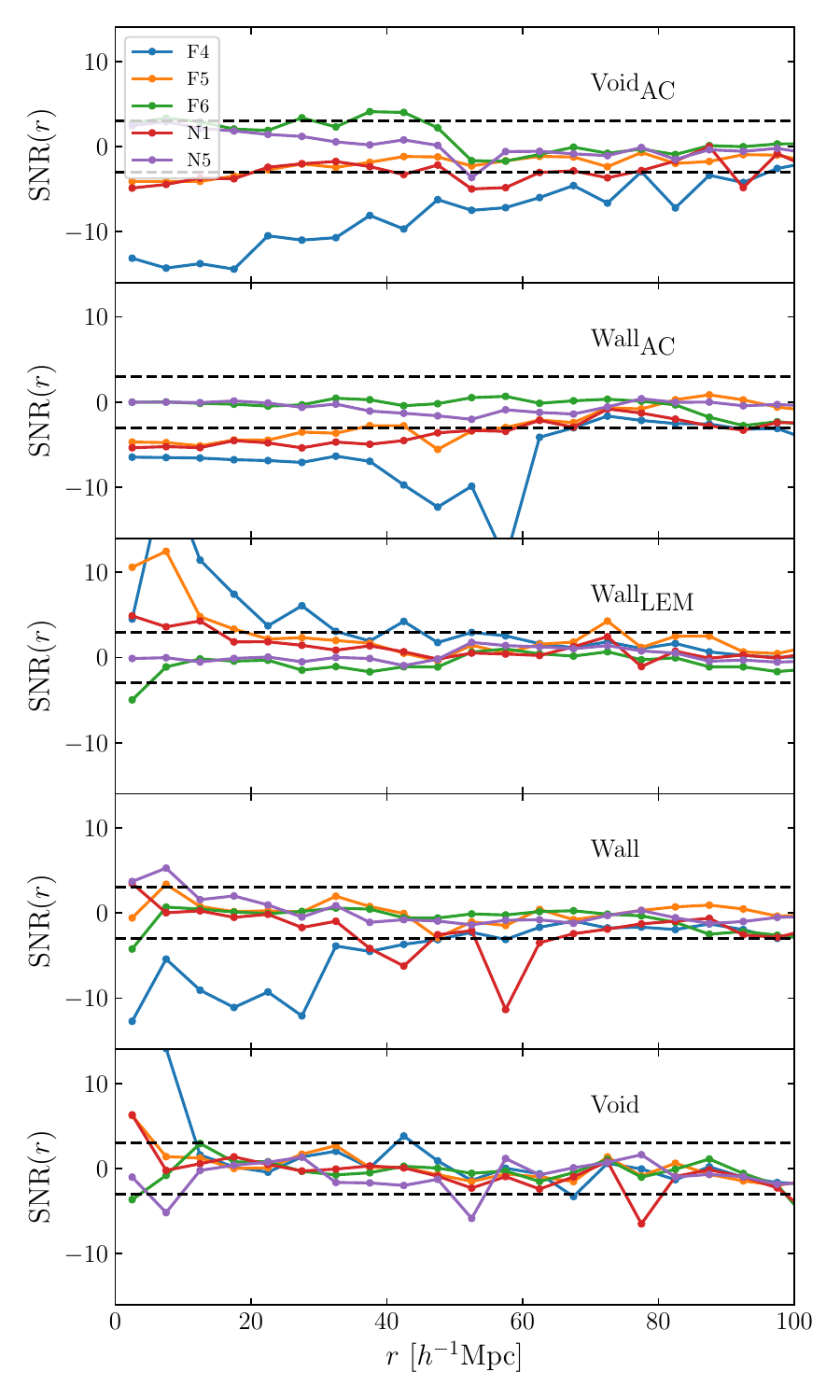}
    \caption{Summary of the SNR measured in the ELEPHANT suite for the marked correlation functions using marks based on the large-scale environment as introduced in Section~\ref{sec:what_mark}. Colours refer to MG simulations and the panels show different marks as indicated on the labels.
    The horizontal dashed lines in black indicate a SNR of $\pm3$.
    }
    \label{fig:performance_env_marks}       
\end{figure}

In Figure~\ref{fig:performance_env_marks} we present the SNR as defined in Eq.~\eqref{eq:signal_to_noise} for marks based on the environmental classification introduced in Section~\ref{sec:what_mark}.
While marks like the Void and Wall correlation function only exhibit significant differences on small scales below 10$\Mpc$ and 40$\Mpc$, respectively, the marks introducing anti-correlation produce significant differences up to scales of around 80$\Mpc$.
Particularly the weaker modifications of gravity like F5 and F6 appear to profit from anti-correlation as the $\textrm{Void}_{\textrm{AC}}$ mark has an SNR of 3 up to around 40$\Mpc$ and the $\textrm{Wall}_{\textrm{AC}}$ mark shows a similar behaviour for F5.
Furthermore N1 can be well distinguished with the $\textrm{Wall}_{\textrm{AC}}$ mark up to scales of around 60$\Mpc$.
The $\textrm{Void}_{\textrm{LEM}}$ mark performs well on the F4 simulations up to scales around $\sim$60$\Mpc$ but the SNR is only around 3 from 30$\Mpc$ onwards.
N1 and F5 show a SNR of around 3 or larger only for scales up to around 20$\Mpc$ and 30$\Mpc$, respectively.

\begin{figure}
    \centering    
     \includegraphics[width=0.9\columnwidth]{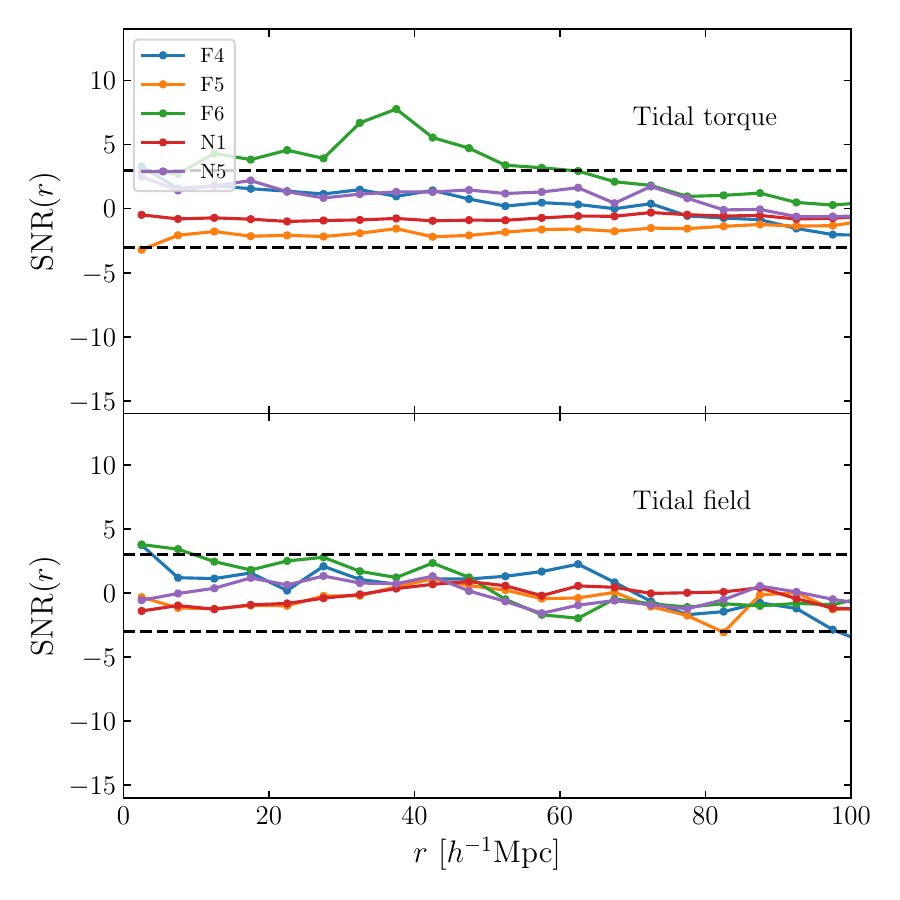}
    \caption{Same as Figure~\ref{fig:performance_env_marks} but for marks based on the tidal field and tidal torque.}
    \label{fig:performance_tidal_field_torque_marks}       
\end{figure}

The situation is different for marks using the tidal field/torque as depicted in Figure~\ref{fig:performance_tidal_field_torque_marks}.
Using the tidal field as it is does not seem to yield any significant difference for the investigated MG theories.
Interestingly, by taking just a linear function of the tidal torque, we can report significant differences for F6 up to scales of $\sim$60$\Mpc$.
As the tidal torque is small for symmetric large-scale environments we basically upweight filaments and walls by taking the tidal torque as a mark.
Since there are many galaxies located in walls this appears to compensate for the fact that MG effects are expected to be more screened in walls compared to voids and lead to significant differences to GR when used as a mark.

Although shot noise is expected to decrease at higher scales and that they might not affect differences between two measurements strongly, there is no guarantee that the observed significant differences seen in Figure~\ref{fig:performance_env_marks} and \ref{fig:performance_tidal_field_torque_marks} are still present after a correction for shot noise.
In principle we could assess the overall impact of shot noise on these marks by looking at differences in the corresponding measurements for the Covmos catalogues as we have done it in Figure~\ref{fig:White_Mark_rel_difference_Covmos} for the White mark.
However, due to the peculiar way the Covmos catalogues have been set up we do not expect that they also represent large-scale structure environments in a realistic way.
Furthermore, qualitative differences are not enough to assess precisely how the SNR will change after a proper correction.
We relegate, therefore, a thorough investigation of shot-noise effects for these marks to future work for which high resolution full N-body simulations of MG are necessary.

\subsection{Performance of the White mark}
\begin{figure}
    \centering
    \includegraphics[width=\columnwidth]{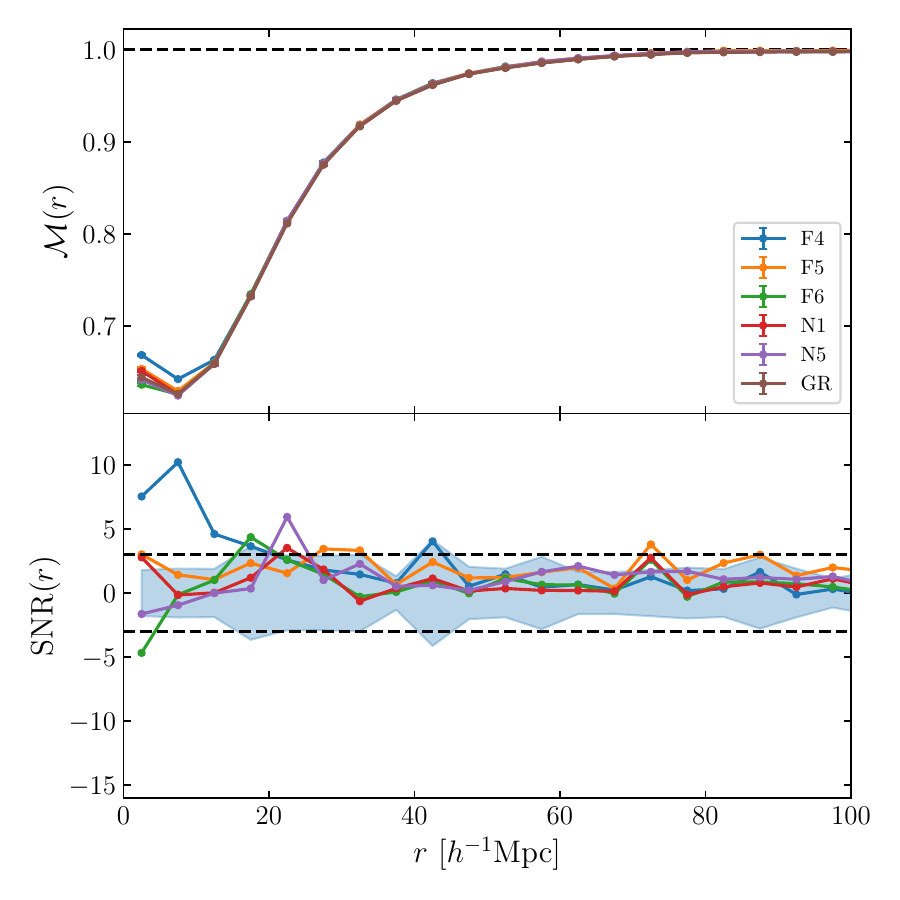}
    \caption{Performance of the White mark for fixed parameters $(\rho_*,p)=(10^{-6},1.0)$.
    The upper panel displays the mean marked correlation function taken over 5 realisations together with errorbars estimated as the mean standard deviation.
    Colours refer to the different gravity simulations.
    The lower panel shows the SNR as introduced in Eq.~\eqref{eq:signal_to_noise} and the blue shaded region refers to the error of a single measurements divided by the error of the mean difference (see Eq.~\eqref{eq:alpha}) for F4.}
    \label{fig:performance_white_mark}
\end{figure}

In Figure~\ref{fig:performance_white_mark} we present the performance of the White mark, corrected for shot noise, to be compared with the other marks to follow.
We used 64 grid cells per dimension and a PCS MAS to obtain the density field on the grid and the parameters were fixed to $(\rho_{*},p)=(10^{-6}, 1.0)$.
As described in Section~\ref{sec:imapct_sn}, we are using third and second order polynomials to fit the shot-noise dependency of $w_f(\mathbf{r})$ and $\bar{m}_f$, respectively.
Overall, the amplitude of the marked correlation function does not differ much from unity implying a minor impact of the mark.
Although we can see differences from unity up to scales of around 70$\Mpc$, the signal is very similar between GR and MG for most of the scales except below 20$\Mpc$.
This can be deduced from the lower plot as well where we show the SNR, directly quantifying the difference between GR and MG.
The SNR lays inside the $3\sigma$ region with only occasional peaks outside that range, which can be accounted to sample variance.
Only for F4 we can report significant differences for the first four bins in $r$ ranging up to $\sim20$$\Mpc$.
The SNR for the case $(\rho_*, p)=(4.0, 10.0)$, although not shown here, exhibit similar overall structure as the presented case.
This makes the White mark with those configurations in real space not particularly powerful in distinguishing MG from GR as possible differences only show up at very low scales.

\begin{figure*}
    \captionsetup[subfigure]{labelformat=empty}
    \centering    
     \includegraphics[width=0.45\textwidth]{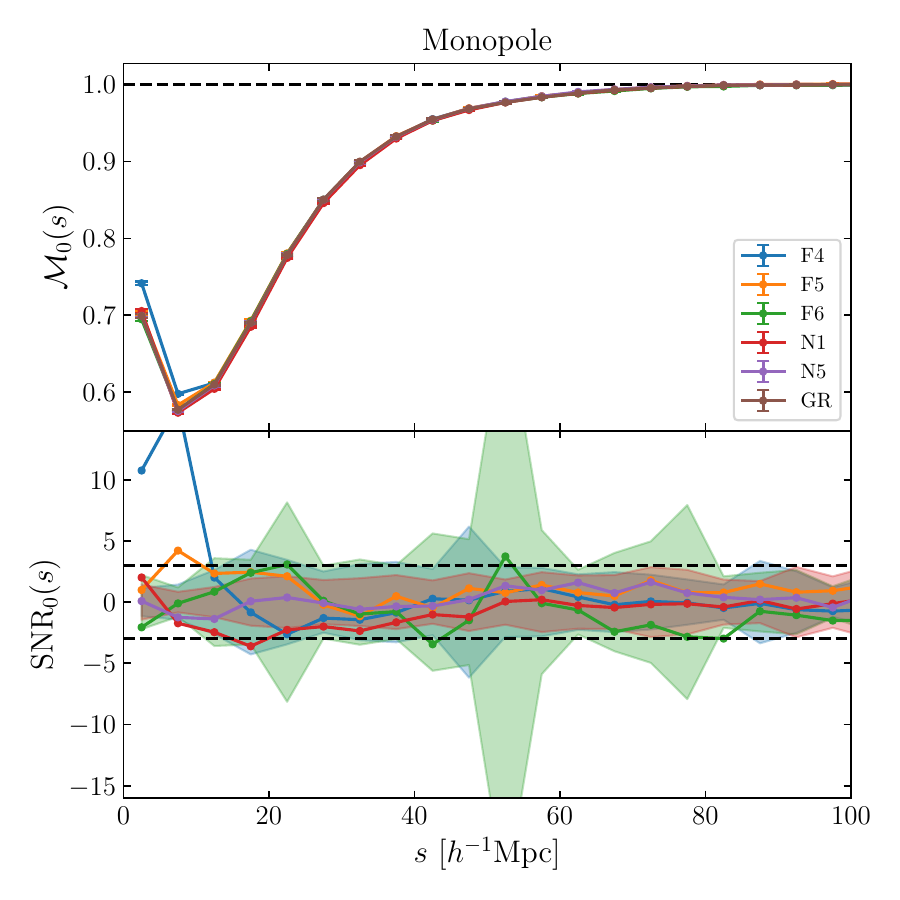}
     \includegraphics[width=0.45\textwidth]{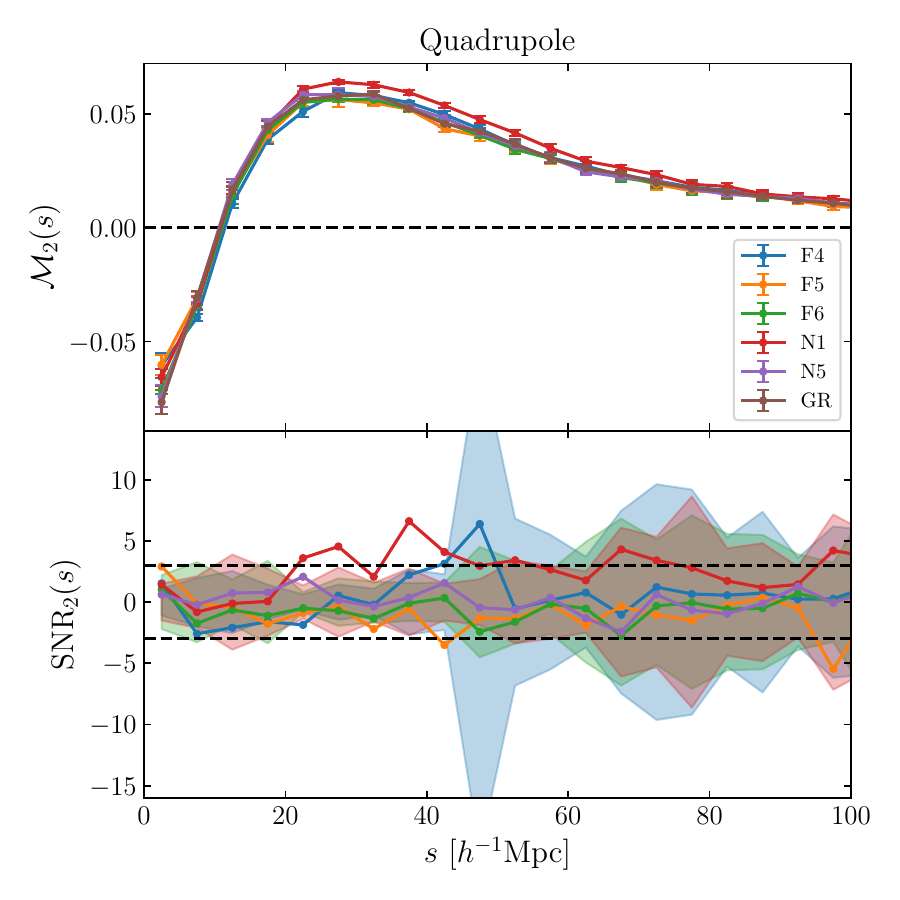}
    \caption{Redshift-space monopole (left side) and quadrupole (right side) of the marked correlation function for the White mark with parameters fixed to $(\rho_{*},p)=(10^{-6},1.0)$ and corrected for shot noise.
    Upper panels show the multipoles themselves with colours referring to different gravity simulations.
    Displayed is the mean over five realisations with respective mean standard deviations as the error bars.
    For the monopole the horizontal dashed line in black marks an amplitude of 1 while for the quadrupole it marks an amplitude of 0.
    Lower panels show the signal to noise ratio, as in Eq.~\eqref{eq:signal_to_noise}, with according colour for the different MG models.
    Shaded regions refer to the error of a single measurement divided by the mean error of the difference as introduced in Eq.~\eqref{eq:alpha}.
    Dashed black lines in the lower panels indicate a SNR of $\pm3$.
    }
    \label{fig:white_RSS}
\end{figure*}

In Figure~\ref{fig:white_RSS} we present the monopole and quadrupole of the White mark in redshift space with the same parameter configuration as before, meaning $(\rho_{*},p)=(10^{-6},1.0)$.
The left panel depicts the monopole, exhibiting a very similar amplitude and shape as the real space measurements in Figure~\ref{fig:performance_white_mark}.
The monopole does converge to unity at higher scales which is expected as, even in redshift space, the marks should become uncorrelated at high scales leading to $\mathcal{M}(s, \mu)=1$.
The SNR in the monopole shows no significant difference over all scales except for F4 below $20\Mpc$ that we have also seen in real space.
The quadrupole (right panel) has much weaker signal compared to the monopole and converges to zero at high scales.
This can be explained with the same reasoning as to why the monopole converges to one.
$\mathcal{M}(s, \mu)$ is independent of $\mu$ on large scales and therefore does not possess higher-order multipoles.
Even though the amplitude is very small we nevertheless see interesting differences for N1 in the SNR on moderate to high scales.
However, the SNR is barely above 3 and the bin-to-bin variance is fairly high.
Lastly, we did not find significant differences for the configuration $(\rho_{*},p) = (4.0, 10.0)$ over extended scales.
Similar to what we have seen in real space, the White mark in these configurations is overall not really promising in redshift space.

\subsection{Performance of the \texorpdfstring{$\tanh$}{tanh}-mark}\label{sec:tanh_mark}
\begin{figure*}
    \centering    
     \includegraphics[width=0.45\textwidth]{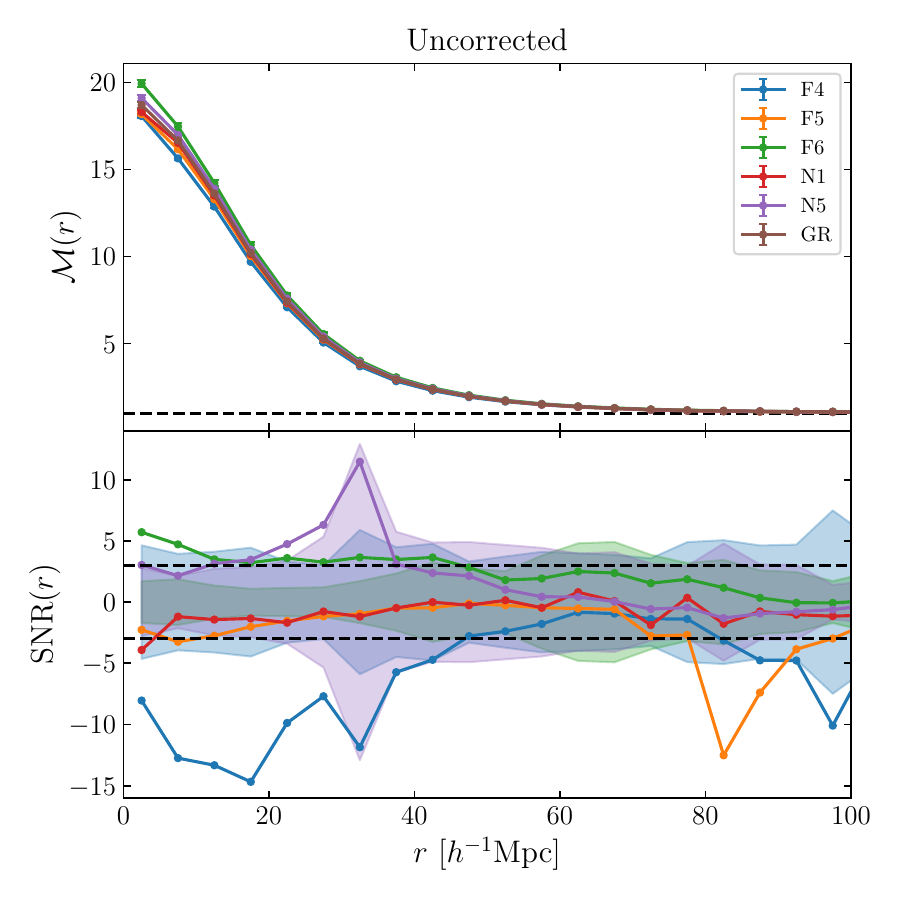}
     \includegraphics[width=0.45\textwidth]{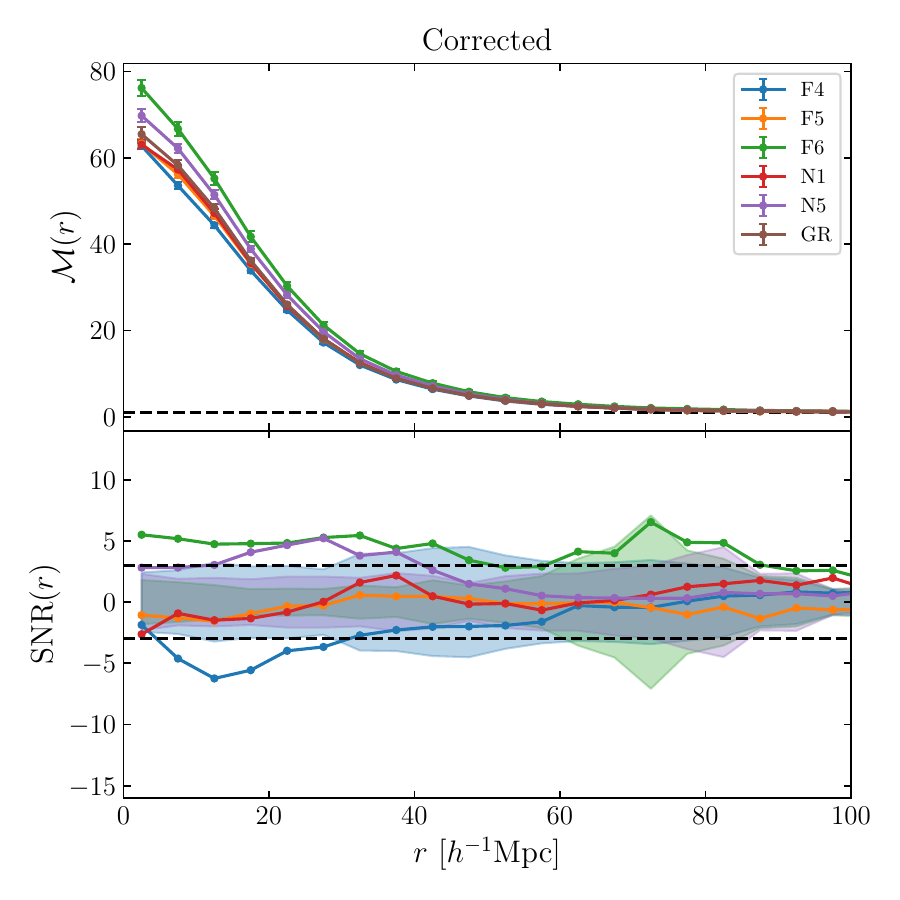}
    \caption{Marked correlation function for the $\tanh$-mark with $(a,b)=(0.6, -0.5)$.
    The left panels depict the case where we do not apply any correction for shot noise while in the right panels we do apply our shot-noise correction methodology as described in Section~\ref{sec:imapct_sn}.
    Upper panels show the marked correlation functions with colours encoding the different MG models. 
    Displayed is the mean over 5 realisations and errorbars are obtained by taking the mean standard deviation.
    The lower panel displays the SNR  where the black-dashed line indicates a difference of $\pm3$.
    Shaded areas mark the error of a single measurement divided by the mean error of the difference.
    The black dashed line in the upper panels indicates an amplitude of 1.}
    \label{fig:tanh_0.6_-0.5_realspace}       
\end{figure*}

In Figure~\ref{fig:tanh_0.6_-0.5_realspace} we present both the marked correlation function before and after shot-noise correction in the left and right panel, respectively, for the $\tanh$-mark as introduced in Section~\ref{sec:AC_delta}.
The configuration is set to $(a,b)=(0.6,-0.5)$ for which we showed in Section~\ref{sec:imapct_sn} that our correction algorithm can be safely applied.
We use polynomials of order three and two for correcting $w_f(\mathbf{r})$ and $\bar{m}_f$, respectively.
The marked correlation function is largely featureless and converges to 1 on large scales regardless if a correction for shot noise is applied or not. 
This convergence is also present in the White mark in Figure~\ref{fig:performance_white_mark} and is due to the mark getting uncorrelated at large scales.
It is striking how the amplitudes differ on smaller scales between the uncorrected (left panel) and corrected case (right panel) underlining again the importance of applying a shot-noise correction in order to measure correct amplitudes.
In the lower panels we are showing the SNR as defined in Eq.~\eqref{eq:signal_to_noise}.
The general trend of the SNR for the different MG models appears to be similar regardless if a shot-noise correction is applied or not.
However, e.g. F4 shows much larger significance on small scales in the case of no correction.
Most importantly, the $\tanh$-mark in this configuration leads to significant differences for F6 in the corrected case, up to scales of around 80$\Mpc$.
Worth to notice here that significant differences are also found for F4 and N5 but only for scales smaller than roughly 30 and 40$\Mpc$, respectively.
While differences on these scales might be sufficient to be grasped by theoretical models appropriately, it has yet to be tested as modelling becomes increasingly more challenging at small scales. 
In Section~\ref{sec:discussion} we are discussing our results on the $\tanh$-mark in the light of current constraints in the literature on the $f_{R0}$ parameter.

Since our SNR as defined in Eq.~\eqref{eq:signal_to_noise} computes the error over 5 realisations any significant differences can only be claimed for a volume of 5 realisations.
It is therefore instructive to elaborate on how the difference compares to the error of a single measurement as indicated by the shaded area in the plot.
It appears that particularly at higher scales the difference is of the same size as the error itself rendering a detection at the current volume of around $1\,h^{-3}\textrm{Gpc}^3$ with only one measurement at hand impossible.
However, this could be alleviated by considering simulations in larger volumes.
Assuming that the error of the single measurement is Gaussian, it scales like $\propto 1/\sqrt{V}$, where $V$ is the volume of the survey/simulation.
An increase in volume by a factor of 9 for F6 would be necessary in order to detect the difference with just one measurement at scales between $60\Mpc$ and $80\Mpc$.
This increase in volume translates in a larger box side length by a factor of around 2.1.
On smaller scales, below around $40\Mpc$, the reported differences would be significant even with only a single measurement.

\begin{figure}   
    \centering
    \includegraphics[width=\columnwidth]{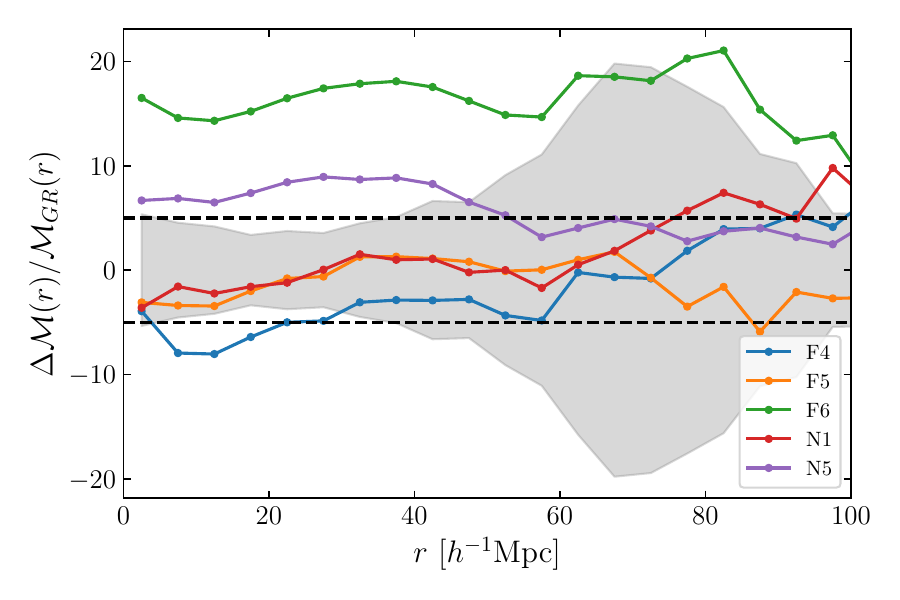}
    \caption{Relative differences between GR and MG for the $\tanh$-mark with parameters fixed to $(a,b)=(0.6, -0.5)$. We show the mean relative difference over five realisations and the shaded area corresponds to the relative standard deviation (relative error of single measurement) for the GR realisations.
    Grey dashed lines indicate relative differences of $\pm$5\%.}
    \label{fig:tanh_0.6_-0.5_relative_differences}
\end{figure}

In order to better compare our results with the literature where often only relative differences between GR and MG are reported \citep{Aguayo2018MNRAS, Armijo2018MNRAS, Alam2021JCAP} we show a corresponding plot in Figure~\ref{fig:tanh_0.6_-0.5_relative_differences}.
Large relative differences beyond 5\% are reached for F4 only on smaller scales below around 20$\Mpc$ while N5 shows larger relative differences all the way up to around 50$\Mpc$.
This is topped by F6 exhibiting relative differences above 15\% over almost all scales decreased only above $\sim$80$\Mpc$.
However, the relative error for the GR simulation increases towards those scales rendering the detection with only one realisation at those high scales unfeasible.
As shown in Figure~\ref{fig:tanh_0.6_-0.5_realspace} more volume is needed to shrink the uncertainties to a level at which the large reported differences between GR and F6 at high scales can be taken advantage of.
These results can be compared with the Fig. 5 in the work of \citet{Aguayo2018MNRAS} and Fig. 15 in \citet{Alam2021JCAP} where marks based on the Newtonian gravitational potential were used and appear to be the most performant in terms of relative difference between GR and MG.
While certainly performing very good for F4 and to some extend for F5, we can report much larger relative differences for F6, reaching up to higher scales if a $\tanh$-mark in the configuration $(a,b)=(0.6,-0.5)$ is used.
Furthermore, our mark is very easy to compute and does not need information from halos as is the case if the Newtonian potential is to be computed in the way defined in the mentioned studies.

\begin{figure*}
    \centering    
     \includegraphics[width=0.45\textwidth]{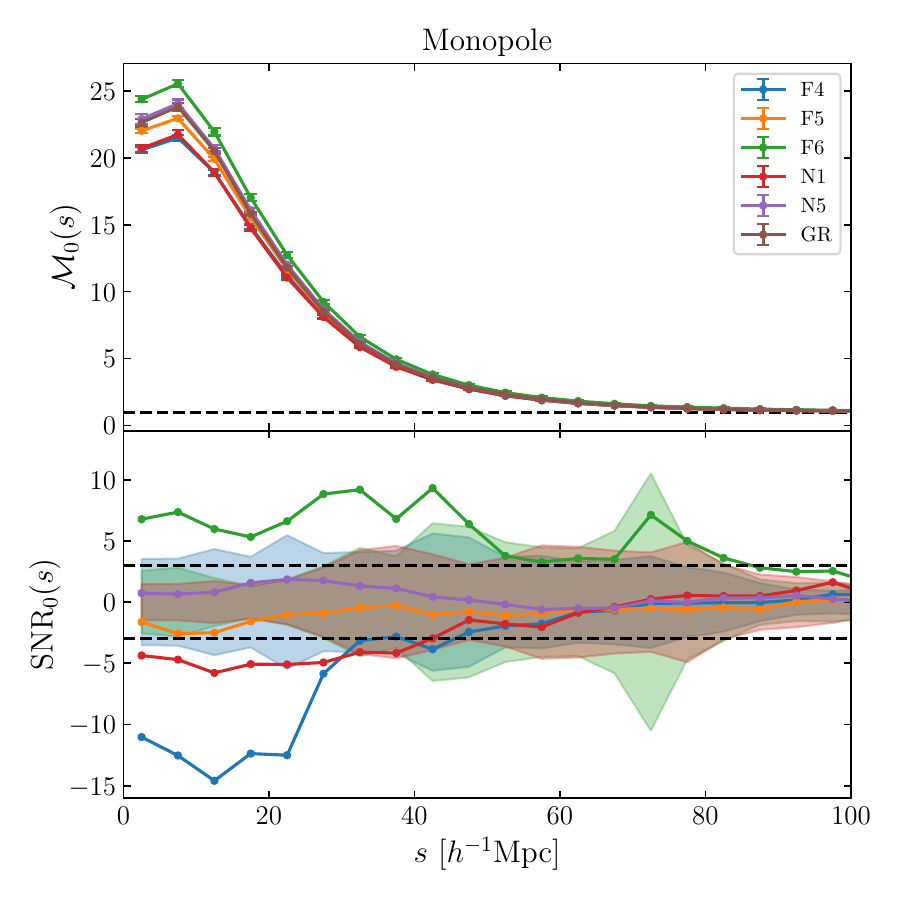}
     \includegraphics[width=0.45\textwidth]{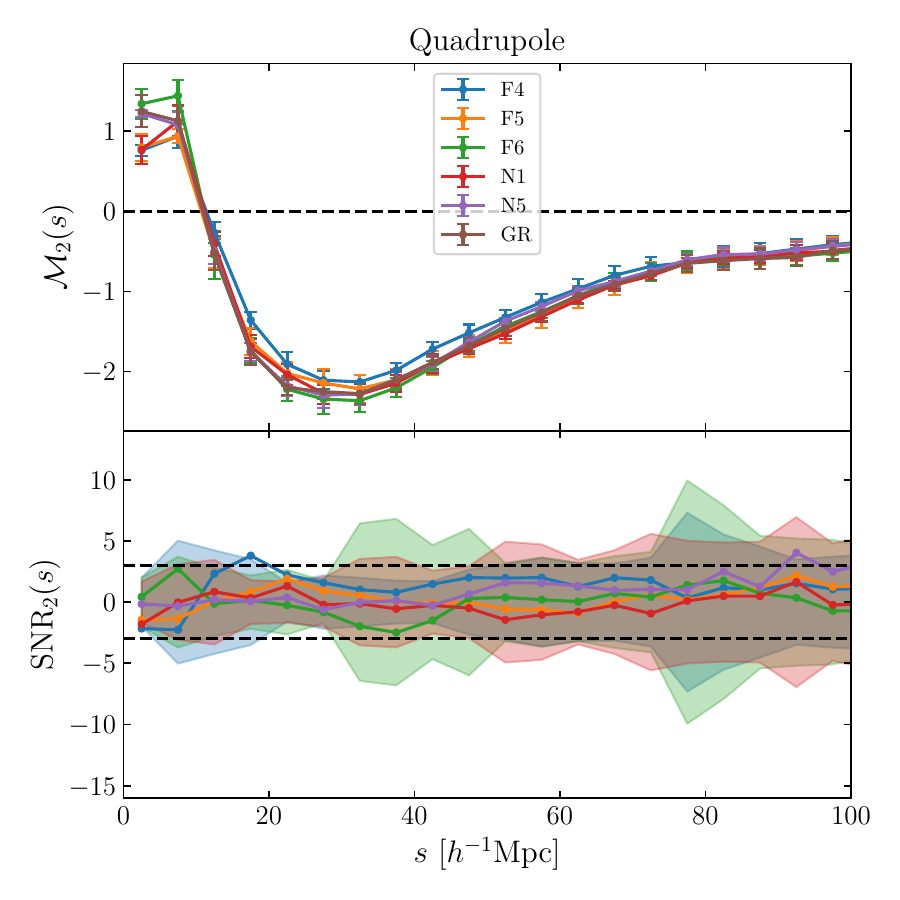}
    \caption{Redshift-space monopole (left side) and quadrupole (right side) of the marked correlation function for the $\tanh$-mark with parameters fixed to $(a,b)=(0.6,-0.5)$ and corrected for shot noise. Upper panels show the multipoles themselves with colours referring to different gravity simulations. Displayed is the mean over five realisations with respective mean standard deviations as the errorbars. 
    For the monopole the horizontal dashed line in black indicates an amplitude of 1 and for the quadrupole an amplitude of 0.
    Lower panels show the SNR, as in Eq.~\eqref{eq:signal_to_noise}, with according colours for the different MG models. Shaded regions refer to the error of a single measurement divided by the mean error of the difference and dashed lines in the lower panels indicate a SNR of $\pm3$.
    }
    \label{fig:tanh_RSS}
\end{figure*}

Finally, in Figure~\ref{fig:tanh_RSS} we present the shot-noise-corrected monopole and quadrupole of the marked correlation function in redshift space for the $\tanh$-mark with parameters fixed to $(a,b)=(0.6,-0.5)$.
In general, compared to the White mark in Figure~\ref{fig:white_RSS}, the amplitude of both the monopole and the quadrupole is much larger but the large-scale behaviour is very similar.
Looking at the monopole in the left panel, similarities with the real space measurements in Figure~\ref{fig:tanh_0.6_-0.5_realspace} are striking both in the shape of the measurements as well as the SNR. 
However, N5 has a reduced SNR in redshift space while conversely N1 has now significant differences at scales lower than 40$\Mpc$.
F6 and F4 look largely the same as in real space, most importantly the former still showing significant differences up to around 80$\Mpc$.
The shaded region seems to increase in size for F6 making an even larger volume necessary to detect the difference with a single observation only.
The amplitude in the quadrupole in the upper right panel is smaller and also the SNR in the lower right panel stays within 3$\sigma$ rendering the quadrupole not suitable as a statistic to detect MG with this mark.

\section{Discussion}\label{sec:discussion}

We have found the $\tanh$-mark to be particularly promising regarding distinguishing MG from GR. Of course, studying possible differences between $f(R)$ theories and GR has to be done in the light of constraints on $f_{R0}$ in the literature. A somewhat older compilation of constraints can be found in Tab. 1 in the work of \citet{2014AnPLombriser}, where the strongest limit comes from dwarf galaxies and the solar system imposing $|f_{R0}| \leq 10^{-7}-10^{-6}$. In a more recent analysis, \citet{2021PhRvDLiu} found similar limits using Fisher forecasts on cluster abundances and galaxy clustering. Even tighter constraints, $f_{R0} < 1.4\times 10^{-8}$, are found in the work of \citet{2020PhRvDDesmond} using galaxy morphology.
Although F6 might not be a viable MG theory after all and has to be replaced with weaker modifications like F7 or F8, finding significant differences for F6 makes the $\tanh$-mark promising to distinguish also weaker models.

\begin{figure}
    \centering
    \includegraphics[width=\columnwidth]{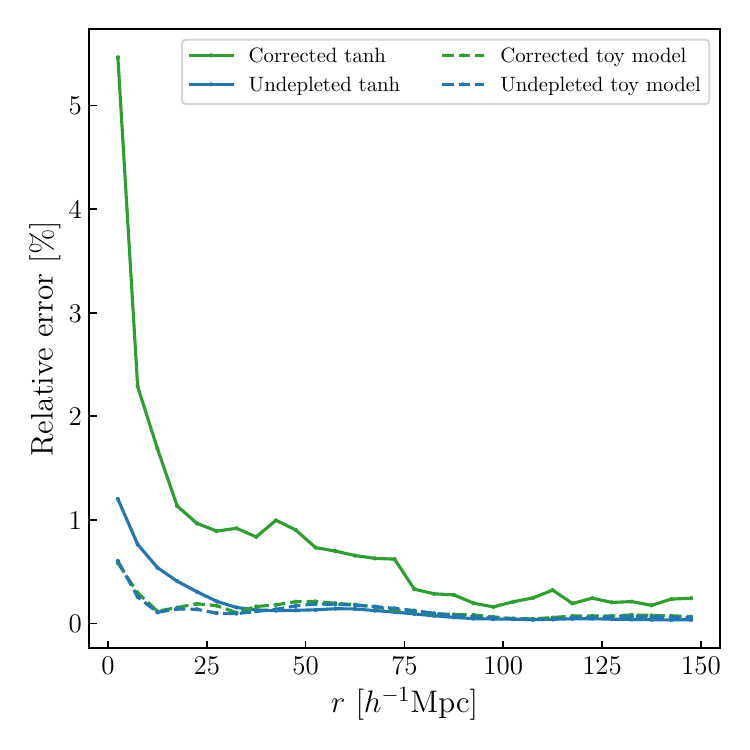}
    \caption{Relative error of the weighted correlation function in Covmos. The mark is set to the $\tanh$ with parameters fixed to $(a,b)=(0.6,-0.5)$ for the solid line and the toy model for the dashed line. Green colours refer to the correction using 30 realisations of depletions and blue colours stand for the undepleted case. The relative error is computed as the mean standard deviation over 5 realisations divided by the mean.}
    \label{fig:SN_rel_error}
\end{figure}

In Section~\ref{sec:imapct_sn} we presented a robust technique to correct for shot-noise effects for general marks, where the mark function can be expressed as a Taylor expansion in powers of the density contrast. Since the error on the measurements plays a crucial role in our performance metric in Eq.~\eqref{eq:signal_to_noise}, it is instructive to discuss how the relative error of the measurements is impacted when the shot-noise correction is applied. In particular due to the approximations made in estimating the shot-noise-corrected signal, additional uncertainties might be introduced. In Figure~\ref{fig:SN_rel_error} we present the relative error for the undepleted case and the corrected case, both for the $\tanh$-mark with $(a,b)=(0.6, -0.5)$ and toy model. Since we can capture the shot-noise behaviour in the toy model very accurately, the relative error stays almost the same and does not significantly change. However, in the case of the $\tanh$-mark where the correction is only approximate, the relative error does increase to around 1\% on scales larger than $25~\Mpc$. Although, at very large scales the shot-noise correction does not greatly impact the uncertainty since the overall contribution of shot noise on those scales is marginal. It is evident from the figure that most of the uncertainty from the correction is introduced on scales below $25~\Mpc$. Here we are within the smoothing radii and shot noise is expected to be the strongest.
It is important to note that this does not capture the effect on the relative error between not applying a correction at all and applying a correction. We can only conclude that while we might be able to accurately recover the true signal, the step of applying the correction via the fitting introduces additional uncertainties that increase towards smaller scales. This uncertainty is expected to be smaller the higher $\bar{N}$ is, since the fitting process should be less prone to uncertainties. Intuitively, this means that the points that we need to fit as depicted in Figure~\ref{fig:SN_toy_model_analytic_correction} and \ref{fig:SN_tanh_fit_results} are distributed closer to $1/\bar{N}=0$ and hence the extrapolation to the $y$-axis is more robust.

\begin{figure*}
    \captionsetup[subfigure]{labelformat=empty}
    \centering
    \begin{subfigure}[b]{0.45\textwidth}
     \includegraphics[width=\columnwidth]{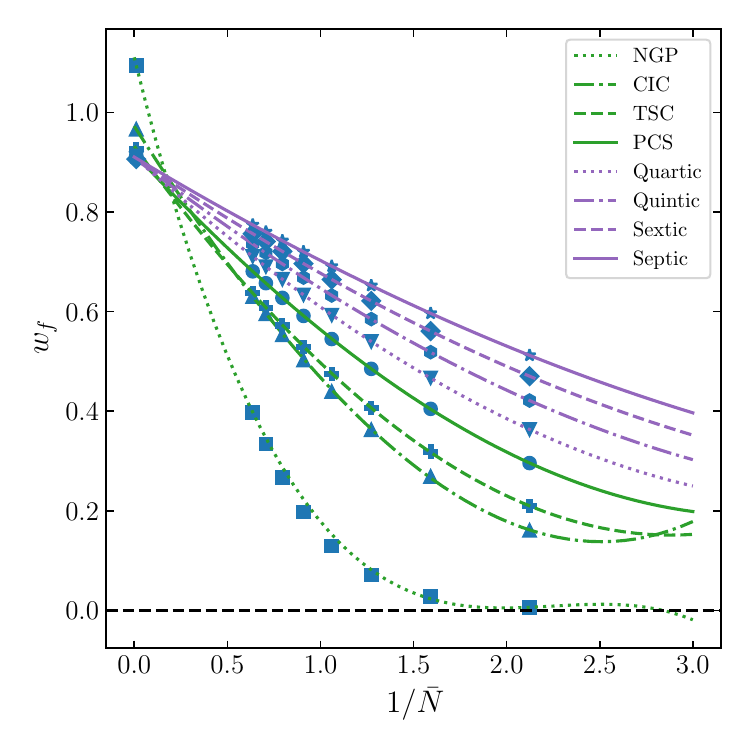}
    \caption{}
    \label{fig:white_full_fit_subplot00}       
    \end{subfigure}
    \begin{subfigure}[b]{0.45\textwidth}
     \includegraphics[width=\columnwidth]{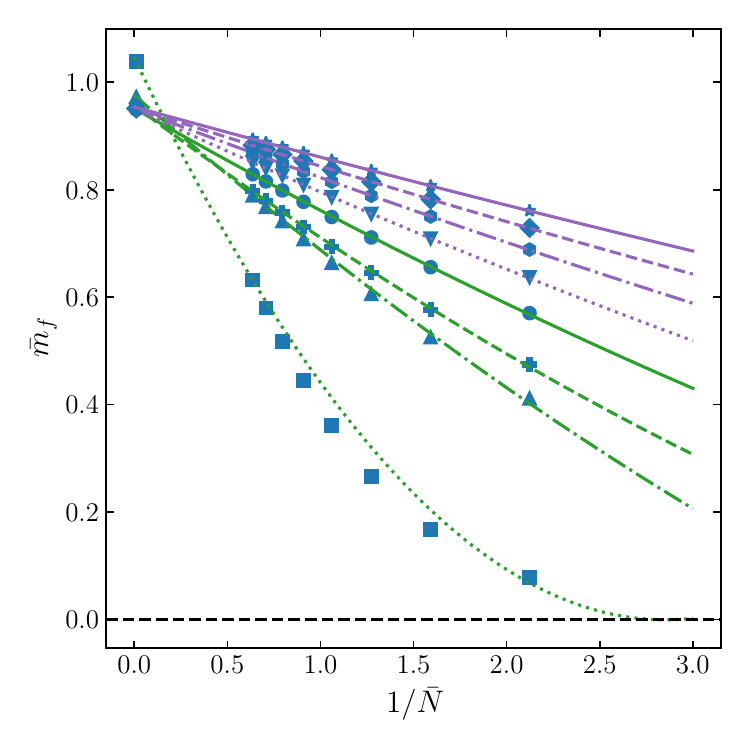}
    \caption{}
    \label{fig:white_full_fit_subplot01}       
    \end{subfigure} 
    \caption{Impact of the smoothing scale on the polynomial behaviour of shot noise. The mark is fixed to the White mark with $(\rho_{*}, p) = (4.0, 10.0)$. Left and right panels show the fit of $w_f$ and $\bar{m}_f$ for one realisation of Covmos, respectively. Depleted measurements are obtained via taking the mean of 30 realisations of depletions. Linestyles refer to different MAS used for obtaining the density field. The error of the undepleted case and 1.7\% depletion are computed as 10\% of the smallest uncertainty of the other depletions.}
    \label{fig:SN_smoothing_scale_fits}
\end{figure*}

\begin{figure*}
    \captionsetup[subfigure]{labelformat=empty}
    \centering
    \begin{subfigure}[b]{0.45\textwidth}
     \includegraphics[width=\columnwidth]{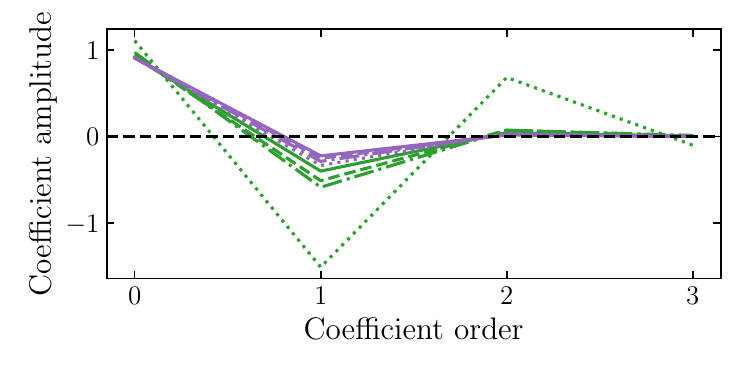}
    \caption{}
    \label{fig:white_full_fit_subplot10}       
    \end{subfigure}
    \begin{subfigure}[b]{0.45\textwidth}
     \includegraphics[width=\columnwidth]{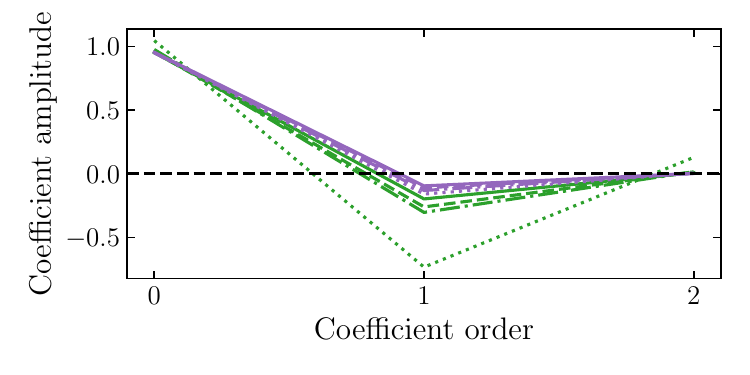}
    \caption{}
    \label{fig:white_full_fit_subplot11}       
    \end{subfigure}
    \caption{Coefficient amplitudes of the polynomial fits corresponding to Figure~\ref{fig:SN_smoothing_scale_fits}.}
    \label{fig:SN_smoothing_scale_coeff}
\end{figure*}

We have briefly touched upon the impact of the MAS kernel on our methodology in Section~\ref{sec:SN_white} with Figure~\ref{fig:White_Mark_rel_difference_Covmos} and it is crucial to assess this in more detail.
One can show that the shot-noise correction is smaller the higher the order of the MAS is, which can be understood as a larger smoothing scale as higher-order MAS kernels are more extended in configuration space.
In Figure~\ref{fig:SN_smoothing_scale_fits} and \ref{fig:SN_smoothing_scale_coeff} we present an illustration of how the smoothing scale as well as the shape of the MAS affect the behaviour of shot noise, particularly on the amplitude of the different powers of $1/\bar{N}$.
We show the result of the standard fitting procedure for obtaining the shot-noise-corrected signal and fits do include the undepleted catalogue, which is not accessible in real data. This figure illustrates the contributions from different powers of $1/\bar{N}$ in the shot-noise polynomial. Having included the undepleted case enables an accurate estimate of the shot-noise behaviour and identifying any breakdown at a given polynomial order. As we can already seen from the fits in Figure~\ref{fig:SN_smoothing_scale_fits}, once an NGP MAS is employed the low orders for the polynomials do not seem to be sufficient anymore to describe the data. This is further underlined by Figure~\ref{fig:SN_smoothing_scale_coeff} showing an increase in amplitude for the polynomial coefficients when lowering the MAS order. This means that the higher the smoothing scale the less contributions are coming from higher-order shot-noise expressions, and the better we can fit the dependency with a low-order polynomial.
Intuitively this makes sense if we hypothetically increase the smoothing scale to infinity. In that case, the obtained density field would be a constant in space and therefore all galaxies will have the same weight.
In that scenario, the weighted correlation function will reduce to the unweighted correlation function that is only affected by shot noise at zero-lag. Hence in our measurements there would be no contamination and the polynomial fits would just be a constant.
Such a trend can also be seen to some extent in Figure~\ref{fig:SN_smoothing_scale_fits}, where the curve becomes more and more linear and converges to a vertical line the higher the order of the MAS.

Using two different types of catalogues for gauging the shot-noise correction has its limitations, which we discuss in the following. The Covmos catalogues were absolutely necessary in the first place in order to have an almost noise-free signal to test the method. An inherent difference, which has to be kept in mind when interpreting our findings, is the fact that Covmos realisations are not simply an upscaled version of the ELEPHANT suite, rather a completely different set of catalogues. First and foremost, Covmos catalogues consist of dark matter particles instead of galaxies and second, the redshift and assumed cosmology are different compared to ELEPHANT. Most importantly, the Covmos catalogues are not extracted from full N-body simulations. This means that the features and the general shape of the weighted correlation functions look different between Covmos and ELEPHANT simulations. Nevertheless, the functional form of the Taylor expansion of the mark stays the same and having access to realisations with very large number densities of points served the purpose of validating the method for estimating the corrected signal.
Care has to be taken, however, in the choice of the mark to comply with the convergence criterion of the Taylor expansion as described in Section~\ref{sec:limits_SN}, which is universally valid for both sets of catalogues. By construction, the ELEPHANT suite is limited by the the small number of galaxies in the catalogues. This results in the shot-noise correction to introduce larger uncertainties than would do in larger catalogues.
Having only twice as many objects in the simulations would already half the distance to extrapolate from the undepleted case to $1/\bar{N}=0$ on a linear scale. The ELEPHANT simulations are therefore a particularly difficult case and we expect better results if applied to state-of-the-art N-body simulations with higher densities. Nevertheless, as long as the mark is chosen appropriately, a robust recovery of the true signal is possible. For future analysis we recommend to use simulations with higher resolution in order to mitigate the impact of shot noise.

We conclude the discussion with a brief summary of the main points raised in this section as well as Section~\ref{sec:imapct_sn} to be considered when applying our methodology to correct for shot noise in marked correlation functions:
\begin{itemize}
    \item[--] Enough realisations of depletion should be used in order to get converged depleted point sets. In our setting 30 depletions appeared to be enough.
    \item[--] The polynomial order should be chosen accordingly such that the shot noise dependency is well modelled without overfitting.
    \item[--] A higher-order MAS (e.g. TSC/PCS) is preferred to reduce possible bias due to shot noise at low scales.
    \item[--] In order to apply our methodology to correct for shot noise, a Taylor expansion of the mark function in powers of $\delta_{Rf}$ has to exist.
    \item[--] The convergence radius $B$ of the Taylor series should be larger than $
    \sqrt{\langle \delta_{Rf}^2 \rangle_{\rho}} $ in order to satisfy the assumption of a convergent Taylor expansion.
    \item[--] In addition to the criterion, as given in the last bullet point, also the fraction of points inside the convergence radius can be checked, which is more agnostic about the actual PDF of $\delta_{Rf}$ in the catalogues.
    \item[--] A Taylor expansion with non-zero and decreasing coefficients both for odd and even powers of $\delta_{Rf}$ are preferred in order to allow for a robust estimation of shot noise.
    \item[--] We do not recommend to use marks which have a mean mark very close to zero after correcting for shot noise as measurements get very unstable and uncertainties diverge.
\end{itemize}

\section{Conclusions}\label{sec:conclusion}

In this work, we have studied marked correlation functions in the context of detecting MG and how discreteness effects from estimating marks on a finite point set propagate into the measurement of marked correlation functions. We utilised the Covmos realisations \citep{Baratta2023A&A} that have a particularly high density of points, making them the most suitable for this purpose, and of ELEPHANT simulations with HOD galaxies \citep{Alam2021JCAP} to investigate the discriminating power of marked correlation functions between MG and GR. The latter is comprised of several realisations of GR as well as of $f(R)$ and nDGP gravity theories.
These are two particularly interesting modifications to GR to be studied with marked correlation functions because they exhibit screening mechanisms making the fifth force dependent on the environment.
We proposed several marks based on large-scale environments using the T-Web formalism as well as local density. This includes marks that creates anti-correlation between galaxies in different environments or between galaxies in low- and high-density regions.

For the first time we undertook a thorough investigation of a possible bias due to shot noise in marked correlation functions.
We showed on a toy model that the effect of shot noise can be treated analytically by computation of a small amount of terms. We were able to recover the signal $w_f(\mathbf{r}) = (1+W_f(\mathbf{r}))\bar{m}_f$ from the undepleted catalogue to sub-percent precision.
For general marks, under the assumption that the mark function can be Taylor-expanded in powers of the density contrast, we showed that an analytic treatment is hopeless due to the necessity of computing an infinity of higher-order correlators.
Instead, we developed a methodology for estimating the shot-noise-corrected signal from measuring the weighted correlated function and mean mark in catalogues depleted to different densities.
This is possible by noting the resummation behaviour of shot-noise contributions to $w_f(\mathbf{r}) = (1+W_f(\mathbf{r}))\bar{m}_f^2$ and $\bar{m}_f$ as a power series of the reciprocal of the mean number of points per grid cell $1/\bar{N}$. 
By applying our method to the $\tanh$-mark in the Covmos realisations, we were able to recover an unbiased signal of $1+W_f(\mathbf{r})$ within 5\% accuracy for all tested scales.
We proceeded with an extension of the formalism in redshift space, where we found the same method to be applicable.
Furthermore, we derived a measurable criterion based on the work of \citet{Philcox2020PhRvD} to assess the validity of assuming a Taylor expansion of the mark and provide guidelines for the application of our methodology for shot-noise correction.
We found effects of shot noise mostly on scales below 20-30$\Mpc$ when using the White mark, which might be important for the modelling of marked correlation functions, although the impact on the relative difference between GR and MG appears to be mild.
Moreover, we found that the NGP MAS to give biased results due to higher-order terms in the  $1/\bar{N}$ series being non-negligible. This makes the NGP MAS a sub-optimal choice compared to higher-order schemes, such as PCS.

Equipped with a robust method to recover the true signal in measured marked correlation function, we tested the performance of the previously proposed marks on the ELEPHANT simulations. 
Concerning marks based on the local density, we did not find the White mark to be particularly powerful on large scales.
Only on the very lowest scales, below $20~\Mpc$, we reported significant differences.
In redshift space, the situation changes slightly for the N1 model where we found differences in the quadrupole at $s>20~\Mpc$.
We found that the novel $\tanh$-mark that we introduced is very effective. 
It allows significant differences for $f(R)$-gravity with $\log(|f_{R0}|)=-6$ compared to GR, and uniquely up to scales of 80$\Mpc$.
At those scales, modelling the weighted correlation function is more tractable; making this mark an excellent candidate to test for deviations from GR in real surveys. Furthermore, we found promising results when using the tidal torque as a mark, with significant differences up to scales of $r\approx 60~\Mpc$. The use of anti-correlation together with large-scale environments, as in the $\textrm{Void}_{\textrm{AC}}$- and $\textrm{Wall}_{\textrm{AC}}$-mark, exhibits significant discriminating power for several MG theories on moderate scales.
However, these findings have to be tested with high-density simulations to assess if they are biased by discreteness effects, as our correction method cannot be applied straightforwardly to those types of mark.

In summary, this work demonstrates that correcting for shot noise in marked correlation functions is of paramount importance to measure unbiased amplitudes without being plagued by shot noise, and in turn
to be able to distinguish MG from GR. This is also particularly important for the modelling of the weighted correlation function in the same way as it is for the power spectrum. Generally, we found shot noise to have the strongest impact on small to intermediate scales.
Marks that incorporates an anti-correlation between objects in high- and low-density regions by switching signs in the mark are found to be the most effective for distinguishing between GR from MG, also when using scales beyond $20~\Mpc$. In the future, extending the concept of anti-correlation in weighted correlation functions to different marks could alleviate the current constraint due to the convergence radius of the Taylor expansion as is the case for the $\tanh$-mark.
In general, a consolidation of the $\tanh$-mark performances on improved MG and GR simulations with higher densities would be desired. Moreover, a thorough investigation of a kind of model-independent shot-noise effect on general marked correlation functions would enable an appropriate correction for marks based on large-scale environments or the tidal torque/field, which we showed to be interesting candidates.
A future study should assess the capability of the Lagrangian perturbation theory model to capture the behaviour of our novel mark based on the local density on intermediate scales.
The $\tanh$-mark could then serve as an optimal choice for a weighted clustering analysis in current and future galaxy surveys since no accurate modelling of small scales is required. Having a relevant model of marked correlation functions together with a high-performance mark should add a powerful observable to find the needle in the haystack of gravity theories.

\begin{acknowledgements}
The project leading to this publication has received funding from the Excellence Initiative of Aix-Marseille University - A*MIDEX, a French "Investissements d'Avenir" programme (AMX-19-IET-008 - IPhU).
M. Kärcher is funded by the Excellence Initiative of Aix-Marseille University - A*MIDEX, a French "Investissements d'Avenir" programme (AMX-19-IET-008 - IPhU).
This research made use of \texttt{matplotlib}, a Python library for publication quality graphics \citep{Hunter:2007_matplotlib}.
\end{acknowledgements}

\bibliographystyle{aa}
\bibliography{refs}

\appendix

\section{Marks and cross-correlation}\label{sec:app_cross}
We assume a population of $N$ galaxies that can be split up into a $1$-population and a $2$-population with respective numbers $N_1$ and $N_2$, summing up to $N=N_1+N_2$. 
This split could have been done by using the $
\textrm{Void}_{\textrm{AC}}$ mark where we assign a mark of -1 to all galaxies residing in voids and +1 otherwise.
We assume boxes with periodic boundary conditions, as in the main text, and $RR_n$ denotes the normalised $RR$ counts such that $RR_n=RR/(N_R(N_R-1))$.
Next we define the correlation functions for each population of galaxies as well as the cross-correlation among the two sub-populations to be 
\begin{equation}
\begin{split}
    \xi &= \frac{DD}{N(N-1)RR_n} -1, \\
    \xi_{11} &= \frac{D_1D_1}{N_1(N_1-1)RR_n} -1, \\
    \xi_{22} &= \frac{D_2D_2}{N_2(N_2-1)RR_n} -1, \\
    \textrm{and} \quad \xi_{12} &= \frac{D_1D_2}{N_1N_2RR_n} -1.
\end{split}
\end{equation}
We note here that when we cross-correlate we do not assume double counting hence the normalisation by the total number of possible pairs is only $N_1N_2$.
Terms of the form $D_iD_j$ refer to unnormalised counts of pairs between the $i$- and $j$-population.

The total double counted pair counts can be split up into contributions like 
\begin{equation}
    DD = D_1D_1 + D_2D_2 + 2D_1D_2,
\end{equation}
where the factor of 2 is necessary as the cross-counts are not double counted on their own.
This allows for the following split of the total correlation function
\begin{equation}
\begin{split}
    \xi = & \frac{D_1D_1}{N(N-1)RR_n} + \frac{D_2D_2}{N(N-1)RR_n} \\
    & + 2\frac{D_1D_2}{N(N-1)RR_n} -1 \\
    = & \frac{N_1(N_1-1)}{N(N-1)}\frac{D_1D_1}{N_1(N_1-1)RR_n} + \frac{N_2(N_2-1)}{N(N-1)}\frac{D_2D_2}{N_2(N_2-1)RR_n} \\
    & + 2\frac{N_1N_2}{N(N-1)}\frac{D_1D_2}{N_1N_2RR_n} -1 \\
    = & f_{11} (\xi_{11}+1) + f_{22} (\xi_{22}+1) + 2f_{12}(\xi_{12}+1) -1,
\end{split}
\end{equation}
where we defined
\begin{equation}
    \begin{split}
        f_{11} &= \frac{N_1(N_1-1)}{N(N-1)}, \\
        f_{22} &= \frac{N_2(N_2-1)}{N(N-1)}, \\
        \textrm{and}\quad f_{12} &= \frac{N_1N_2}{N(N-1)}.
    \end{split}
\end{equation}
By realising that $f_{11} + f_{12} + 2f_{12} = 1$ we can finally write 
\begin{equation}
    \xi = f_{11}\, \xi_{11} + f_{22} \,\xi_{22} + 2f_{12} \,\xi_{12}.
\end{equation}

Completely analogous can be proceeded if we consider weighted correlation functions where the mark can only take two values, like in the $\textrm{Void}_{\textrm{AC}}$ mark.
We define the individual weighted correlation functions to be
\begin{equation}
    \begin{split}
        W & = \frac{WW}{(\sum m_i)^2 - \sum m_i^2}\frac{1}{RR_n} - 1, \\
        W_{11} & = \frac{W_1 W_1}{(\sum_1 m_i)^2 - \sum_1 m_i^2}\frac{1}{RR_n} - 1, \\
        W_{22} & = \frac{W_2W_2}{(\sum_2 m_i)^2 - \sum_2 m_i^2}\frac{1}{RR_n} - 1, \\
        \textrm{and} \quad W_{12} & = \frac{W_1W_2}{\sum_1 m_i \sum_2 m_i}\frac{1}{RR_n} - 1,
    \end{split}
\end{equation}
for which we used the shorthand notation 
$\sum = \sum_1 + \sum_2$ to indicate sums over weights belonging solely to galaxies from either population 1 or population 2. The total sum $WW$ over products of weights can be split up into contributions from $W_1W_1$, $W_2W_2$ and $W_1W_2$ in an analogous way as the $DD$ counts in the unweighted case.
Defining prefactors as
\begin{equation}
    \begin{split}
        f^W_{11} & = \frac{(\sum_1m_i)^2 - \sum_1 m_i^2}{(\sum m_i)^2 - \sum m_i^2},\\
        f^W_{22} & = \frac{(\sum_2 m_i)^2 - \sum_2 m_i^2}{(\sum m_i)^2 - \sum m_i^2},\\
        \textrm{and} \quad f^W_{12} & = \frac{\sum_1 m_i \sum_2 m_i}{(\sum m_i)^2 - \sum m_i^2},
    \end{split}
\end{equation}
and also noting that $(\sum w_i)^2 - \sum w_i^2 = (\sum_1 w_i)^2 - \sum_1 w_i^2 + (\sum_2 w_i)^2 - \sum_2 w_i^2 + 2\sum_1 w_i \sum_2 w_i$ we arrive at
\begin{equation}
    W = f^W_{11}\, W_{11} + f^W_{22}\, W_{22} + 2f^W_{12}\, W_{12}.
\end{equation}
This shows that for specific marks the weighted correlation function can be split up into a sum of auto-correlations and a cross-correlation. This can be generalised if the mark e.g. takes three or more different values and the result will include contributions from all possible auto- and cross-correlations.

One particularly interesting case is the $\textrm{Void}_{\textrm{AC}}$ mark where the two values the mark can take is simply $-1$ and $+1$. Let us assume that the 1-population has $-1$ as a mark and the 2-population has $+1$.
First of all we realise that in that case $W_1W_1 = D_1D_1$ as well as $W_2W_2=D_2D_2$ because the sum of pair-product weights will simply be a sum of 1's.
Furthermore $W_1W_2 = -D_1D_2$ as the product of two weights will always be -1 for pairs in the cross-correlation.
The normalisation also simplifies yielding $(\sum_1 w_i)^2 - \sum_1 w_i^2 = N_1(N_1-1)$ and analogous for the 2-population. For the cross-correlation we get $\sum_1 w_i \sum_2 w_i = -N_1N_2$.
Hence the individual auto- and cross-correlations are the same between the weighted and unweighted case $W_{11} = \xi_{11}$, $W_{22} = \xi_{22}$ and $W_{12} = \xi_{12}$. This implies that the total weighted correlation function has the same individual contributions as the unweighted one but with different prefactors
\begin{equation}
    W = f^W_{11} \, \xi_{11} + f^W_{22} \, \xi_{22} + 2 f^W_{12} \, \xi{12},
\end{equation}
since the prefactors simplify to
\begin{equation}
    \begin{split}
        f^W_{11} & = \frac{N_1(N_1-1)}{(N_1 - N_2)^2-N}, \\
        f^W_{22} & = \frac{N_2(N_2-1)}{(N_1 - N_2)^2-N},\\
        \textrm{and} \quad f^W_{12} & = -\frac{N_1N_2}{(N_1 - N_2)^2-N}.
    \end{split}
\end{equation}
If we define the constant $\mathcal{C} = ((N_1+N_2)^2-N)/(N(N-1))$ then we can write the marked correlation function for the $\textrm{Void}_{\textrm{AC}}$-mark as 
\begin{equation}
    \mathcal{M} = \frac{1+\frac{f_{11}}{\mathcal{C}}\, \xi_{11} + \frac{f_{22}}{\mathcal{C}}\, \xi_{22} - 2 \frac{f_{12}}{\mathcal{C}} \, \xi_{12}}{1+f_{11}\, \xi_{11} + f_{22}\, \xi_{22} - 2 f_{12}\,  \xi_{12}}.
\end{equation}
This illustrates the fact that in this case the marked correlation function is nothing else than a specific combination of unweighted auto and cross-correlations.

\section{Convolution of the density contrast}\label{sec:app_SN}
In this section we show how an additional convolution of the already smoothed density contrast can be treated simply as a single convolution with a higher-order smoothing kernel.
Let us start with
\begin{equation}
    \delta_{RR}(\mathbf{x}) = \frac{1}{a^6} \int_{\mathbf{x}''}  \left[\int_{\mathbf{x}'} F\left(\frac{\mathbf{x}''-\mathbf{x}'}{a}\right) \delta(\mathbf{x}') \, \text{d}^3 x'\right] G\left(\frac{\mathbf{x}-\mathbf{x}''}{a}\right) \, \text{d}^3 x''
\end{equation}
where we can identify the smoothed density contrast inside the square brackets and two smoothing kernels $F$ and $G$.
This can be rewritten as an integral over $\mathbf{x}''$ involving only the two kernels with a coordinate transformation $\mathbf{y}=\mathbf{x}-\mathbf{x}''$
\begin{equation}
\begin{split}
    \delta_{RR}(\mathbf{x}) &= \frac{1}{a^6}  \int_{\mathbf{x}'}\left[ \int_{\mathbf{x}''} F\left(\frac{\mathbf{x}-\mathbf{x}'-\mathbf{y}}{a}\right) G\left(\frac{\mathbf{y}}{a}\right) \, \text{d}^3 y\right]  \delta(\mathbf{x}')\, \text{d}^3 x' \\
                & = \frac{1}{a^3} \int_{\mathbf{x}'} \delta(\mathbf{x}') H\left(\frac{\mathbf{x}-\mathbf{x}'}{a}\right) \, \text{d}^3 x',
\end{split}
\end{equation}
where we identified in the first equality that after the coordinate change the two kernels are convolved in the variable $\mathbf{y}$ resulting in a new kernel $H$ at location $\mathbf{x}-\mathbf{x}'$.
Hence the density field is only convolved once with the $H$-kernel that is the convolution of both $G$ and $F$.
Now, if the two kernels are a PCS and NGP kernel, respectively, then the $H$-kernel would be a quartic kernel.

\label{lastpage}
\end{document}